\documentclass{scrartcl}
\usepackage[utf8]{inputenc}
\usepackage{amsmath}
\usepackage{amsfonts}
\usepackage{amssymb}
\usepackage{xcolor}
\usepackage{graphicx}
\usepackage{bbm}
\usepackage{comment}
\usepackage{xcolor}
\usepackage{comment}
\usepackage{bm}
\usepackage{booktabs}
\usepackage[colorlinks=true,linkcolor=blue]{hyperref}
\usepackage{placeins}
\usepackage[ruled,vlined]{algorithm2e}
\usepackage{subcaption}
\usepackage{multirow}

\date{}
\SetKwInput{KwInput}{Input}                
\SetKwInput{KwOutput}{Output}              

\usepackage[backend=bibtex, bibstyle=authoryear, citestyle=authoryear, sorting=nyt,
	natbib=true,
	doi=false,
	isbn=false,
	url=false,
	eprint=false, 
	giveninits=true,
	date=year,
	maxnames = 2, 
	maxbibnames = 99]{biblatex}

\newcommand{\bmu}{\boldsymbol{\mu}}
\newcommand{\bpi}{\boldsymbol{\pi}}
\newcommand{\bomega}{\boldsymbol{\omega}}
\newcommand{\bSigma}{\boldsymbol{\Sigma}}

\title{A Two-Stage Bayesian Semiparametric Model for Novelty Detection with Robust Prior Information\footnote{\url{https://doi.org/10.1007/s11222-021-10017-7}}}

\author{
Francesco Denti\\University of California Irvine\\ 
\texttt{fdenti@uci.edu}
\and 
Andrea Cappozzo\\Politecnico di Milano\\ \texttt{andrea.cappozzo@polimi.it}
\and 
Francesca Greselin\\
University of Milano-Bicocca\\ \texttt{francesca.greselin@unimib.it} 
}

\begin{document}

\maketitle

\begin{abstract}
Novelty detection methods aim at partitioning the test units into already observed and previously unseen patterns. However, two significant issues arise: there may be considerable interest in identifying specific structures within the novelty, and contamination in the known classes could completely blur the actual separation between manifest and new groups. Motivated by these problems, we propose a two-stage Bayesian semiparametric novelty detector, building upon prior information robustly extracted from a set of complete learning units. We devise a general-purpose multivariate methodology that we also extend to handle functional data objects.  We provide insights on the model behavior by investigating the theoretical properties of the associated semiparametric prior. From the computational point of view, we propose a suitable $\boldsymbol{\xi}$-sequence to construct an independent slice-efficient sampler that takes into account the difference between manifest and novelty components. We showcase our model performance through an extensive simulation study and applications on both multivariate and functional datasets, in which diverse and distinctive unknown patterns are discovered.
\par\vskip\baselineskip\noindent
\end{abstract}
\noindent

\section{Introduction}
\label{SEC:Intro}
Supervised classification techniques aim at predicting a qualitative output for a test set by learning a classifier on a fully-labeled training set. To this extent, classical methods assume that the labeled units are realizations from each and every sub-groups in the target population. 
However, many real datasets contradict these assumptions. As an example, one may think about an evolving ecosystem in which novel species are likely to appear over time. In other words, basic classifiers cannot handle the presence of previously unobserved - or hidden - classes in the test set. Novelty detection methods, also known as \emph{adaptive} methods, address this issue by modeling the presence of classes in the test set that have not been previously observed in the training. Relevant examples of this type of data analysis include, but are not limited to, radar target detection \citep{carpenter1997artmap}, detection of masses in mammograms \citep{tarassenko1995novelty}, handwritten digit recognition \citep{tax1998outlier} and e-commerce \citep{manikopoulos2002network}, for which labeled observations may not be available for every group. 

Within the model-based family of classifiers, adaptive methods recently appeared in the literature. \cite{Miller2003} pioneer a mixture model methodology for class discovery. 
\cite{Bouveyron} introduces an adaptive classifier in which two algorithms, based respectively on transductive and inductive learning, are devised for inference. More recently, \cite{fop2018unobserved} extend the original work of \cite{Bouveyron} by accounting for unobserved classes and extra variables in high-dimensional discriminant analysis. 

Classical model-based classifiers are not robust, as they lack the capability of handling outlying observations in the training and in the test set. On the one hand, the presence of outliers in the training set can significantly alter the learning phase, resulting in poorly representative classes and therefore jeopardizing the entire classification process. In the training set, we identify as outliers units with implausible labels and/or values. \cite{Cappozzo2019e} extend the work of \cite{Bouveyron} addressing this problem by using a robust estimator that relies on impartial trimming \citep{Gordaliza1991}. In short, the most unlikely data points under the currently estimated model are discarded. On the other hand, dealing with outliers in the test set is a more delicate task. Ideally, we would like to distinguish between \textit{novelties}, i.e., test observations displaying a common, specific pattern, and \textit{anomalies}, i.e., test observations that can be regarded as noise. While the distinction between novel and anomalous entities is most often apparent in practice, there exist some circumstances under which such separation is vague and somewhat philosophical. Let us go back to the aforementioned evolving ecosystem example. It may happen that at an early instant, a real novelty is mistaken to be mere noise due to its embryonic stage. Contrarily, if we fitted the same model at a later time point, the increased sample size could be sufficient to acknowledge an actual novel species.

To address the discussed challenges, we propose a two-stage Bayesian semiparametric novelty detector. We devise our model to sequentially handle the outliers in the training set and the latent classes in the test set. In the first stage, we learn the main characteristics of the known classes (for example, their mean and variance) from the labeled dataset using robust procedures. 
In the second phase, we fit a Bayesian semiparametric mixture of known groups and a novelty term to the test set. We use the training insights to elicit informative priors for the known components, modeled as Gaussian distributions. The novelty term is instead captured via a flexible Dirichlet Process mixture: this modeling choice reflects the lack of knowledge about its distributional properties and overcomes the problematic and unnatural a priori specification of its number of components.
We call our proposal Bayesian Robust Adaptive model for Novelty Detection, hereafter denoted as Brand.
Essentially, Stage II of Brand is formed by two nested mixtures, which can provide uncertainty quantification regarding the two partitions of interest. 
First, Brand separates the entire test set into known components and  a novelty term. Secondly, the novel data points can be a posteriori clustered  into different sub-components. ``True novelties'' and anomalies may be distinguished, based on clusters cardinality.  

The rest of the article proceeds as follows. In Section \ref{Sec::Model} we present our two-stage methodology for novelty detection. We dedicate Section \ref{sec::Properties} to the investigation of the random measures clustering properties induced by our model. In Section \ref{sec:func_version}, we propose an extension of the multivariate model, delineating a novelty detection method suitable for functional data. Section \ref{Sec::PostInf} discusses posterior inference, while in Section \ref{sec:applications} we present an extensive simulation study and applications to multivariate and functional data. Concluding remarks and further research directions are outlined in Section \ref{sec:conclusions}.

\section{A Two-Stage Bayesian procedure for Novelty Detection}
\label{Sec::Model}
Given a classification framework, consider the complete collection of learning units $\mathbf{X}=\{(\mathbf{x}_n, \mathbf{l}_n)\}_{n=1}^N$, where $\mathbf{x}_n$ denotes a $p$-variate observation and $\mathbf{l}_{n}=j \in \{1,\ldots, J\}$ its associated group label. Both terms are directly available and the distinct values in $\mathbf{l}_{n}$, $n=1,\ldots,N$ represent the $J$ \emph{observed classes} with subset sizes $n_1,\ldots,n_J$. Correspondingly, let $\bm{Y} = \{(\mathbf{y}_m, \mathbf{z}_m)\}_{m=1}^M$ be the test set where, differently from the usual setting in semisupervised learning, the unknown labels $\mathbf{z}_{m}$ could belong to a set that encompasses more elements than \{1,\ldots, J\}. That is, a countable number of extra classes may be ``hidden'' in the test with no prior information available on their magnitude or on their structure. Therefore, it is reasonable to account for the novelty term via a single flexible component from which a dedicated post-processing procedure may reveal circumstantial patterns (see Section \ref{Disentangling}). Both $\mathbf{x}_n$ and $\mathbf{y}_m$ are independent realizations of a continuous random vector (or function, see Section \ref{sec:func_version}) $\mathcal{X}$, whose conditional distribution varies according to the associated class labels. In the upcoming Sections, we assume that each observation in class $j$ is independent multivariate Gaussian, having density $\phi\left(\cdot|\boldsymbol{\Theta}_j\right)$ with location-scale parameter $\boldsymbol{\Theta}_j=\left(\boldsymbol{\mu}_j,\boldsymbol{\Sigma}_j\right)$, where $\boldsymbol{\mu}_j\in \mathbb{R}^{p}$ denotes the mean vector and $\mathbf{\Sigma}_j$ the corresponding covariance matrix. This allows for the automatic implementation of standard powerful methods in the training information extraction (see Section \ref{sec:stageI}). Notwithstanding, the proposed methodology is general enough that it can be easily extended to deal with different component distributions.

\begin{figure*}[ht!]
    \centering
    \includegraphics[scale=.5]{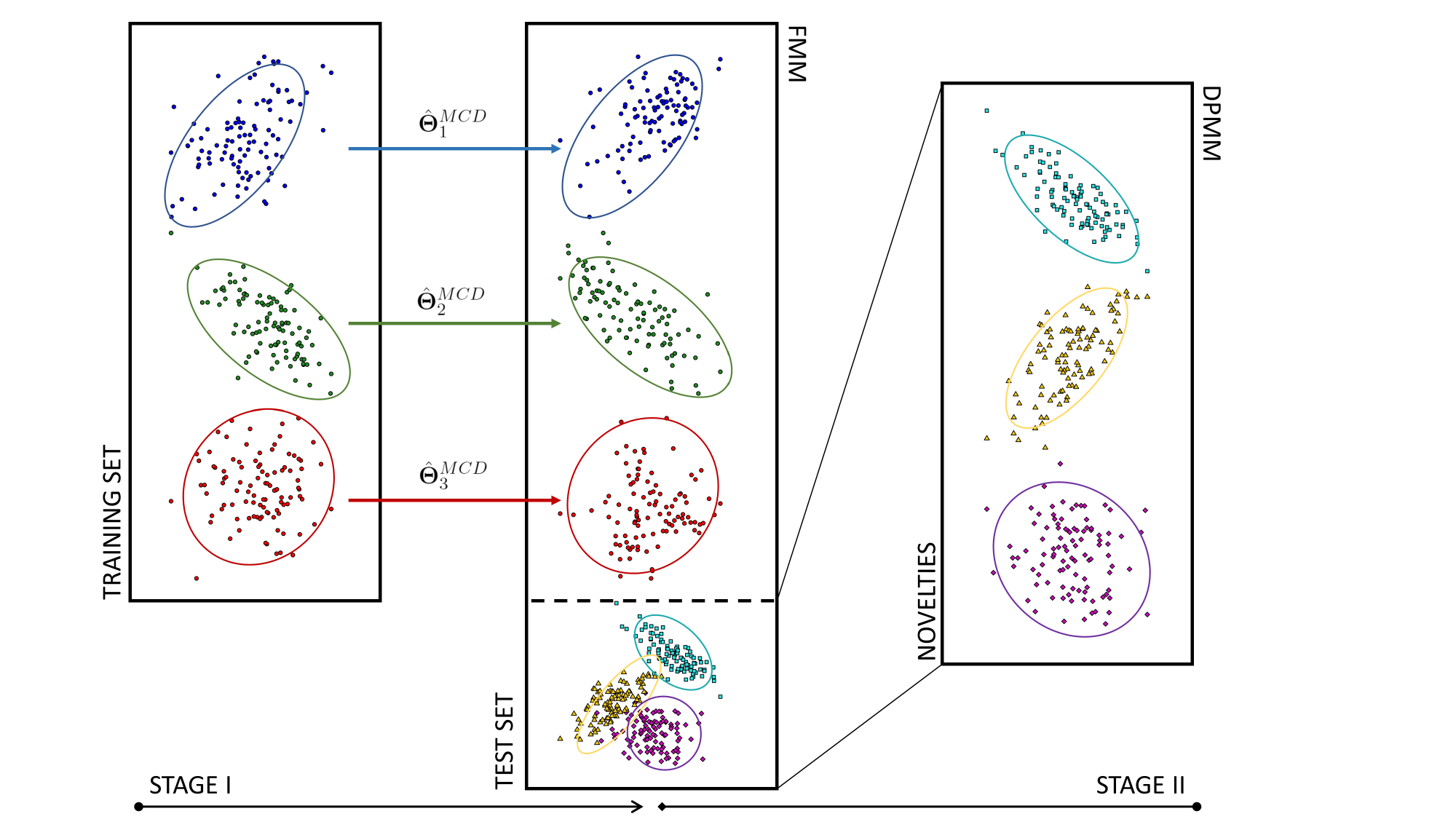}
    \caption{A diagram that summarizes Brand two-stage structure. In Stage I, robust information extraction (via the MCD estimator) is performed and subsequently used to elicit the priors for the mixture model in Stage II. In the second stage, a finite mixture model (FMM) is fitted to the data, distinguishing among known components and novelties. The novelty term is modeled with a Dirichlet Process mixture model (DPMM).}
    \label{fig:diagram}
\end{figure*}

Our modeling purpose is to classify the data points of the test set either into one of the $J$ observed classes or into the novel component. At the same time, we investigate the presence of homogeneous groups in the novelty term, discriminating between unseen classes and outliers. To do so, we devise a two-stage strategy. The first phase, described in Section \ref{sec:stageI}, relies on a class-wise robust procedure for extracting prior information from the training set. Then,  we fit a Bayesian semiparametric mixture model to the test units. A full account of its definition is reported in Section \ref{sec:phase_II}. A diagram summarizing our modeling proposal is reported in Figure \ref{fig:diagram}.

\subsection{Stage I: Robust extraction of prior information} \label{sec:stageI}

The first step of our procedure is designed to obtain reliable estimates $\hat{\boldsymbol{\Theta}}_j$ for the parameters of the observed class $j$, $j=1\ldots,J$, from the learning set. To this aim, one could employ standard methods as the maximum likelihood estimator, or different posterior estimates under the Bayesian framework. Nonetheless, these standard approaches are not robust against contamination, and the presence of only a few outlying points could entirely bias the subsequent Bayesian model, should the informative priors be improperly set. We report a direct consequence of this undesirable behavior in the simulation study of Section \ref{sec:sim_study}. Therefore, we opt for more sophisticated alternatives to learn the structure of the known classes, employing methods that can deal with outliers and label noise. Particularly, the selected methodologies involve the Minimum Covariance Determinant (MCD) estimator \citep{rousseeuw1984least, Hubert2018} and, when facing high-dimensional data (as in the functional case of Section \ref{sec:func_application}), the Minimum Regularized Covariance Determinant (MRCD) estimator \citep{Boudt2020}. Clearly, at this stage, one can use any robust estimators of multivariate scatter and location for solving this problem: see, for instance, the comparison study reported in \citet{Maronna2017} for a non-exhaustive list of suitable candidates.

We decide to rely on the MCD and MRCD for their well-established efficacy in the classification framework \citep{Hubert2004} and direct availability of fast algorithms for inference, readily implemented in the \texttt{rrcov R} package \citep{Todorov2009a}. We briefly recall the main MCD and MRCD features in the remaining part of this section. For a thorough treatment the interested reader is referred to \cite{hubert2010minimum} and \cite{Boudt2020}, respectively.  

The MCD is an affine equivariant and
highly robust estimator of multivariate location and
scatter, for which a fast algorithm is available \citep{Driessen1999}. The raw MCD estimator with parameter $\eta^{MCD} \in [0.5,1]$ such that $\lfloor (n + p + 1)/2\rfloor \leq \lfloor \eta^{MCD} N\rfloor \leq N$ defines the
following location and dispersion estimates:
\begin{itemize}
\item $\hat{\bmu}^{MCD}$ is the mean of the $\lfloor \eta^{MCD} N\rfloor$ observations for which
the determinant of the sample covariance matrix
is minimal
\item $\hat{\bSigma}^{MCD}$ is the  corresponding covariance matrix,
multiplied by a consistency factor $c_0$ \citep{Croux1999}
\end{itemize} 
with $\lfloor \cdot \rfloor$ denoting the floor function. The MCD is a consistent, asymptotically normal and highly robust estimator with bounded influence function and breakdown value equal to $(1-\lfloor \eta^{MCD} N\rfloor/N) \%$ \citep{Butler1993, Cator2012}. However, a major drawback is its inapplicability when the data dimension $p$ exceeds the subset size $\lfloor \eta^{MCD} N\rfloor$ as the covariance matrix of any $\lfloor \eta^{MCD} N\rfloor$-subset becomes singular. This situation appears ever so often in our context, as the MCD is group-wise applied to the observed classes in the training set, such that it is sufficient to have 
\begin{equation*}
p>\min_{n_j,j=1,\ldots,J}\lfloor \eta^{MCD} n_j\rfloor
\end{equation*}
for the MCD solution to be ill-defined. To overcome this issue, \cite{Boudt2020} introduced the MRCD estimator. The main idea is to replace the subset-based covariance estimation with a regularized one, defined as a weighted average of the sample covariance on the $\lfloor  \eta^{MCD} N\rfloor$-subset and a predetermined positive definite target matrix. The MRCD estimator is defined as the multivariate location and regularized scatter based on the $\lfloor  \eta^{MCD} N\rfloor$-subset that makes its overall determinant the smallest. The MRCD preserves the good breakdown properties of its non-regularized counterpart, and besides, it is applicable in high-dimensional problems where $\lfloor  \eta^{MCD} N\rfloor$ is possibly smaller than $p$.

The first phase of our two-stage modeling thus works as follows: considering the available labels $\mathbf{l}_{n}$, $n=1,\ldots,N$ we apply the MCD (or MRCD) estimator within each class to extract  $\hat{\bmu}_j^{ MCD}$ and $\hat{\bSigma}_j^{MCD}$, $j=1\ldots,J$. For ease of notation, we use superscript `MCD' for the robust estimates even when we consider its regularized version. In general, if the sample size is large enough, the MCD solution is preferred. There is no reason for $\eta^{MCD}$ to be the same in all observed classes. If a group is known a priori to be particularly outliers-sensitive, one should set its associated MCD subset size to a smaller value than the remaining ones. However, since this type of information is seldom available, we subsequently let $\eta_j^{MCD}=\eta^{MCD}$ for all classes in the learning set. This concludes the first stage: the retained estimates are then incorporated in the Bayesian model for the second stage, presented in Section \ref{sec::SubSections2}. The robust knowledge extracted from $\mathbf{X}$ is treated as a source of reliable prior information, eliciting informative hyperparameters. 
In this way, outliers and label noise that might be present in the labeled units will not bias the initial beliefs for the known groups in the second stage, which is the main methodological contribution of the present manuscript.

\subsection{Stage II: BNP novelty detection in test data} \label{sec:phase_II}
\label{sec::SubSections2}
We assume that each observation in the test set is generated according to a mixture of $J+1$ elements: $J$ multivariate Gaussians  $\phi(\cdot| \boldsymbol{\Theta}_j)$ that have been observed in the learning set, and an extra term $f^{\,nov}$ called \emph{novelty} component. 
In formulas:

\begin{equation}
	\mathbf{y}_m|\boldsymbol{\pi},\boldsymbol{\Theta}_j,f^{nov} \sim \sum_{j=1}^J \pi_j \phi(\cdot| \boldsymbol{\Theta}_j) + \pi_0 f^{\,nov}.
\end{equation}
We define $\boldsymbol{\pi}=\{\pi_j\}_{j=1}^J$, where $\pi_j$ denotes the prior probability of the observed class $j$ (already present in the learning set), while $\pi_0$ is the probability of observing some novelty. Of course, $\sum_{j=0}^J\pi_j=1$. To reflect our lack of knowledge on the novelty component $f^{\,nov}$, we employ a Bayesian nonparametric specification. In particular, we resort to the Dirichlet Process mixture model (DPMM) of Gaussians densities \citep{Lo1984,Escobar1995} imposing the following structure:
\begin{equation}
	f^{\,nov} = \int \phi(\cdot| \boldsymbol{\Theta}^{nov}) G(d\boldsymbol{\Theta}^{nov}), \quad  G \sim DP(\gamma,H),
\end{equation} 
where $DP(\gamma,H)$ denotes a Dirichlet Process with concentration parameter $\gamma$ and base measure $H$ \citep{Ferguson1973}. Adopting Sethuraman's Stick Breaking construction \citep{Sethuraman1994a}, we can express the likelihood as follows:
\begin{align}
\begin{split}
\mathcal{L}(\mathbf{y}|\bpi,\bmu, \bSigma,\bomega)=&\prod_{m=1}^M \left[\sum_{j=1}^J \pi_j \phi\left(\mathbf{y}_m|\bmu_j,\bSigma_j \right)+\pi_0\sum_{h=1}^\infty \omega_h \phi\left(\mathbf{y}_m|\bmu^{nov}_h,\bSigma^{nov}_h\right)\right].
\end{split}
\label{eq::lik}
\end{align}
The term $\sum_{h=1}^\infty \omega_h \phi\left(\cdot|\mathbf{\Theta}^{nov}_h \right)$ represents a Dirichlet Process realization convoluted with a Normal kernel, for flexibly modeling a potentially infinite number of hidden classes and/or outlying observations. The following prior probabilities for the parameters complete the Bayesian model specification:  
\begin{align} \label{eq:prior_spec}
\begin{split}
\bm{\Theta}_j &= \left(\bmu_j, \bSigma_j \right) \sim P_j^{Tr},\quad\quad\quad\quad j=1,\ldots,J,\\ 
\bm{\Theta}^{nov}_h&=\left(\bmu^{nov}_h, \bSigma^{nov}_h \right) \sim H, \quad\quad\quad h=1,\ldots,\infty, \\
\bpi &\sim Dir\left(a_0,a_1,\ldots,a_J \right), \quad \quad
\bomega \sim SB\left(\gamma\right).
\end{split}
\end{align}
Values $a_1, \ldots, a_J$ are the hyper-parameters of a Dirichlet distribution on the known classes. We can exploit the learning set to determine reasonable values of such hyper-parameters by setting $a_j=n_j/N$. 
The quantity $a_0$ determines the initial prior belief on how much novelty we are expecting to discover in the test set. Generally, the parameter controlling the novelty proportion  $a_0$ is a priori considered to be small. We adopt a conjugate Normal-inverse-Wishart (NIW) prior for both the location-scale parameters of the manifest and the novel classes. For each of the known groups, we assume that 
$$ P_j^{Tr} \equiv NIW\left(\hat{\bmu}_j^{MCD},\lambda^{Tr},\nu^{Tr},\hat{\bSigma}_j^{MCD} \right), \quad j=1,\ldots,J
$$
where $\hat{\bmu}_j^{MCD}$ and $\hat{\bSigma}_j^{MCD}$ are the MCD robust estimates obtained in Stage I.
At the same time, the precision parameter ${\lambda}^{Tr}$ and the degrees of freedom ${\nu}^{Tr}$ are treated as tuning parameters to enforce high mass around the robust estimates. By letting these two parameters go off to infinity, we can also recover the degenerate case $P_j^{Tr}=\delta_{\hat{\bm{\Theta}}_j}$ where the Dirac's delta denotes a point mass centered in $\hat{\bm{\Theta}}_j$. That is, the prior beliefs extracted from the training set can be flexibly updated by gradually transitioning from transductive to inductive inference by increasing ${\lambda}^{Tr}$ and ${\nu}^{Tr}$ \citep{Bouveyron}.
Similarly, we set $H \equiv NIW\left(\bm{m}_0,{\lambda_0},{\nu_0},\bm{S}_0 \right),$ where the hyperparameters are chosen to induce a flat prior for the novel components. Lastly, with $\bomega \sim SB\left(\gamma\right)$ we denote the vector of Stick-Breaking weights, composed of elements defined as
\begin{equation}
w_k=v_k\prod_{l<k}(1-v_l), \quad v_k \sim Beta(1,\gamma).
\label{eq::SB}
\end{equation}
It is well known that, under the DP specification, the expected number of clusters induced in the novelty term grows as $\gamma \log M$. We choose the DP mostly for computational convenience: if more flexibility is required, Brand can easily be adapted to accommodate 
different nonparametric priors, such as the Pitman-Yor process \citep{Pitman1995,Pitman1997} or the geometric process and its extensions \citep{DeBlasi2020}.
To facilitate posterior inference given the specification in \eqref{eq:prior_spec}, we consider the following \textit{complete likelihood:}
\begin{equation}
\begin{aligned}
\mathcal{L}(\mathbf{y},\boldsymbol{\alpha}, \boldsymbol{\beta}|\bpi,\boldsymbol{\Theta},\bomega)=
&\prod_{m=1}^M \left[ \pi_{\alpha_m}\mathbbm{1}_{\{\alpha_m > 0 \:\cap\: \beta_m =0\}}+\right.\\&\left.+
\pi_{0}\mathbbm{1}_{\{\alpha_m = 0 \:\cap\: \beta_m >0)\}}\omega_{\beta_m}
\right]\times \\ &\times \phi\left(\mathbf{y}_m|\mathbf{\Theta}^*_{m} 
\right),\\
\mathbf{\Theta}^*_{m}=&\mathbf{\Theta}_{\alpha_m}\mathbbm{1}_{\{\beta_m=0\}} + \mathbf{\Theta}^{nov}_{\beta_m}\mathbbm{1}_{\{\alpha_m=0\}}
\end{aligned}
\label{eq::model}
\end{equation}
where $\alpha_m \in \{0,\ldots,J \}$ and $\beta_m \in \{0,\ldots,\infty \}$ are latent variables identifying the unobserved group membership for $\mathbf{y}_m$, $m=1,\ldots, M$. In details, the former identifies whether observation $m$ is a novelty $(\alpha_m=0)$ or not $(\alpha_m>0)$, whereas the latter defines, within the novelty subset, the resulting data partition $(\beta_m>0)$. To complete the specification, we set $\omega_0=1$.\\
Lastly, we want to underline that there might be some cases where the number of novelty groups is known to be bounded and does not grow with the sample size as in the DP case. In those situations, an appealing alternative to the DPMM is the Sparse Mixture Model, studied by \citet{Rousseau2011} and recently investigated in
\citet{Malsiner-Walli2016,}.

\subsection{Distinguishing novelties from anomalies}
\label{Disentangling}
The advantage of employing a DPMM for the novelty part is twofold: on the one hand, all the data coming from unseen components are modeled with a unique, flexible density. On the other hand, the clustering naturally induced by the DPMM favors the separation of the novelty component into actual unseen classes and outlying units.
More specifically, since the concept of an outlier does not possess a rigorous mathematical definition \citep{Ritter2015}, the estimated sample sizes of the discovered classes act as an appropriate feature for discriminating between scattered outlying units and actual hidden groups. 
That is, if a component $\phi\left(\cdot|\boldsymbol{\Theta}^{nov}_h \right)$ fits only a small number of data points, we can regard those units as outliers. Similarly, we assume to have discovered an extra class whenever it possesses a substantial structure. In real applications, domain-expert supervision will always be crucial for class interpretation when extra groups are believed to have been detected. While the mixture between known and novel distributions is identifiable and not subjected to the label switching problem, the same cannot be said about the DP component modeling the novelty density. To recover a meaningful estimate for the partition of points regarded as novel ($\beta_m >0$) we first compute the pairwise coclustering matrix $\mathcal{P}=\{p_{m,m'}\}$, whose entry $p_{m,m'}$ denotes the probability that $\bm{y}_m$ and $\bm{y}_{m'}$ belong to the same cluster. We then retrieve the best partition minimizing the Variation of Information (VI) criterion, as suggested in \citet{Wade2018}. More details on how to post-process the MCMC output are given in Section \ref{Sec::PostInf}.

\section{Properties of the proposed semiparametric prior}
\label{sec::Properties}

We now investigate the properties of the underlying random mixing measure induced by the model specification we presented in the previous section. All the proofs are deferred to the Supplementary Material. We start by noticing that model in \eqref{eq::lik}-\eqref{eq:prior_spec} can be generalized in the following hierarchical form, which highlights the dependence on a discrete random measure $\tilde{p}$:
\begin{equation}
\begin{aligned}    
\bm{y}_m|\boldsymbol{\Theta}_m &\sim N(\boldsymbol{\Theta}_m) \quad \boldsymbol{\Theta}_m|\tilde{p} \overset{i.i.d.}{\sim} \tilde{p}\\
    \tilde{p}&= \sum_{j=1}^J \pi_j \delta_{\boldsymbol{\Theta}_j}+\pi_0\left[\sum_{h=1}^{+\infty}\omega_h \delta_{\boldsymbol{\Theta}^{nov}_h}\right]\\
    \left(\pi_0,\pi_1,\ldots,\pi_J\right) &\sim Dir(a_0,a_1,\ldots,a_J) \quad \bm{\omega}\sim SB(\gamma)\\
    \boldsymbol{\Theta}_j &\sim P_j^{Tr} \quad \quad \boldsymbol{\Theta}_h^{nov} \sim H.
\end{aligned}
\label{eq:modptilde}
\end{equation}
From \eqref{eq:modptilde} we can see how our model is an extension of the contaminated informative priors proposed in \citet{Scarpa2009}, where the authors propose to juxtapose a single atom to a DP.
To simplify the exposition of the results in this section, without loss of generality, we assume that both $\boldsymbol{\Theta}_j$ and $\boldsymbol{\Theta}_h^{nov}$ are univariate random variables. Consequently, we suppose that each $P_j^{Tr}$ is a probability distribution with mean $\mu_{j},$ second moment $\mu_{j,2}$ and variance $\sigma^2_{j}$, $j=1,\ldots,J$. Similarly, let $\mathbb{E}\left[ \boldsymbol{\Theta}^{nov}_h \right] = \mu_0$,
$\mathbb{V}\left[ \boldsymbol{\Theta}^{nov}_h \right] = \sigma^2_0 \:\: \forall h \geq1$ and $a=\sum_{j=0}^Ja_j$.
For all $m \in \{1,\ldots,M\}$, we can prove that
\begin{equation*}
\begin{aligned}
    \mathbb{E}\left[\boldsymbol{\Theta}_m\right]=\sum_{j=0}^J\frac{a_j}{a}\mu_{j},\quad \quad
    \mathbb{V}\left[\boldsymbol{\Theta}_m\right]=
    \sum_{j=0}^J
    \frac{a_j}{a}\left(\mu_{j,2}-\frac{a_j}{a}\mu_{j}^2\right)-2\sum^J_{l>j\geq 0}\frac{a_ja_l}{a^2}\mu_l\mu_j.
\end{aligned}
\end{equation*}
The overall variance can also be written in terms of variances of every observed mixture components:
\begin{equation*}
    \mathbb{V}\left[\boldsymbol{\Theta}_m\right]=
    \sum_{j=0}^J
    \frac{a_j}{a}\left(
    \sigma_j^2+\left(1-\frac{a_j}{a}\right)\mu_j^2\right)
    -2\sum^J_{l>j\geq 0}\frac{a_ja_l}{a^2}\mu_l\mu_j.
\end{equation*}
The previous expressions are important to compute the covariance between the two random elements $\boldsymbol{\Theta}_{m}$ and $\boldsymbol{\Theta}_{m'}$, which helps to understand the behavior of $\tilde{p}$. Consider a vector $\bm{\varrho}=\{\varrho_j\}_{j=0}^{J}$, with the first entry equal to $\frac{1}{1+\gamma}$ and the remaining entries equal to $1$. Then, 
\begin{equation}
\begin{aligned}
Cov_{\gamma}(\boldsymbol{\Theta}_m,\boldsymbol{\Theta}_{m'})&=\sum_{j=0}^J\left(\frac{a_j(a_j+1)}{a(a+1)}\varrho_j \mu_{j,2}-\frac{a^2_j}{a^2}\mu_j^2\right)+\\ &-\frac{2}{a^2(a+1)}\sum^J_{j>l\geq0}a_ja_l\mu_j\mu_l+\\ &+\frac{a_0(a_0+1)}{a(a+1)}\frac{\gamma}{1+\gamma}\mu_0^2.
\end{aligned}
\label{eq:cov}
\end{equation}
The covariance is composed of three terms. In the first two, the seen and unseen components have the same influence. The last term is non-negative and entirely determined by quantities linked to the novel part of the model. Notice that if $\gamma\rightarrow 0$ the covariance becomes
\begin{align*}
Cov_{0}(\boldsymbol{\Theta}_m,\boldsymbol{\Theta}_{m'})&=\sum_{j=0}^J\left(\frac{a_j(a_j+1)}{a(a+1)} \mu_{j,2}-\frac{a^2_j}{a^2}\mu_j^2\right)- \frac{2}{a^2(a+1)}\sum^J_{j>l\geq0}a_ja_l\mu_j\mu_l 
\end{align*}
which is the same covariance we would obtain if $\tilde{p}=\tilde{p}_0 \equiv \sum_{j=0}^J \pi_j \delta_{\boldsymbol{\Theta}_j}$, i.e., if we were dealing with a ``standard'' mixture model with $J+1$ components. This implies that \eqref{eq:cov} can be rewritten as
\begin{align*}
Cov_{\gamma}(\boldsymbol{\Theta}_m,\boldsymbol{\Theta}_{m'})=Cov_{0}(\boldsymbol{\Theta}_m,\boldsymbol{\Theta}_{m'})-
\frac{a_0(a_0+1)}{a(a+1)}\frac{\gamma}{1+\gamma}\sigma_0^2,
\end{align*}
which leads to a nice interpretation. The introduction of novelty atoms decreases the ``standard'' covariance. This effect gets stronger as the prior weight given to the novelty component $a_0$, the dispersion of the base measure $\sigma_0^2$ and/or the concentration parameter $\gamma$ increases.

Given the discrete nature of $\tilde{p}$, we can expect ties between realizations sampled from this measure, say $\boldsymbol{\Theta}_m$ and $\boldsymbol{\Theta}_{m'}$. Therefore, we can compute the probability of obtaining a tie as:
\begin{equation}
\begin{aligned}
\mathbb{P}\left(\boldsymbol{\Theta}_{m}=\boldsymbol{\Theta}_{m'}\right)=&
\sum_{j= 1}^{J}
\frac{a_j(a_j+1)}{a(a+1)}+\frac{a_0(a_0+1)}{a(a+1)}\cdot\frac{1}{1+\gamma}
=\sum_{j= 0}^{J}
\frac{a_j(a_j+1)}{a(a+1)}\varrho_j,
\label{eq::tie}
\end{aligned}
\end{equation}
where the contribution to this probability of the novelty terms is multiplicatively reduced by a factor that depends on the inverse of the concentration parameter. 
If a priori we expect a large number of clusters in the novelty term (large $\gamma$), the probability of a tie reduces. Indeed, some noticeable limiting cases arise:
\begin{align*}
    \lim_{\gamma\rightarrow+\infty} \mathbb{P}\left(\boldsymbol{\Theta}_{m}=\boldsymbol{\Theta}_{m'}\right)=\sum_{j= 1}^{J}
\frac{a_j(a_j+1)}{a(a+1)}, \quad \quad
    \lim_{\gamma\rightarrow 0} \mathbb{P}\left(\boldsymbol{\Theta}_m=\boldsymbol{\Theta}_{m'}\right) =\sum_{j=0}^{J}
\frac{a_j(a_j+1)}{a(a+1)}.
\end{align*}
If $\gamma \rightarrow 0$ we obtain a finite mixture of $J+1$ components. Conversely, $\gamma\rightarrow+\infty$ leads to the case of a DP with numerous atoms characterized by similar probability, hence annihilating the contribution to the probability of the novelty term. Moreover, suppose we rewrite the distribution of $\bm{\pi}$ as $Dir\left(\frac{a_0}{J+1},
\frac{\tilde a}{J+1},\ldots,\frac{\tilde a}{J+1}
\right)$. In this case, the hyperparameters relative to the observed groups are assumed equal to $\tilde{a}$. Then, we obtain $a= \frac{a_0 + J\tilde a }{J+1}$, and

\begin{equation}
\begin{aligned}
\mathbb{P}\left(\boldsymbol{\Theta}_m=\boldsymbol{\Theta}_{m'}\right)=&
\frac{
  \frac{J\tilde a}{J+1} 
  \left(
  \frac{\tilde a}{J+1}+1
  \right)}{
  \frac{a_0 + J\tilde a }{J+1}
  \left(
  \frac{a_0 + J\tilde a }{J+1}+1
  \right)
  }+\frac{\frac{a_0}{J+1}\left(\frac{a_0}{J+1}+1\right)}{  \frac{a_0 + J\tilde a }{J+1}
  \left(
  \frac{a_0 + J\tilde a }{J+1}+1
  \right)}\cdot\frac{1}{1+\gamma}.
\label{tieJ}
\end{aligned}
\end{equation}
As $J$ increases, the second part of \eqref{tieJ} vanishes. Accordingly, if we suppose an unbounded number of observed groups letting $J\rightarrow\infty$, then we have \[\mathbb{P}\left(\boldsymbol{\Theta}_m=\boldsymbol{\Theta}_{m'}\right)=1/(1+\tilde a)\]
as in the classical DP case, and the model loses its ability to detect novel instances.\\

\section{Functional Novelty Detection} \label{sec:func_version}

The modeling framework introduced in Section \ref{Sec::Model} is very general and can be easily modified to handle more complex data structures. In this section, we develop a methodology for functional classification that allows novelty functional detection, building upon model \eqref{eq::lik}-\eqref{eq:prior_spec}. We hereafter assume that our training and test instances are error-prone realizations of a univariate stochastic process $\mathcal{X}(t)$, $t \in \mathcal{T}$ with $\mathcal{T} \subset \mathbb{R}$.

Recently, numerous authors have contributed to the area of Bayesian nonparametric functional clustering \citep[see, for example ][]{Bigelow2009,Petrone2009,Rodriguez2014,Rigon}.
\citet{canale2017} propose a Pitman-Yor mixture with a spike-and-slab base measure to effectively model the daily basal body temperature in women by including the a priori known distinctive biphasic trajectory that characterizes healthy beings. Instead of modifying the base measure of the nonparametric process, \citet{Scarpa2009} address the same problem by contaminating a point mass with a realization from a DP.  As such, part of our method can be seen as a direct extension of the latter, where $J\geq 1$ different atoms centered in locations learned from the training set are contaminated with a DP.

Let $\boldsymbol{\Theta}_m(t) = \left(f_m(t),\sigma^2_m(t)\right)$ denote the vector comprising the smooth functional mean $f_m: \mathcal{T}\rightarrow \mathbb{R}$ and the measurement noise $\sigma^2_m:\mathcal{T}\rightarrow \mathbb{R}^{+}$ for a generic curve $m$ in the test set, evaluated at the instant $t$. Then the Brand model, introduced in Section \ref{sec:phase_II} for multivariate data, can be modified as follows:

\begin{equation} \label{eq:func_model}
\begin{aligned}
y_m(t)|\boldsymbol{\Theta}_m(t) &= f_m(t) + \varepsilon_m(t);  \quad  \varepsilon_m(t) \sim N(0,\sigma_m^2(t))\\
\boldsymbol{\Theta}_m(t)|\tilde{p} &\sim \tilde{p},  \quad
\tilde{p} = \sum_{j=1}^J \pi_j \delta_{\boldsymbol{\Theta}_j}+\pi_0\left[\sum_{h=1}^{+\infty}\omega_h \delta_{\boldsymbol{\Theta}^{nov}_h}\right],\\
\left(\pi_0,\pi_1,\ldots,\pi_J\right) &\sim Dir(a_0,a_1,\ldots,a_J), \quad \bm{\omega}\sim SB(\gamma),\\
\boldsymbol{\Theta}_j &\sim P_j^{Tr}, \quad \quad \boldsymbol{\Theta}_h^{nov} \sim H,
\end{aligned}
\end{equation}
where all the distributions $P_j^{Tr}$ and the base measure $H$ model the functional mean and the noise independently. We propose the following informative prior for $\bm{\Theta}_j=\left(f_{j}(t),\sigma^2_{j}(t)\right)$:
\begin{equation}
    \begin{aligned}
    f_{j}(t)\overset{ind.}{\sim}& N\left(\bar{f}_{j}(t), \varphi_{j} \right),\\
    \sigma^2_{j}(t) \overset{ind.}{\sim}& IG\left(2+ \frac{\left(\bar{\sigma}^2_{j}(t)\right)^2}{v_j}, \bar{\sigma}^2_{j}(t)\left(1+\frac{\left(\bar{\sigma}^2_{j}(t)\right)^2}{v_j} \right)\right).
    \end{aligned}
\end{equation}
We denote the estimates obtained from the training set of the mean and variance functions, as $\bar{f}_{j}$ and $\bar{\sigma}^2_{j}$, respectively, for each observed class $j$, with $j=1,\ldots,J$. The hyper-parameters $\varphi_j$ define the degree of confidence we a priori assume for the information extracted from the learning set, while the Inverse Gamma ($IG$) specification ensures that $\mathbb{E}\left[\sigma^2_{j}(t)\right] = \bar{\sigma}^2_{j}(t)$ and $Var\left[\sigma^2_{j}(t)\right] = v_j$. It remains to define how we compute $\bar{f}_{j}$ and $\bar{\sigma}^2_{j}$, that is, how the robust extraction of prior information is performed in this functional extension. Applying standard procedures in Functional Data Analysis \citep{RamsayJamesSilverman}, we first smooth each training curve $x_n(t)$ via a weighted sum of $B$ basis functions
\begin{equation*}
x_n(t)\approx \sum_{b=1}^{B} \rho_{nb} \phi_{b}(t) \quad n=1,\ldots,N
\end{equation*}
where $\phi_{b}(t)$ is the $b$-th basis evaluated in $t$ and $\rho_{nb}$ its associated coefficient. Given the acyclic nature of the functional objects treated in Section \ref{sec:func_application}, we will subsequently employ B-spline bases \citep{Boor2001}. Clearly, depending on the problem at hand, other basis functions may be considered.
After such representation has been performed, we are left with $J$ matrices of coefficients each of dimension $n_j \times B$. By treating them as multivariate entities, as done for example in \cite{Abraham2003}, we resort to the very same procedures described in Section \ref{sec:stageI}, and we set
\begin{equation*}
\begin{aligned}
\bar{f}_{j}(t) &= \sum_{b=1}^{B} \hat{\rho}^{MCD}_{jb} \phi_{b}(t),\quad \quad  \bar{\sigma}^2_{j}(t) = \frac{1}{n_j-1}\sum_{n: \mathbf{l}_{n}=j\: \cap \:\mathcal{I}^{(j)}_{MCD}} \left(x_n(t)-\bar{f}_{j}(t) \right)^2
\end{aligned}
\end{equation*}
where $\hat{\rho}^{MCD}_{jb}$ is the robust location estimate on the $n_j \times B$ matrix of coefficients, and $\mathcal{I}^{(j)}_{MCD}$ denotes the subset of untrimmed units resulting from the MCD/MRCD procedure in group $j$, $j=1,\ldots,J$.
On the other hand, more flexibility is needed to specify the base measure $H$ for $ \boldsymbol{\Theta}_h^{nov} = \left( f^{nov}_h (t),\sigma^{2\: nov}_h(t)\right)$. Therefore, via the same smoothing procedure considered for the training curves, we build a hierarchical specification for the quantities involved in the novelty term:
\begin{equation}
 \label{eq:prior_H_func_model}
\begin{aligned}
&f_{h}^{nov}(t)= \sum_{b=1}^B \rho_{hb}^{nov} \phi_b(t), \quad\quad 
\rho_{hb}^{nov}\sim N(\psi_h,\tau_h^2),\\ 
&\psi_h  \sim N(0,s^2), \\
&\tau_h^2  \sim IG(a_{\tau},b_{\tau}), \quad \sigma_h^{2\: nov}(t) \sim IG (a_H, b_H).
\end{aligned}
\end{equation}
The first line of \eqref{eq:prior_H_func_model} can be rewritten as $$f_h^{nov}(t) \sim   N\left(    \psi_h \sum_{b=1}^B \phi_b(t), \tau_h^2 \sum_{b=1}^B\phi^2_b(t) \right).$$
We call this model functional Brand: it provides a powerful extension for functional novelty detection. A successful application is reported in Section \ref{sec:func_application}.

\section{Posterior Inference}
\label{Sec::PostInf}
The posterior distribution $p(\bm{\pi},\bm{\omega},\bm{a},\bm{\beta},\bm{\Theta},\bm{\Theta}^{nov}|\mathbf{y})$ is analytically intractable, therefore we rely upon MCMC techniques to carry out posterior inference. An easy sampling scheme can be constructed mimicking the blocked Gibbs sampler of \citet{Ishwaran2001a}, where the infinite series in \eqref{eq::lik} is truncated at a pre-specified level $L<\infty$. However, this approach leads to a non-negligible truncation error if $L$ is too small, and to computational inefficiencies if $L$ is set too high. Instead, we propose a modification of the $\boldsymbol{\xi}$-sequence of the Independent Slice-efficient sampler \citep{Kalli2011}, another well known conditional algorithm to sample from the exact posterior. To adapt the algorithm to our framework, we start from the following alternative reparameterization of the model in \eqref{eq::lik}-\eqref{eq:prior_spec}: 
\begin{equation}
    \begin{aligned}
    \boldsymbol{y}_m|&\tilde{\boldsymbol{\Theta}},\zeta_m \sim N\left(\tilde{\boldsymbol{\Theta}}_{\zeta_m}\right) \quad \quad
    \zeta_m | \tilde{\bm{\pi}} \sim \sum_{k=1}^\infty \tilde{\pi}_k\delta_{k}(\cdot)\\
    \tilde{\pi}_k &= \pi_k^{\mathbbm{1}_{\{0<k\leq J\}}}\cdot \left(\pi_0\cdot \omega_{k-J}\right)^{\mathbbm{1}_{\{k\geq J+1\}}}\quad \text{for } k\geq 1 \\
    \tilde{\boldsymbol{\Theta}}_k &= \boldsymbol{\Theta}_k^{\mathbbm{1}_{\{0<k\leq J\}}}\cdot \left(\boldsymbol{\Theta}^{nov}_{k-J}\right)^{\mathbbm{1}_{\{k\geq J+1\}}}\quad \quad \text{for } k\geq 1
    \end{aligned}
    \label{Oneline}
\end{equation}

where $\tilde{\boldsymbol{\Theta}}$ is obtained by concatenating $\boldsymbol{\Theta}$ and $\boldsymbol{\Theta}^{nov}$, $\delta_k$ is the usual Dirac delta function, the weights $\bm{\pi}$ and $\bm{\omega}$ are defined as in Equation \eqref{eq:modptilde}, and $\zeta_m$ is a membership label which maps each observation to its corresponding atom $\tilde{\boldsymbol{\Theta}}_{\zeta_m}$. 

Trivially, there is a one-to-one correspondence between the membership vectors $\left(\alpha_m,\beta_m\right)$ of model \eqref{eq::model} and $\zeta_m$
\begin{equation}    
\zeta_m = l \iff \alpha_m = l \cdot {\mathbbm{1}_{\{\zeta_m\leq J\}}},\quad \beta_m = (l-J) \cdot {\mathbbm{1}_{\{\zeta_m > J\}}}.
\label{eq:CsiAlphaBeta}
\end{equation}
However, we prefer the form of model \eqref{eq::model} thanks to the direct interpretation of the membership latent variables $\bm{\alpha}$ and $\bm{\beta}$, which associate each observation to the known or novel classes, respectively. 
We introduce two sequences of additional auxiliary parameters: a stochastic sequence $\mathbf{u}=\{u_m\}_{m=1}^M$ of uniform random variables and a deterministic sequence $\boldsymbol{\xi}=\{\xi_l\}_{l\geq 1}$. The introduction of these two latent variables allows for a stochastic truncation at each iteration of the sampler. The stochastic threshold, called $L$, is given as $L = \max L_m$ and $L_m$ is the largest integer such that $\xi_{L_m}>u_m$. This threshold establishes a finite number of mixture components needed at each MCMC iteration, making computations feasible. Then, we can rewrite model \eqref{eq::model} as
\begin{equation}
\begin{split}
\mathcal{L}
\left(\mathbf{y},\boldsymbol{\zeta},\mathbf{u}|\tilde{\bm{\pi}},\bmu, \bSigma\right)= 
& \prod_{m=1}^M 
\left[ \frac{\tilde{\pi}_{\zeta_m}}{\xi_{\zeta_m}} \mathbbm{1}_{ \{u_m < \xi_{\zeta_m}\} } \phi\left(\mathbf{y}_m|\tilde{\mathbf{\Theta}}_{\zeta_m} \right)\right].
\end{split}
\end{equation}
\begin{figure}[t!]
  \centering
  \includegraphics[width=\linewidth, keepaspectratio]{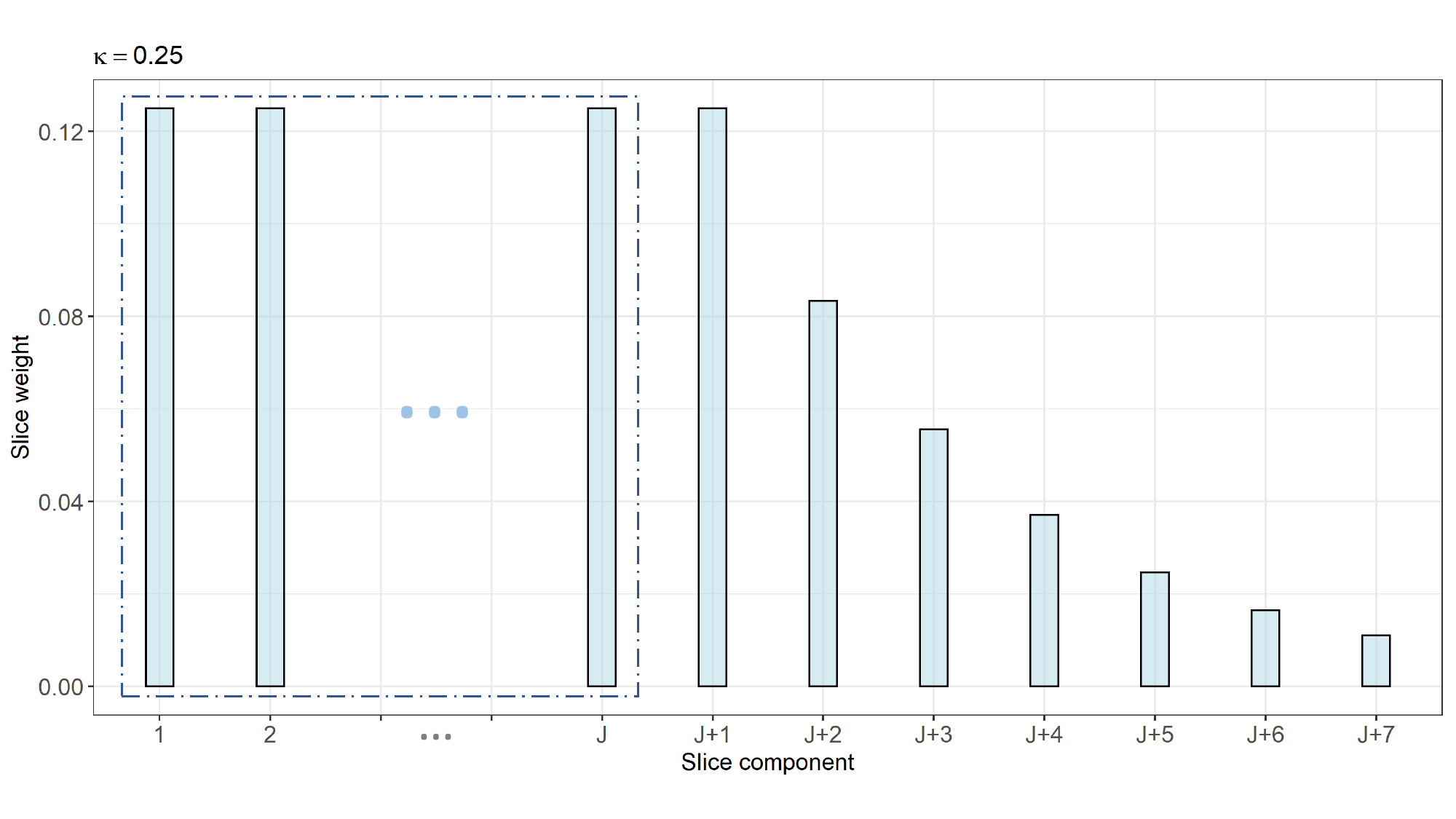}
  \caption{Example of deterministic sequence defined according to \eqref{eq::kappa}, with $\kappa=0.25$. The blue rectangle highlights the weights relative to the known components.}
  \label{fig:kappa}
\end{figure}
In the definition of a dedicated deterministic sequence $\bm{\xi}$, it is crucial to take into account the difference between the manifest and the novel components. Usually, a very common choice is $\xi_l = (1-\kappa)\kappa^{l-1}$, with $\kappa\in\left(0,1\right)$. This option allows to compute each $L_m$ analytically, being the smallest integer such that
\[L_m< 1+ \frac{\log(u_m)-\log(1-\kappa)}{\log(\kappa)}.\] 
However, the default choice of a geometrically decreasing $\boldsymbol{\xi}$-sequence is inappropriate in this context, since
we are dealing with a mixture where not all the components are conceptually equivalent. The default $\boldsymbol{\xi}$-sequence tends to favor components that come first in the mixture specification (in our case, the known ones). To overcome this issue, we propose the following intuitive modification.
Given a value for $\kappa\in(0,1)$, we equally divide the $(1-\kappa)\%$ of the mass into the first $J+1$ elements of the sequence. We then induce a geometric decay in the remaining ones to split the residual fraction $\kappa$. We force the element in position $J+1$ to have the same mass given to the manifest components, to avoid an under-representation of the novelty part. To do so, we define
\begin{equation}
    \xi_l = \begin{cases}
    \frac{1-\kappa}{J+1} & \text{if } l\leq J \\
    \frac{1-\kappa}{J+1}\left(\frac{(J+1)\kappa}{JK+1}\right)^{l-J-1} & \text{if } l> J+1 \\
    \end{cases}
    \label{eq::kappa}
\end{equation}
It is easy to prove that $\sum_{l=1}^{+\infty}\xi_l=1$. According to \eqref{eq::kappa}, the first $J+1$ elements of the sequence have masses equal to $(1-\kappa)/(J+1)$. The truncation threshold $L^*$ changes accordingly, becoming the largest integer such that
\begin{equation} L^* < J + 1 + \frac{\log(\min(\bm{u}))-\log\left(\frac{1-\kappa}{J+1}\right)}{\log\left(\frac{(J+1)\kappa}{J\kappa+1}\right)}.
\label{eq:truncate}
\end{equation}
Inequality \eqref{eq:truncate} states that the truncation threshold $L^*$ can be only greater or equal to $J+1$, ensuring that the MCMC always takes into consideration the creation of at least one cluster in the novel distribution.
A representation of the modified $\boldsymbol{\xi}$-sequence is depicted in Figure \ref{fig:kappa}.  
We report the pseudo-code for the devised Gibbs sampler in the Appendix. The algorithm for the functional extension is not included for conciseness. However, its structure closely follows the one outlined for the multivariate case.

Once the MCMC sample is collected, we first compute the a posteriori probability of being a novelty for every test unit $m$, $PPN_m=\mathbbm{P}\left[\mathbf{y}_m \sim f^{nov}|\mathbf{Y} \right]$, that is estimated according to the ergodic mean:
\begin{equation} \label{eq:PPN}
\hat{PPN_m}= \frac{\sum_{i=1}^I \mathbbm{1}_{\{a^{(i)}_m=0)\}}}{I},   
\end{equation} 
where $\alpha^{(i)}_m$ is the value assumed by the parameter $\alpha_m$ at the $i$-th iteration of the MCMC chain and $I$ is the total number of iterations. We remark that the inference on $\bm{\alpha}$ can be conducted directly, since the mixture between the $J$ observed components and $f^{nov}$ is not subjected to label switching. In contrast, we need to take this problem into account when dealing with $\bm{\beta}$. To perform valid inference, one possibility is to rely on the posterior probability coclustering matrix (PPCM) as indicated in Section \ref{Disentangling}. Each entry of this matrix $p_{m,m'}=\mathbb{P}\left[ \mathbf{y}_m \text{ and } \mathbf{y}_{m'} \text{ belong to the same novelty class}\right]$ is estimated as
\begin{equation}
    \hat{p}_{m,m'} = \frac{\sum_{i=1}^I \mathbbm{1}_{\{\beta^{(i)}_m=\beta^{(i)}_{m'}\}}}{I}.
\end{equation}
Once we obtain the PPCM, we employ it to estimate the best partition (BP) in the novelty subset. Indeed, one can recover the BP by minimizing a loss function defined over the space of partitions, which can be computed starting from the PPCM. A famous loss function was proposed by \citet{Binder1978}, and investigated in a BNP setting by \citet{Lau2007}. 
However, the so-called Binder loss presents peculiar asymmetries, preferring to split clusters over merging. These asymmetries could result in a number of estimated clusters higher than needed. Therefore, we adopt the Variation of Information \citep[VI -][]{MeilaM2007} as loss criterion. The associated loss function, recently proposed by \citet{Wade2018}, is known to provide less fragmented results.

Finally, once the BP for the novelty component has been estimated, we can rely on a heuristic based on the cluster sizes to discriminate anomalies from actual new classes. Let us suppose that the BP consists of $S$ novel clusters. Denote the number of instances assigned to cluster $s\in\{1,\ldots,S\}$ with $m^{nov}_s$. A cluster $s$ is considered to be an agglomerate of outlying points if its cardinality $m^{nov}_s$ is sufficiently small in comparison to the entire novelty sample size, otherwise it is regarded as a proper novel group.

\section{Applications} \label{sec:applications}
\subsection{Simulation Study} \label{sec:sim_study}
In this section, we present a simulation study aimed at highlighting the capabilities of the new semiparametric Bayesian model in performing novelty detection and we compare it with existing methodologies. We consider different scenarios varying the sample sizes of the hidden classes and the adulteration proportions in the training set. At the same time, we evaluate the importance of the robust information extraction phase and how it affects the learning procedure.   
\subsubsection{Experimental setup}
\begin{figure}
\includegraphics[scale=.45]{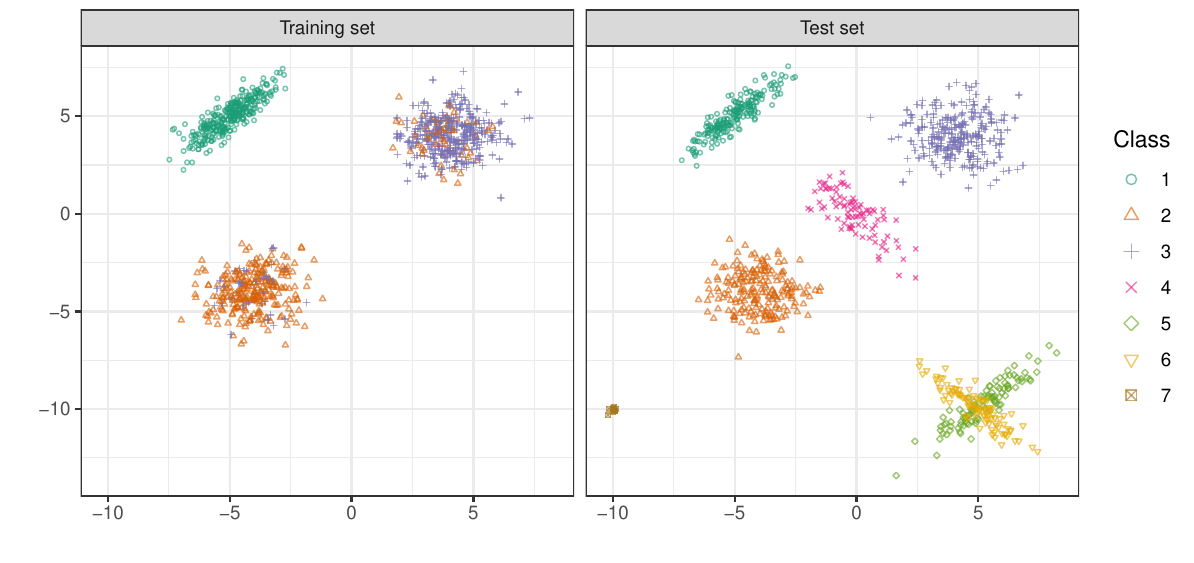}
\centering
\caption{Simulated data for the \texttt{Label noise = True}, \texttt{Novelty size = Not small} scenario. Classes $4,\ldots,7$ are not observed in the learning set.}
\label{fig:example_sim_study}
 \end{figure}

We consider a training set formed by $J = 3$ observed classes, each distributed according to a bivariate Normal density $\mathcal{N}_2(\boldsymbol{\mu}_j, \boldsymbol{\Sigma}_j)$, $j=1,2,3$, with the following parameters:

\[  \boldsymbol{\mu}_1=(-5, 5)', \quad \boldsymbol{\mu}_2=(-4, -4)', \quad \boldsymbol{\mu}_3=(4, 4)'\]
 \[ \boldsymbol{\Sigma}_1 = \begin{bmatrix}
    1       & 0.9\\
    0.9      & 1
    \end{bmatrix} \quad
    \boldsymbol{\Sigma}_2 = \begin{bmatrix}
    1       & 0\\
    0       & 1
    \end{bmatrix} \quad
    \boldsymbol{\Sigma}_3 = \begin{bmatrix}
    1       & 0\\
    0       & 1
    \end{bmatrix}.
    \]
The class sample sizes are, respectively, equal to $n_1=300$, $n_2=300$ and $n_3=400$. The same groups are also present in the test set, together with four previously unobserved classes. We  generate the new classes via bivariate Normal densities with parameters:
\[  
\begin{aligned}
\boldsymbol{\mu}_4&=(0,0)', \quad \boldsymbol{\mu}_5=(5, -10)', \\ \boldsymbol{\mu}_6&=(5, -10)', \quad \boldsymbol{\mu}_7=(-10, -10)',
\end{aligned}
\]
 \[ 
 \begin{aligned}
 \boldsymbol{\Sigma}_4 &= \begin{bmatrix}
    1       & -0.75\\
   -0.75       & 1
    \end{bmatrix},
    \boldsymbol{\Sigma}_5 = \begin{bmatrix}
    1       & 0.9\\
    0.9       & 1
    \end{bmatrix},\\
    \boldsymbol{\Sigma}_6 &= \begin{bmatrix}
    1       & -0.9\\
  -  0.9       & 1
    \end{bmatrix},
    \boldsymbol{\Sigma}_7 = \begin{bmatrix}
    0.01       & 0\\
    0       & 0.01
    \end{bmatrix}.
    \end{aligned}
\]
The test set encompasses a total of $7$ components: 3 observed and 4 novelties. 
Starting from the above-described data generating process, we consider four different scenarios varying:\begin{itemize}
\item Data contamination level
\begin{itemize}
\item No contamination in the training set (\texttt{Label noise = False})
\item $12\%$ label noise between classes $2$ and $3$ (\texttt{Label noise = True})
\end{itemize}
\item Test set sample size
\begin{itemize}
\item Novelty subset size equal to slightly more than $30\%$ of the test set (\texttt{Novelty size = Not small})
$$
    \begin{aligned}
&m_1=200, \: m_2=200, \: m_3=250, \: m_4=90,\\  &m_5=100, \: m_6=100, \: m_7=10    
    \end{aligned}
$$
\item Novelty subset size equal to $15\%$ of the test set (\texttt{Novelty size = Small})
$$
    \begin{aligned}
&m_1=350, \: m_2=250, \: m_3=250, \: m_4=49, \\ &m_5=50, \: m_6=50, \: m_7=1.
   \end{aligned}
$$
\end{itemize}
\end{itemize}
Figure \ref{fig:example_sim_study} exemplifies the experiment structure displaying a realization from the \texttt{Label noise = True}, \texttt{Novelty size = Not small} scenario. As it is evident from the plots, the label noise is strategically included to cause a more difficult identification of the fourth class, should the parameters of the second and third classes be non-robustly learned. Further, notice that the last group presents limited sample size and variability: it could easily be regarded as pointwise contamination (i.e., an anomaly) rather than an actual new component. Nonetheless, following the reasoning outlined in the introduction, we are interested in evaluating the ability of the nonparametric density to capture and discriminate these types of peculiar patterns as well.
For each combination of contamination level and test set sample size, we simulate $B=100$ datasets. Results are reported in the following subsection.

\subsubsection{Simulation results}
\begin{figure*}
\begin{subfigure}{1\textwidth}
  \centering
\includegraphics[scale=.75]{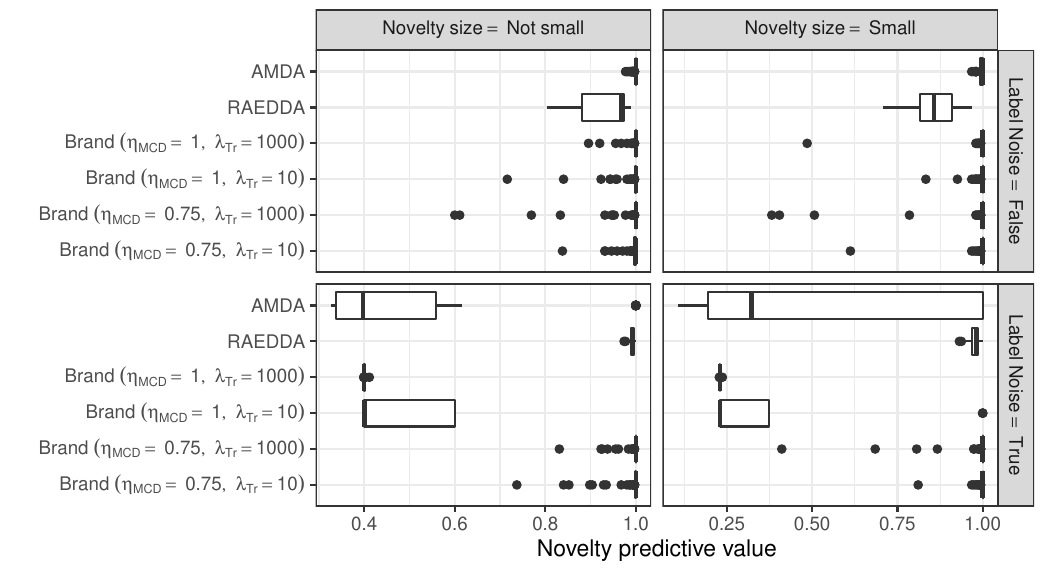}\\\vspace{.25cm}
\end{subfigure}
\begin{subfigure}{1\textwidth}
  \centering
\includegraphics[scale=.75]{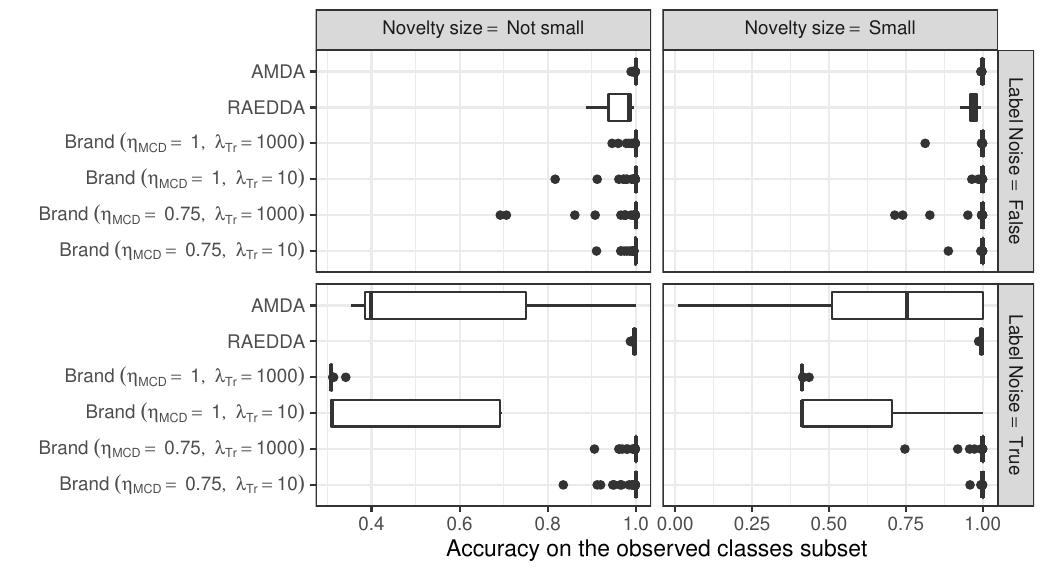}\\\vspace{.25cm}
\end{subfigure}
\begin{subfigure}{1\textwidth}
  \centering
\includegraphics[scale=.75]{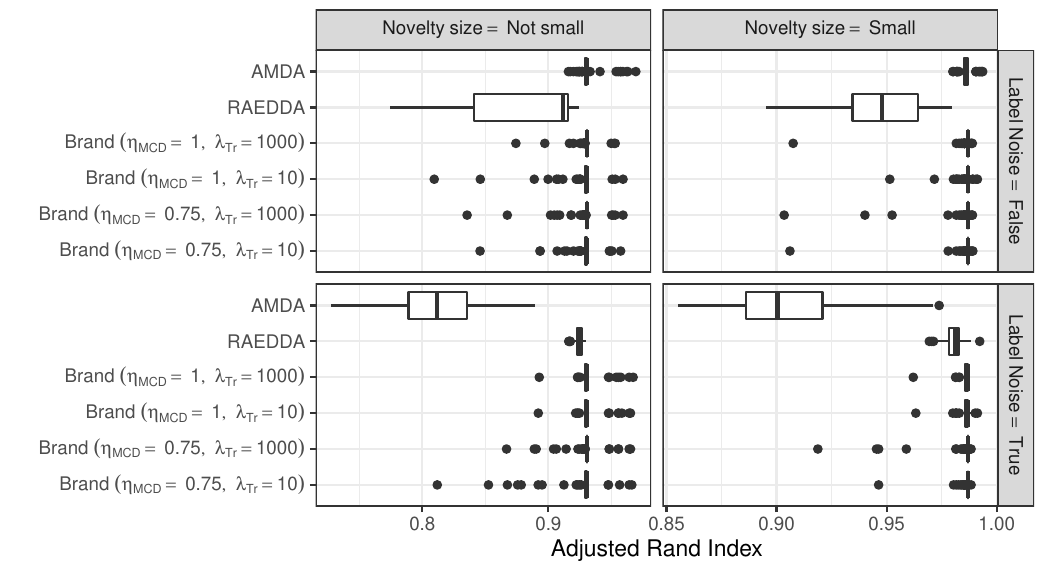}\\\vspace{.25cm}
\end{subfigure}
\caption{Box plots for (from top to bottom) novelty predictive value, accuracy on the known classes and ARI 
 metrics for $B = 100$ repetitions of the simulated experiment, varying data contamination level and test set sample size.}
\label{fig:boxplot_sim_study}
\end{figure*}
\begin{figure*}[t]
  \centering
  \includegraphics[scale=.4]{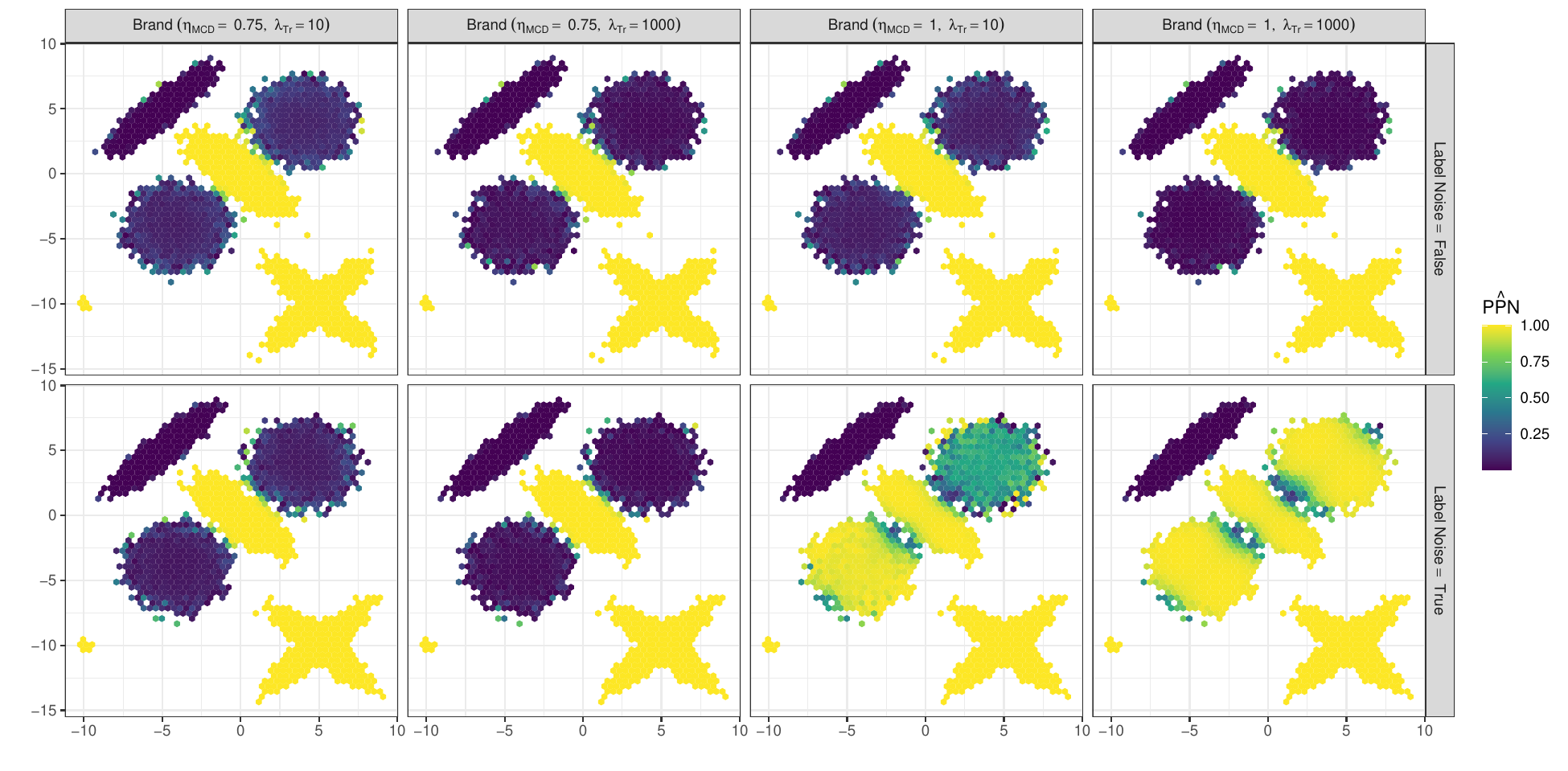}
  \caption{Hex plots of the average estimated posterior probability of being a novelty, according to formula \eqref{eq:PPN}, for $B = 100$ repetitions of the simulated experiment, varying data contamination level and Brand hyper-parameters, \texttt{Not small} novelty subset size. The brighter the color the higher the probability of belonging to $f^{nov}$.}
  \label{fig:hex_plot}
\end{figure*}
We compare the performance of the Brand model with different hyper-parameters specifications:
\begin{itemize}
\item the information from the training set is either non-robustly ($\eta_{MCD}=1$) or robustly ($\eta_{MCD}=0.75$) extracted, 
\item the precision parameter associated with the training prior belief is either very high ($\lambda_{Tr}=1,000$) or moderately low ($\lambda_{Tr}=10$).
\end{itemize}
In addition, two model-based adaptive classifiers are considered in the comparison, namely the inductive RAEDDA model \citep{Cappozzo2019e} with labeled and unlabeled trimming levels respectively equal to $0.12$ and $0.05$, and the inductive AMDA model \citep{Bouveyron}. For each replication of the simulated experiment, a set of four metrics is recorded from the test set:
\begin{itemize}
\item \textit{Novelty predictive value (Precision)}: the proportion of units marked as novelties by a given method truly belonging to classes $4,\ldots,7$,
\item \textit{Accuracy on the observed classes subset}: the classification accuracy of a given method within the subset of groups already observed in the training set,
\item \textit{Adjusted Rand Index} \citep[\textit{ARI},][]{Rand1971}: measuring the similarity between the partition returned by a given method and the underlying true structure,
\item \textit{PPN}: a posteriori probability of being a novelty, computed according to Equation \eqref{eq:PPN} (Brand only).
\end{itemize}

We run $40,000$ MCMC iterations and discard the first $20,000$ as a burn-in phase. Apart from the hyper-parameters for the training components, fairly uninformative priors are employed in the base measure $H$, with $\bm{m}_0=(0,0)', \: \lambda_0=0.01, \: \nu_0=10$ and $\bm{S}_0=10\boldsymbol{I}_2$. Lastly, a Gamma DP concentration parameter is considered with prior rate and scale hyper-parameters both equal to $1$. 

Figure \ref{fig:boxplot_sim_study} reports the results for $B=100$ repetitions of the experiment under the different simulated scenarios. A Table containing the values on which this plot is built is deferred to the Supplementary Material. The \textit{Novelty predictive value} metric highlights the capability of the model to correctly recover and identify the previously unseen patterns. As expected, in the adulteration-free scenarios, all methodologies succeed well enough in separating known and hidden components. The worst performance is exhibited by the RAEDDA model for which, due to the fixed trimming level, a small part of the group-wise most extreme (but still genuine) observations is discarded, thus slightly overestimating the novelties percentage (the same happens for the ARI metric). 
Different results are displayed in scenarios wherein the label noise complicates the learning process. Robust procedures efficiently cope with the adulteration present in the training set, while the AMDA and the Brand methods when $\eta_{MCD}=1$ tend to largely overestimate the novelty component. Particularly, the harmful effect caused by the mislabeled units is exacerbated in the Brand model that sets high confidence in the priors $(\lambda_{Tr}=1,000)$, while a partial mitigation, albeit feeble, emerges when $\lambda_{Tr}$ is set equal to $10$. This consequence is even more apparent in the hex plots of Figure \ref{fig:hex_plot}, where we see that the latter model tries to modify its prior belief to accommodate the (outlier-free) test units, while the former, forced to stick close to its prior distribution by the high value of $\lambda_{Tr}$, incorporates the second and third class in the novelty term. The final output, as displayed in the \textit{Accuracy on the observed classes subset} boxplots, has an overall high misclassification error when it comes to identifying the test units belonging to the previously observed classes. Differently, setting robust informative priors prevents this undesirable  behavior, as it is shown by both the high level of accuracy and the associated low posterior probability of being a novelty in the feature space wherein the observed groups lie. 
On the other hand, the true partition recovery, assessed by the Adjusted Rand Index,  does not seem to be influenced by the label noise, with our proposal always outperforming the competing methodologies regardless of which hyper-parameters are selected. As previously mentioned, for $Brand(\eta_{MCD}=1, \lambda_{Tr}=10)$ and $Brand(\eta_{MCD}=1, \lambda_{Tr}=1,000)$ cases the second and third classes are assimilated into the nonparametric component in the \texttt{Label Noise = True} scenario. This is due to the fact that the mislabeled units prevent Brand from correctly learning the true structures of groups two and three in Stage I. As a consequence, no correspondence between these improperly estimated classes in the training is found in the test set, so much so that the DP prior creates them anew within the novelty term. Clearly, this is a sub-optimal behavior as the separation of what is known from what is novel is completely lost, yet it may raise suspicion on dealing with a contaminated learning set, suggesting the need of a robust prior information extraction.

Additional simulated experiments, involving a high-dimensional scenario and novelty detection problem under model misspecification are included in the Supplementary Material. 
\subsection{X-ray images of wheat kernels}
Sophisticated and advanced techniques like X-rays, scanning microscopy and laser technology are increasingly employed for the automatic collection and processing of images. Within the domain of computer vision studies, novelty detection is generally portrayed as a one-class classification problem. There, the aim is to separate the known patterns from the absent, poorly sampled or not well defined remainder \citep{Khan2014}. Thus, there is strong interest in developing methodologies that not only distinguish the already observed quantities from the new entities, but that also identify specific structures within the novelty component.
\begin{figure}[ht]
\centering
\includegraphics[scale=.8]{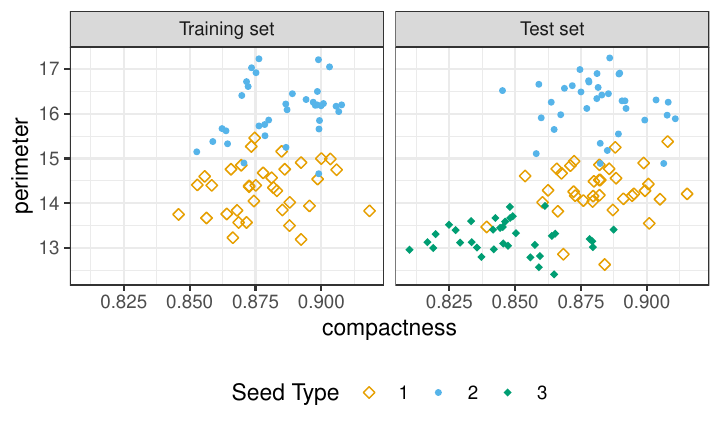}
\caption{Learning scenario (only \texttt{compactness} and \texttt{perimeter} variables displayed) for novelty detection of an unobserved wheat variety, seed dataset.}
\label{fig:seed_learning_scenario}
\end{figure}
The present case study involves the detection of a novel grain type by means of seven geometric parameters, recorded postprocessing X-ray photograms of kernels \citep{Charytanowicz2010}. In more detail, for the $210$ samples belonging to the three different wheat varieties, high quality visualization of the internal kernel structure is detected using a soft X-ray technique and, subsequently, the image files are post-processed via a dedicated computer software package \citep{Strumio1999}. The obtained dataset is publicly available in the University of California, Irvine Machine Learning data repository. This experiment involves the random selection of $70$ training units from the first two cultivars, and a test set of $105$ samples, including $35$ grains from the third variety. The resulting learning scenario is displayed in Figure \ref{fig:seed_learning_scenario}. The aim of the analysis is to employ Brand 
to detect the third unobserved variety, whilst performing classification of the known grain types with high accuracy. Firstly, the MCD estimator with hyper-parameter $h_{MCD}=0.95$ is adopted for robustly learning the training structure of the two observed wheat varieties. In the second stage, our model is fitted to the test set,  discarding $20,000$ iterations for the burn-in phase, and subsequently retaining $10,000$ MCMC samples. 
As usual, fairly uninformative priors are employed in the base measure $H$, with $\bm{m}_0=\bm{0}, \: \lambda_0=0.01, \: \nu_0=10$ and $\bm{S}_0=\boldsymbol{I}_7$, where $\mathbf{0}$ denotes the 7-dimensional zero vector. For the training components, mean and covariance matrices of the Normal-inverse-Wishart priors are directly determined by the MCD output of the first stage, while $\nu^{Tr}$ and $\lambda^{Tr}$ are specified to be respectively equal to $250$ and $1,000$. The latter value indicates that after having robustly extracted information for the two observed classes, high trust is placed in the prior distributions of the known components. 
Model results are reported in Figure \ref{fig:seed_BNP_ggplots}, where the posterior probability of being a novelty $PPN_m=\mathbbm{P}\left[\mathbf{y}_m \sim f^{nov}|\mathbf{Y} \right]$, $m=1,\ldots,M$, displayed in the plots below the main diagonal, are estimated according to the ergodic mean in  \eqref{eq:PPN}. The a posteriori classification, computed via majority vote, is depicted in the plots above the main diagonal, where the water-green solid diamonds denote observations belonging to the novel class. The confusion matrix associated with the estimated group assignments is reported in Table \ref{tab:seed_confu_mat}, where the third group variety is effectively captured by the flexible process modeling the novel component. 

\begin{table}[th!]
 \caption{Confusion matrix for the semiparametric Bayesian classifier on the test set, seeds dataset. The label ``New'' indicates observations that are estimated to have arisen from the novelty component.}
 \centering
\begin{tabular}{ cccc } 
\toprule
&  & Truth & \\
Classification & 1 & 2 & 3 \\ 
  \midrule
  1   &  30 &   0 &   7 \\  
  2   &   2 &  35 &   0 \\  
  New &   3 &   0 &  28 \\
\bottomrule
\end{tabular}
\label{tab:seed_confu_mat}
\end{table}
All in all, the promising results obtained with this multivariate dataset may foster the employment of our methodology in automatic image classification procedures that supersede the one-class classification paradigm, allowing for a much more flexible anomaly and novelty detector in computer vision applications.
\begin{figure*}[ht!]
\centering
\includegraphics[width=\textwidth, keepaspectratio]{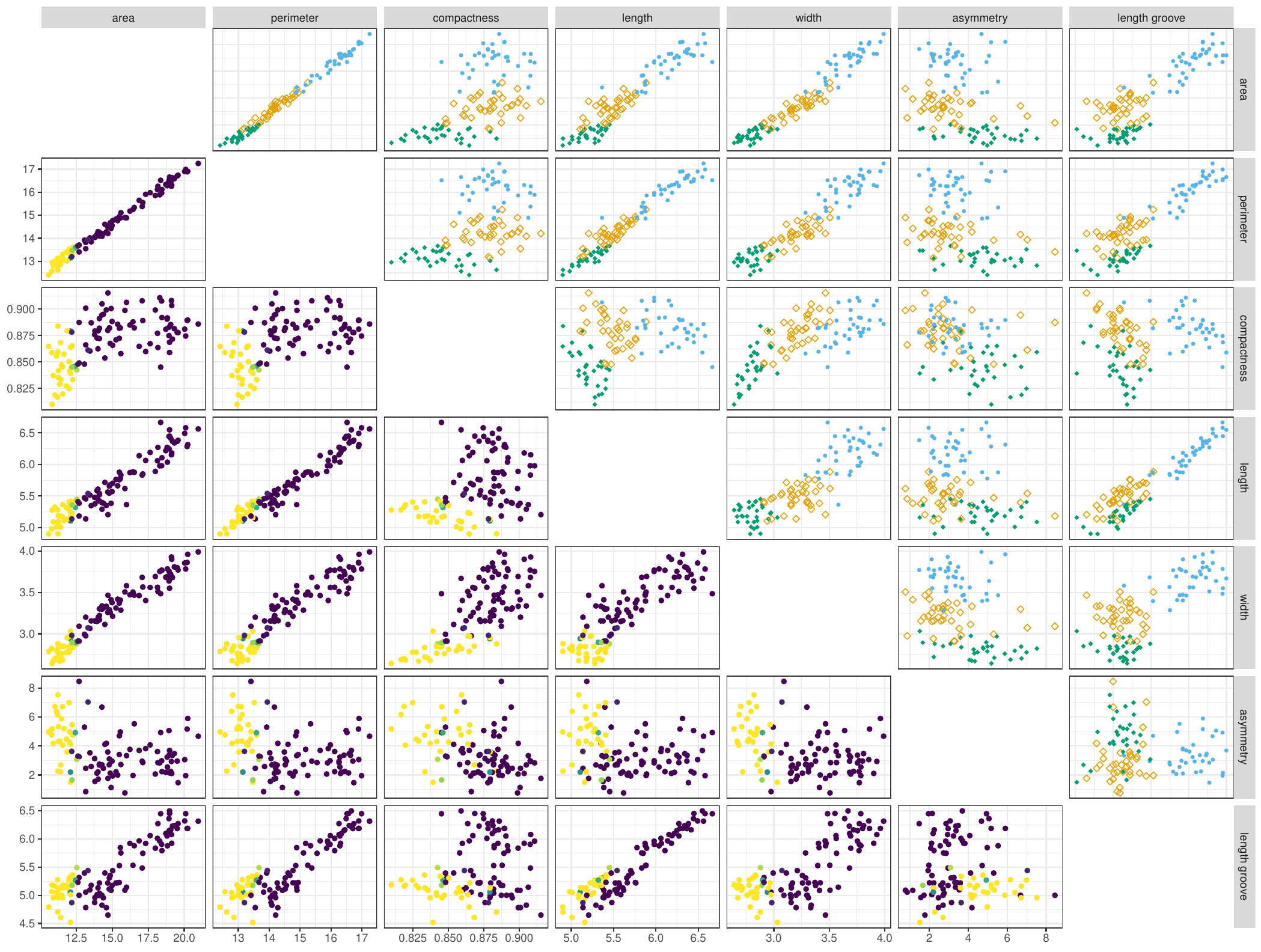}
\caption{Test set for the considered experimental scenario, seeds dataset. Plots below the main diagonal represent the estimated posterior probability of being a novelty. The brighter the color the higher the probability of belonging to $f^{nov}$. Plots above the main diagonal display the associated group assignments, where the water-green solid diamonds denote observations classified as novelties.}
\label{fig:seed_BNP_ggplots}
\end{figure*}

\subsection{Functional novelty detection of meat variety} \label{sec:func_application}

In recent years, machine learning methodologies have experienced an ever-growing interest in countless fields, including food authentication research \citep{Singh2019}. An authenticity study aims to characterize unknown food samples, correctly identifying their type and/or provenance. Clearly, no observation is to be trusted in a context wherein
the final purpose is to detect potentially adulterated units, in which, for example, an entire subsample may belong to a previously unseen pattern. Motivated by a dataset of Near Infrared Spectra (NIR) of meat varieties, we employ the functional model introduced in Section \ref{sec:func_version} to perform classification and novelty detection when having a hidden class and four manually adulterated units in the test set.
The considered data report the electromagnetic spectrum for a total of $231$ homogenized meat samples, recorded from
$400-2498$ $nm$ at intervals of $2$ $nm$ \citep{McElhinney1999}. The units belong to five different meat types, with $32$ beef, $55$ chicken, $34$ lamb, $55$ pork, and $55$ turkey records. The amount of light absorbed at a given wavelength is recorded for each meat sample: $A = log_{10}(1/R)$ where $R$ is the reflectance value. The visible part of the electromagnetic spectrum ($400-780$ $nm$) accounts for color differences in the meat types, while their chemical composition is recorded further along the spectrum.
NIR data can be interpreted as a discrete evaluation of a continuous function in a bounded domain. Therefore, the procedure described in Section \ref{sec:func_version} is a sensible methodological tool for modeling this type of data objects \citep{Crane2003}. We randomly partition the recorded units into labeled and unlabeled sets. The former includes $28$ chicken, $17$ lamb, $28$ pork, and $28$ turkey samples. The latter contains the same proportion of these four meat types with an additional $32$ beef units. The last class is not observed in the test set and needs to be discovered. Also, four validation units are manually adulterated and added to the test set as follows:
\begin{itemize}
\item a shifted version of a pork sample, achieved by removing the first $15$ data points and appending the last $15$ group-mean absorbance values at the end of the spectrum;
\item a noisy version of a pork sample, generated by adding Gaussian white noise to the original spectrum;
\item a modified version of a turkey sample, obtained by abnormally increasing the absorbance value in a single specific wavelength to simulate a spike;
\item a pork sample with an added slope, produced by multiplying the original spectrum by a positive constant.
\end{itemize}
These modifications mimic the ones considered in the ``Chimiom\'{e}trie 2005'' chemometric contest, where participants were tasked to perform discrimination and outlier detection of mid-infrared spectra of four different starches types \citep{FernandezPierna2007a}. In our context, both the beef subpopulation and the adulterated units are previously unseen patterns that shall be captured by the novelty component.

\begin{figure*}[!h]
  \centering
  \includegraphics[scale=.5]{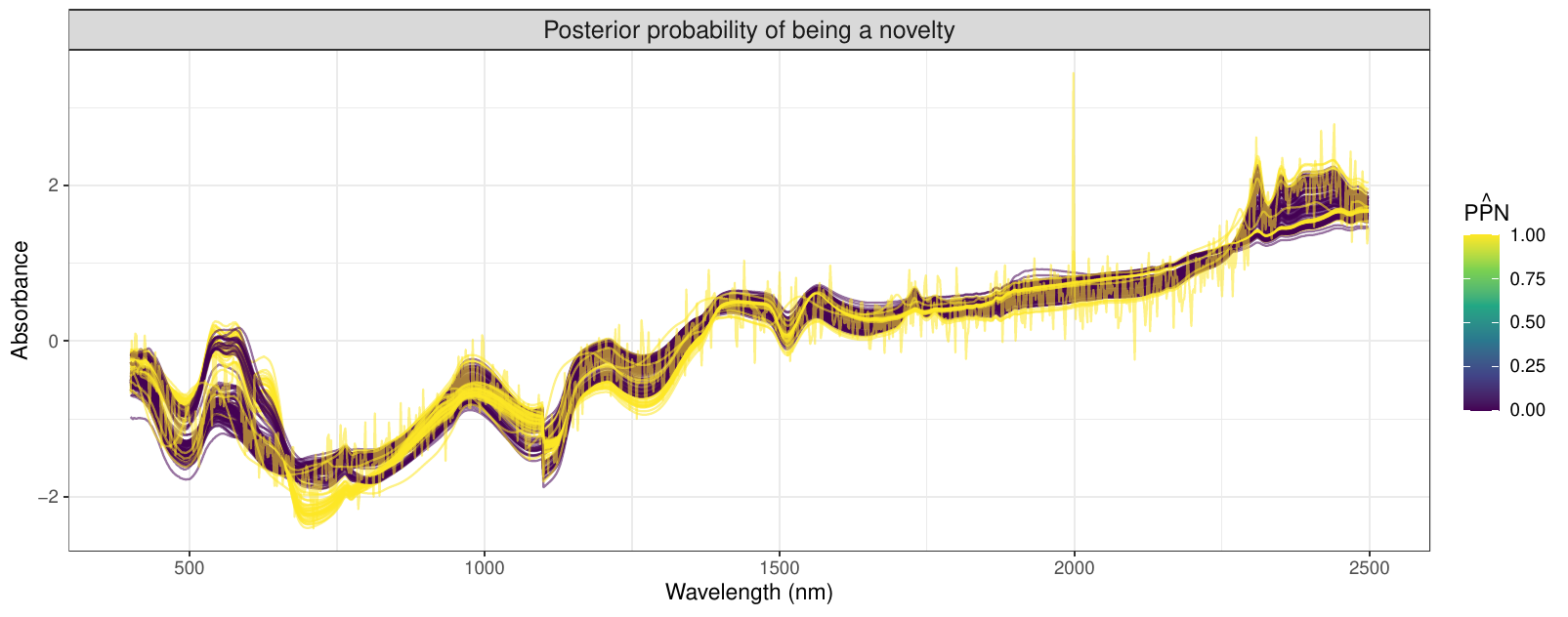}
  \caption{Estimated posterior probability of being a novelty, according to formula \eqref{eq:PPN}, the brighter the color the higher the probability of  belonging to $f^{nov}$.} 
  \label{fig:ppn_clas_meat_data}
\end{figure*}

Firstly, we extract robust prior information from the learning set. Given the spectra non-cyclical nature, we approximate each training unit via a linear combination of $B=100$ B-spline bases, and their associated coefficients are retrieved. Given the high-dimensional nature of the smoothing process, the MRCD is employed to obtain robust group-wise estimates for the splines coefficients. These quantities, which are linearly combined with the B-spline bases, account for the training atoms $\boldsymbol{\Theta}_j$, $j=1,\ldots,4$ specified in Equation \eqref{eq:func_model}.
We adopt a value of $\eta_{MCD}=0.75$ in the first stage, providing functional atoms robust against contamination that may arise in the training set. In this experiment, an inductive approach is considered, for which the training estimates will be kept fixed throughout the subsequent Bayesian learning phase. We further set $a_\tau=3, b_\tau=1, s^2=1, a_H=5,$ and $b_H=1$, inducing low variability on the noise parameters as much as not to compromise the hierarchical structure between known and novelty components. A more detailed discussion on the hyperparameters choice is deferred to the Supplementary Material, where we evaluate alternative effects for different prior settings within a controlled experiment.
Once $\hat{\boldsymbol{\Theta}}_j$, $j=1,\ldots,4$ are retained, the Bayesian model of Section \ref{sec:func_version} is applied to the test units running a total of 20,000 iterations and discarding the first 10,000 as warm-up. Figure \ref{fig:ppn_clas_meat_data} summarizes the results of the fitted model. Each spectrum is colored according to its a posteriori probability of being a novelty, computed as in \eqref{eq:PPN}. The resulting confusion matrix is reported in Table \ref{tab:confu_mat_1}, where it is apparent that the previously unseen class, as well as the adulterated units (labeled as ``Outliers'' in the table), are successfully captured by the novelty component. The obtained classification accuracy is in agreement with the ones produced by state-of-the-art classifiers in a fully-supervised scenario \citep[see, for example,][]{Murphy2012, Gutierrez2014}. That is, our proposal is capable of detecting previously unseen classes and outlying units, whilst maintaining competitive predictive power.
\begin{table*}[!th]
 \centering
\begin{tabular}{ ccccccc } 
\toprule
&  \multicolumn{6}{c}{Truth}\\
Classification & Beef & Chicken & Lamb & Pork & Turkey & Outliers\\ 
  \midrule
Novelty   &  32 &   0 &   0 &   0 &   2 &   4 \\ 
  Chicken &   0 &  21 &   0 &   1 &   12 &   0 \\ 
  Lamb    &   0 &   0 &  17 &   0 &   0 &   0 \\ 
  Pork    &   0 &   4 &   0 &  20 &   3 &   0 \\ 
  Turkey  &   0 &   2 &   0 &   3 &  9 &   0 \\ 
\bottomrule
\end{tabular}
 \caption{Confusion matrix for the semiparametric Bayesian classifier on the test set, meat dataset. The label ``Novelty'' indicates observations that are estimated to have arisen from the $f^{nov}$.}
\label{tab:confu_mat_1}
\end{table*}

\begin{figure*}[th]
    \centering
    \includegraphics[scale=.5]{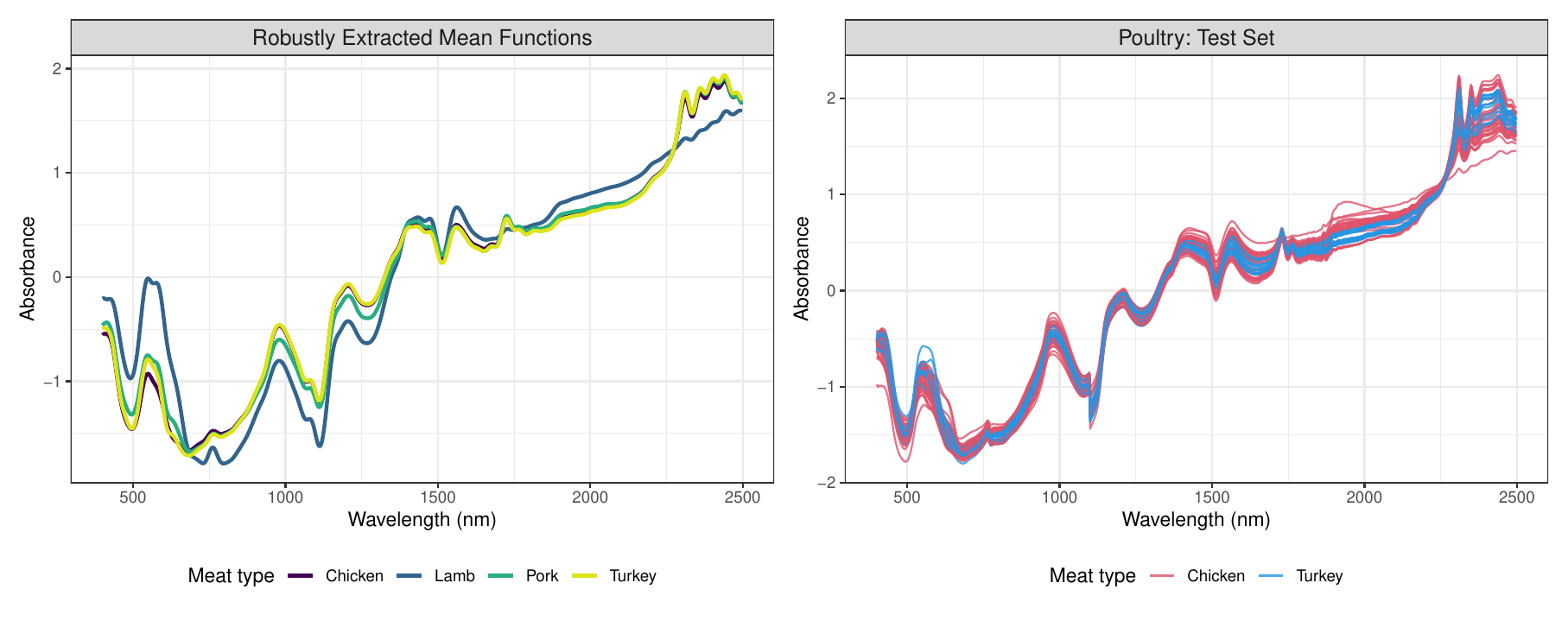}
    \caption{The left panel shows the robust functional means extracted from the training set. The right panel shows the test functional objects containing a type of poultry. }
    \label{fig:convince}
\end{figure*}

Looking at the classification performance, we observe that Brand can correctly recover the underlying data partition, except for the turkey subgroup. Specifically, the model struggles to separate the turkey units from the chicken ones. Figure \ref{fig:convince} provides an explanation for this issue. The left panel shows the robust functional means extracted from the training set. The right panel shows the functional test objects containing the two types of poultry. The overlapping is evident in both cases and it is the main reason why Brand merges the two different sets.

\begin{figure*}[ht!]
  \centering
    \includegraphics[scale=.5]{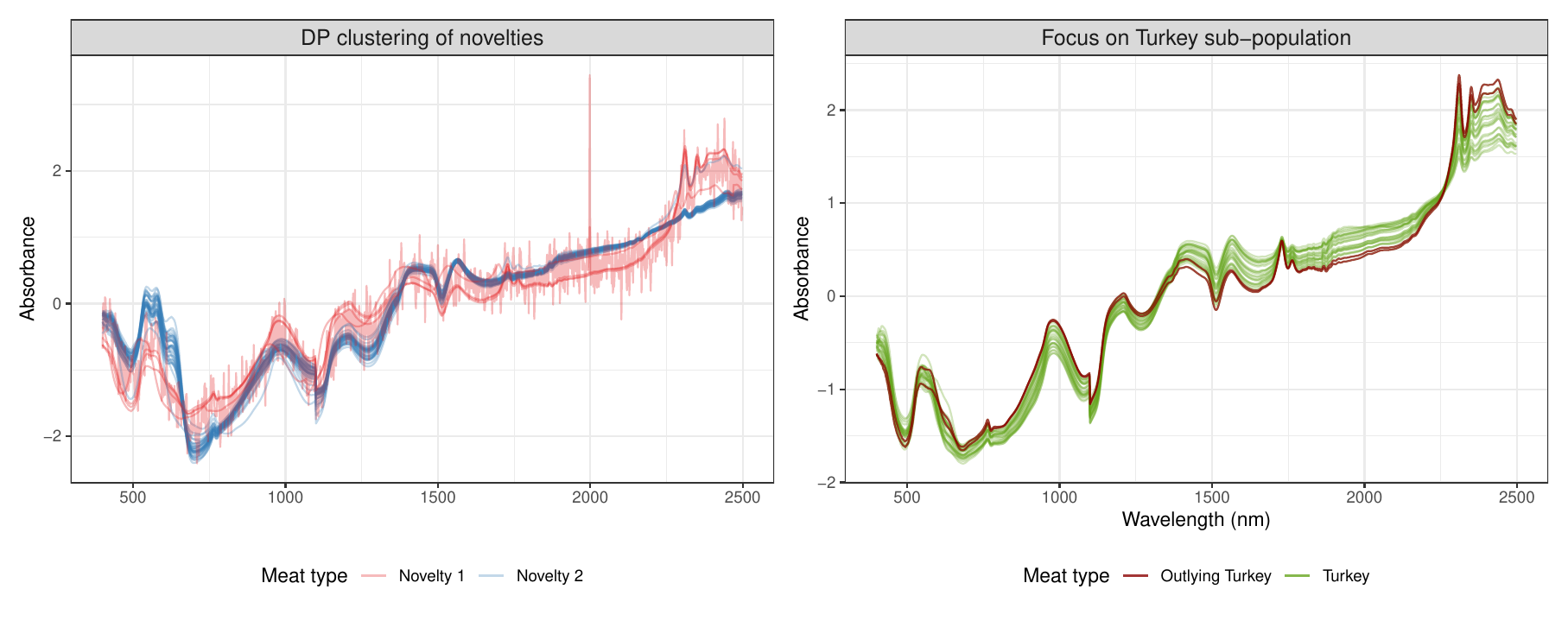}
  \caption{Left panel: best partition of the novelty component recovered by minimizing the Variation of Information loss function. The dashed blue curves are beef samples, while the solid red ones are the manually adulterated units and the two turkeys incorrectly assigned to the novel component. Right panel: true turkey sub-population in the test set, the units incorrectly assigned to the novel component are displayed with solid dark red lines.}
  \label{fig:clas_novelty_meat_data}
\end{figure*}

Focusing on the novelty component, the model entirely captures the beef hidden class and the adulterated units, yet two turkey samples are also incorrectly assigned. The obtained classification for the curves identified to be novelties, resulting by VI minimization, is displayed in the left panel of Figure \ref{fig:clas_novelty_meat_data}, where two distinct clusters are detected. Interestingly, Brand separates the $32$ beef samples (blue dashed lines) from the two turkeys (solid red lines) and classifies three of the four manually adulterated units to the outlying cluster. In contrast, the remaining one is assigned to the beef class, because of its peculiar shape, as it is shown in Figure 13 of the Supplementary Material.

Finally, we investigate why two turkey units are incorrectly assigned to the novel component. A closer look at the turkey sub-population, displayed in the right panel of Figure \ref{fig:clas_novelty_meat_data}, shows how these two samples exhibit a somehow extreme pattern within their group and can, therefore, be legitimately flagged as outlying or anomalous turkeys.
We report two additional figures in the Supplementary Material. The first provides a visual summary of the estimated grouping; the second shows how the turkey test units are partitioned into different clusters.

In this section, we have shown the effectiveness of our methodology in correctly identifying a hidden group in a functional setting, while jointly achieving good classification accuracy and detection of outlying curves. The successful application of the model seems particularly desirable in fields like food authenticity, where generally there is no a priori available information on how many modifications and/or adulteration mechanisms may be present in the samples.

\section{Conclusion and discussion} \label{sec:conclusions}
We have introduced a two-stage methodology for robust Bayesian semiparametric novelty detection.
In the first stage, we robustly extract the observed group structure from the training set. In the second stage, we incorporate such prior knowledge in a contaminated mixture, wherein we have employed a nonparametric component to describe the novelty term. The latter could either correspond to anomalies or actual new groups. This distinction is made possible by retrieving the best partition within the novel subset. We have investigated the properties of the random measure underlying the model and its connections with existing methods. Subsequently, the general multivariate methodology has been extended to handle functional data objects, resulting in a novelty detector for functional data analysis. A dedicated slice-efficient sampler, taking into account the difference between unseen and seen components, has been devised for posterior inference. An extensive simulation study and applications on multivariate and functional data have validated the effectiveness of our proposal, fostering its employment in diverse areas from image analysis to chemometrics.
Brand can represent the starting point for many different research avenues. Future research directions aim at providing a Bayesian interpretation of the robust MCD estimator to propose a unified, fully Bayesian model. More versatile specifications can be adopted for the known components, weakening the Gaussianity assumption. These extensions can be obtained by adopting more flexible distributions while keeping the mean and variance of the resulting densities constrained to the findings in the training set, for example, via centered stick-breaking mixtures \citep{Yang2010}.\\
Similarly, functional Brand can be improved by adopting a more general prior specification via Gaussian Processes \citep{Gpras}. Lastly, it is of paramount interest to develop scalable algorithms, as Variational Bayes \citep{Blei2017} and Expectation-Maximization \citep{Dempster1977}, for inference on massive datasets. Such solutions will offer both increased speed and lower computational cost, which are crucial for assuring the applicability of our proposal in the big data era.

\section{Supporting Information}
The Supplementary Material referenced throughout the article is available with this paper at the Statistics and Computing website. As supporting information, we report the proofs of the theoretical results showed in Section \ref{sec::Properties}. Moreover, to complement the results presented in Section \ref{sec:applications}, we showcase the performance obtained by applying Brand to various challenging simulated data, varying the distributional assumptions and dimensionality. We also discuss an application to a higher dimensional real dataset, the popular benchmark \texttt{Wine} dataset from the UCI dataset repository, considering all its 13 features. Lastly, with the help of a controlled experiment, we guide the reader through the choice of hyperparameters and, more broadly, the whole usage of the model in the functional case. Software routines, including the implementation for both methods, the simulation study, and real data analyses of Section \ref{sec:applications} are openly available at \url{github.com/AndreaCappozzo/brand-public\_repo}.

\section*{Acknowledgement}
The authors want to thank the Editor and the anonymous Reviewers for their suggestions and comments, which significantly improved the scientific value of the manuscript. During the development of this article, F. Denti was funded as a postdoctoral scholar by the NIH grant R01MH115697. Previously, he was also supported as a Ph.D. student by University of Milano - Bicocca, Milan, Italy and Università della Svizzera italiana, Lugano, Switzerland. Andrea Cappozzo and Francesca Greselin’s work was supported by Milano-Bicocca University Fund for Scientific Research, 2019-ATE-0076
\section*{Appendix} \label{sec:appendix_B}
 
\textbf{Gibbs sampling algorithm for model \eqref{eq::lik}-\eqref{eq:prior_spec}}\\
\begin{algorithm}[H]
\SetAlgoLined
\KwInput{Initial values for the MCMC, robust estimates from $\mathbf{X}$.}
\KwOutput{Posterior MCMC sample for the parameters of interest.}
\For{$i=1,\ldots,I$}{
    1. Sample every $u_m$ from a uniform distribution $\mathcal{U}\left(0, \xi_{\zeta_m}\right)$.\\
    2. Compute the stochastic truncation term $L^*$ according to \eqref{eq:truncate}.\\
    3. Let $m_j=\sum_{m=1}^M\mathbbm{1}_{\{\alpha_m=j\}},$ with $j=0,\ldots,J$. Sample $\mathbf{\pi}$ from a conjugate Dirichlet distribution:
    $$\boldsymbol{\pi}\sim Dir(a_0+m_0,a_1+m_1,\ldots,a_J+m_J).$$
    4. Sample the SB variables after integrating out $\bm{u}$: $$
    \begin{aligned}
    v_k|\cdots \sim Beta ( &1+
	\sum_{m=1}^{M}\mathbbm{1}_{\{\alpha_m=0 \:\cap\: \beta_m = k\}} ,\\ &\gamma+\sum_{m=1}^{M}\mathbbm{1}_{\{\alpha_m=0 \:\cap\: \beta_m > k\}} ).
	\end{aligned}$$ \\
    5. Compute the SB weights according to \eqref{eq::SB}\\
    6. Compute the one-line probability weights $\tilde{\boldsymbol{\pi}}$ according to \eqref{Oneline}. \\
    7. Sample the atoms for the observed classes $\mathbf{\Theta}_{j,0}$ exploiting conjugacy between the likelihood and the prior for $j=1,\ldots,J$.\\
    8. Sample the atoms for the novel classes $\mathbf{\Theta}^{nov}_{0,l}$ exploiting conjugacy between the likelihood and the prior for $l=1,\ldots,L^*$.\\
    9. Obtain $\tilde{\boldsymbol{\Theta}}$ concatenating the updated values of $\mathbf{\Theta}$ and $\mathbf{\Theta}^{nov}$.\\
    10. Sample each $\xi_m$ from the following joint discrete distribution:
    \begin{equation*}
    \begin{aligned}
        \mathbb{P}&\left(\zeta_m=l\right)  \propto
        \frac{\tilde{\pi_l}}{\xi_l}\mathbbm{1}_{\{u_m<\xi_l\}} \phi\left(\mathbf{y}_m|\tilde{\mathbf{\Theta}}_{l} \right),
        \quad l= 1,\ldots, L^*, \\
        \mathbb{P}&\left(\text{otherwise}\right)  \propto 0.
    \end{aligned}
    \end{equation*}\\
    11. Recover the values for the membership vectors $\bm{\alpha}$ and $\bm{\beta}$ using \eqref{eq:CsiAlphaBeta}. Divide the elements in $\tilde{\boldsymbol{\Theta}}$ into $\mathbf{\Theta}$ and $\mathbf{\Theta}^{nov}$.
 }
 \caption{Efficient Slice Sampler for the BNP-Novelty detection model}
\end{algorithm}

\printbibliography[heading=bibintoc]

@article{Akhanli2020,
archivePrefix = {arXiv},
arxivId = {2002.01822},
author = {Akhanli, Serhat Emre and Hennig, Christian},
doi = {10.1007/s11222-020-09958-2},
eprint = {2002.01822},
issn = {0960-3174},
journal = {Statistics and Computing},
month = {sep},
number = {5},
pages = {1523--1544},
publisher = {Springer US},
title = {{Comparing clusterings and numbers of clusters by aggregation of calibrated clustering validity indexes}},
url = {https://doi.org/10.1007/s11222-020-09958-2},
volume = {30},
year = {2020}
}

@article{Aeberhard1993,
author = {Aeberhard, Stefan and Coomans, Danny and Vel, Olivier De},
doi = {10.1002/cem.1180070204},
issn = {0886-9383},
journal = {Journal of Chemometrics},
month = {mar},
number = {2},
pages = {99--115},
title = {{Improvements to the classification performance of RDA}},
url = {http://doi.wiley.com/10.1002/cem.1180070204},
volume = {7},
year = {1993}
}

@article{manikopoulos2002network,
	title        = {Network intrusion and fault detection: a statistical anomaly approach},
	author       = {Manikopoulos, Constantine and Papavassiliou, Symeon},
	year         = 2002,
	journal      = {IEEE Communications Magazine},
	publisher    = {IEEE},
	volume       = 40,
	number       = 10,
	pages        = {76--82}
}

@inproceedings{tax1998outlier,
	title        = {Outlier detection using classifier instability},
	author       = {Tax, David MJ and Duin, Robert PW},
	year         = 1998,
	booktitle    = {Joint IAPR international workshops on statistical techniques in pattern recognition (SPR) and structural and syntactic pattern recognition (SSPR)},
	pages        = {593--601},
	organization = {Springer}
}

@article{tarassenko1995novelty,
author = {Tarassenko, L. and Hayton, P. and Cerneaz, N. and Brady, M.},
doi = {10.1049/cp:19950597},
issn = {05379989},
journal = {IEE Conference Publication},
number = {409},
pages = {442--447},
title = {{Novelty detection for the identification of masses in mammograms}},
year = {1995}
}

@inproceedings{carpenter1997artmap,
	title        = {ARTMAP-FD: familiarity discrimination applied to radar target recognition},
	author       = {Carpenter, Gail A and Rubin, Mark A and Streilein, William W},
	year         = 1997,
	booktitle    = {Proceedings of International Conference on Neural Networks (ICNN'97)},
	volume       = 3,
	pages        = {1459--1464},
	organization = {IEEE}
}

@article{hubert2010minimum,
	title        = {Minimum covariance determinant},
	author       = {Hubert, Mia and Debruyne, Michiel},
	year         = 2010,
	journal      = {Wiley interdisciplinary reviews: Computational statistics},
	publisher    = {Wiley Online Library},
	volume       = 2,
	number       = 1,
	pages        = {36--43}
}

@article{rousseeuw1984least,
	title        = {Least median of squares regression},
	author       = {Rousseeuw, Peter J},
	year         = 1984,
	journal      = {Journal of the American statistical association},
	publisher    = {Taylor \& Francis},
	volume       = 79,
	number       = 388,
	pages        = {871--880}
}

@article{Murphy2012,
	title        = {{Variable selection and updating in model-based discriminant analysis for high dimensional data with food authenticity applications}},
	author       = {Murphy, Thomas Brendan and Dean, Nema and Raftery, Adrian E},
	year         = 2010,
	month        = mar,
	journal      = {The Annals of Applied Statistics},
	volume       = 4,
	number       = 1,
	pages        = {396--421},
	doi          = {10.1214/09-AOAS279},
	isbn         = {1932-6157 (Print) 1932-6157 (Linking)},
	issn         = {1932-6157},
	url          = {https://projecteuclid.org/download/pdfview{\_}1/euclid.aoas/1273584460 http://projecteuclid.org/euclid.aoas/1273584460},
	archiveprefix = {arXiv},
	arxivid      = {0910.2585},
	eprint       = {0910.2585},
	pmid         = 20936055
}

@article{forina1986multivariate,
	title        = {{Multivariate data analysis as a discriminating method of the origin of wines}},
	author       = {Forina, M and Armanino, C and Castino, M and Ubigli, M},
	year         = 1986,
	journal      = {Vitis},
	volume       = 25,
	number       = 3,
	pages        = {189--201}
}

@article{Blei2017,
	title        = {{Variational Inference: A Review for Statisticians}},
	author       = {Blei, David M. and Kucukelbir, Alp and McAuliffe, Jon D.},
	year         = 2017,
	journal      = {Journal of the American Statistical Association},
	publisher    = {Taylor {\&} Francis},
	volume       = 112,
	number       = 518,
	pages        = {859--877},
	doi          = {10.1080/01621459.2017.1285773},
	issn         = {1537274X},
	url          = {https://doi.org/10.1080/01621459.2017.1285773}
}

@article{McElhinney1999,
	title        = {{Chemometric processing of visible and near infrared reflectance spectra for species identification in selected raw homogenised meats}},
	author       = {McElhinney, John and Downey, Gerard and Fearn, Tom},
	year         = 1999,
	journal      = {Journal of Near Infrared Spectroscopy},
	volume       = 7,
	number       = 3,
	pages        = {145--154},
	doi          = {10.1255/jnirs.245},
	isbn         = {0967-0335},
	issn         = {09670335}
}

@article{Sethuraman1994a,
	title        = {{A constructive definition of Dirichlet Process prior}},
	author       = {Sethuraman, Jayaram},
	year         = 1994,
	journal      = {Statistica Sinica},
	publisher    = {Institute of Statistical Science, Academia Sinica},
	volume       = 4,
	number       = 2,
	pages        = {639--650},
	issn         = {10170405, 19968507},
	url          = {http://www.jstor.org/stable/24305538}
}

@article{Todorov2009a,
	title        = {{An Object-Oriented Framework for Robust Multivariate Analysis}},
	author       = {Todorov, Valentin and Filzmoser, Peter},
	year         = 2009,
	journal      = {Journal of Statistical Software},
	volume       = 32,
	number       = 3,
	pages        = {1--47},
	doi          = {10.18637/jss.v032.i03},
	issn         = {1548-7660},
	url          = {http://citeseerx.ist.psu.edu/viewdoc/download?doi=10.1.1.150.3305{\&}rep=rep1{\&}type=pdf http://www.jstatsoft.org/v32/i03/}
}

@article{Kalli2011,
	title        = {{Slice sampling mixture models}},
	author       = {Kalli, Maria and Griffin, Jim E and Walker, Stephen G},
	year         = 2011,
	journal      = {Statistics and Computing},
	volume       = 21,
	number       = 1,
	pages        = {93--105},
	doi          = {10.1007/s11222-009-9150-y},
	issn         = {09603174},
	url          = {https://link-springer-com.proxy.unimib.it/content/pdf/10.1007{\%}2Fs11222-009-9150-y.pdf}
}

@article{Wade2018,
	title        = {{Bayesian Cluster Analysis: Point estimation and credible balls (with Discussion)}},
	author       = {Wade, Sara and Ghahramani, Zoubin},
	year         = 2018,
	journal      = {Bayesian Analysis},
	volume       = 13,
	number       = 2,
	pages        = {559--626},
	doi          = {10.1214/17-BA1073},
	issn         = 19316690,
	url          = {https://doi.org/10.1214/17-BA1073},
	archiveprefix = {arXiv},
	arxivid      = {1505.03339},
	eprint       = {1505.03339}
}

@article{FernandezPierna2007a,
	title        = {{Chemometric contest at ‘Chimiom{\'{e}}trie 2005': A discrimination study}},
	author       = {{Fern{\'{a}}ndez Pierna}, Juan Antonio and Dardenne, Pierre},
	year         = 2007,
	month        = apr,
	journal      = {Chemometrics and Intelligent Laboratory Systems},
	volume       = 86,
	number       = 2,
	pages        = {219--223},
	doi          = {10.1016/j.chemolab.2006.06.009},
	issn         = {01697439},
	url          = {https://linkinghub.elsevier.com/retrieve/pii/S0169743906001444}
}

@article{Dempster1977,
	title        = {{Maximum likelihood from incomplete data via the EM algorithm}},
	author       = {Dempster, A and {N. Laird} and Rubin, D},
	year         = 1977,
	journal      = {Journal of the Royal Statistical Society},
	volume       = 39,
	number       = 1,
	pages        = {1--38},
	doi          = {http://dx.doi.org/10.2307/2984875},
	isbn         = {0000000779},
	issn         = {00359246},
	url          = {http://www.jstor.org/stable/2984875},
	archiveprefix = {arXiv},
	arxivid      = {0710.5696v2},
	eprint       = {0710.5696v2},
	pmid         = 9501024
}

@article{Rand1971,
	title        = {{Objective criteria for the evaluation of clustering methods}},
	author       = {Rand, William M},
	year         = 1971,
	month        = dec,
	journal      = {Journal of the American Statistical Association},
	volume       = 66,
	number       = 336,
	pages        = 846,
	doi          = {10.2307/2284239},
	issn         = {01621459},
	url          = {https://www.jstor.org/stable/2284239?origin=crossref}
}

@book{RamsayJamesSilverman,
	title        = {{Functional Data Analysis}},
	author       = {{Ramsay, James, Silverman}, B. W},
	year         = 2005,
	booktitle    = {Springer Series in Statistics},
	publisher    = {Springer-Verlag},
	address      = {New York},
	series       = {Springer Series in Statistics},
	doi          = {10.1007/b98888},
	isbn         = {0-387-40080-X},
	url          = {http://link.springer.com/10.1007/b98888}
}

@article{fop2018unobserved,
archivePrefix = {arXiv},
arxivId = {2102.01982},
author = {Fop, Michael and Mattei, Pierre-Alexandre and Bouveyron, Charles and Murphy, Thomas Brendan},
eprint = {2102.01982},
month = {feb},
title = {{Unobserved classes and extra variables in high-dimensional discriminant analysis}},
url = {http://arxiv.org/abs/2102.01982},
year = {2021}
}

@article{Bouveyron,
	title        = {{Adaptive mixture discriminant analysis for supervised learning with unobserved classes}},
	author       = {Bouveyron, Charles},
	year         = 2014,
	journal      = {Journal of Classification},
	volume       = 31,
	number       = 1,
	pages        = {49--84},
	doi          = {10.1007/s00357-014-9147-x},
	issn         = 14321343,
	url          = {https://link.springer.com/content/pdf/10.1007/s00357-014-9147-x.pdf}
}

@article{Malsiner-Walli2016,
	title        = {{Model-based clustering based on sparse finite Gaussian mixtures}},
	author       = {Malsiner-Walli, Gertraud and Fr{\"{u}}hwirth-Schnatter, Sylvia and Gr{\"{u}}n, Bettina},
	year         = 2016,
	month        = jan,
	journal      = {Statistics and Computing},
	publisher    = {Springer US},
	volume       = 26,
	number       = {1-2},
	pages        = {303--324},
	doi          = {10.1007/s11222-014-9500-2},
	issn         = {0960-3174},
	url          = {http://dx.doi.org/10.1007/s11222-014-9500-2 http://link.springer.com/10.1007/s11222-014-9500-2}
}

@article{Hubert2004,
	title        = {{Fast and robust discriminant analysis}},
	author       = {Hubert, Mia and {Van Driessen}, Katrien},
	year         = 2004,
	month        = mar,
	journal      = {Computational Statistics {\&} Data Analysis},
	volume       = 45,
	number       = 2,
	pages        = {301--320},
	doi          = {10.1016/S0167-9473(02)00299-2},
	issn         = {01679473},
	url          = {https://linkinghub.elsevier.com/retrieve/pii/S0167947302002992}
}

@article{Driessen1999,
	title        = {{A fast algorithm for the minimum covariance determinant estimator}},
	author       = {Rousseeuw, Peter J. and Driessen, Katrien Van},
	year         = 1999,
	month        = aug,
	journal      = {Technometrics},
	volume       = 41,
	number       = 3,
	pages        = {212--223},
	doi          = {10.1080/00401706.1999.10485670},
	issn         = {0040-1706},
	url          = {http://www.tandfonline.com/doi/abs/10.1080/00401706.1999.10485670}
}

@article{Gordaliza1991,
	title        = {{Best approximations to random variables based on trimming procedures}},
	author       = {Gordaliza, Alfonso},
	year         = 1991,
	journal      = {Journal of Approximation Theory},
	volume       = 64,
	number       = 2,
	pages        = {162--180},
	doi          = {10.1016/0021-9045(91)90072-I},
	issn         = 10960430
}

@book{Ritter2015,
	title        = {{Robust Cluster Analysis and Variable Selection}},
	author       = {Ritter, Gunter},
	year         = 2014,
	month        = sep,
	publisher    = {Chapman and Hall/CRC},
	doi          = {10.1201/b17353},
	isbn         = 9781439857977,
	url          = {https://www.taylorfrancis.com/books/9781439857977}
}

@article{Escobar1995,
	title        = {{Bayesian Density Estimation and Inference Using Mixtures}},
	author       = {Escobar, Michael D. and West, Mike},
	year         = 1995,
	month        = jun,
	journal      = {Journal of the American Statistical Association},
	volume       = 90,
	number       = 430,
	pages        = {577--588},
	doi          = {10.1080/01621459.1995.10476550},
	issn         = {0162-1459},
	url          = {http://www.tandfonline.com/doi/abs/10.1080/01621459.1995.10476550}
}

@article{Hennig2015,
	title        = {{What are the true clusters?}},
	author       = {Hennig, Christian},
	year         = 2015,
	journal      = {Pattern Recognition Letters},
	volume       = 64,
	pages        = {53--62},
	doi          = {10.1016/j.patrec.2015.04.009},
	isbn         = {01678655},
	issn         = {01678655},
	url          = {http://dx.doi.org/10.1016/j.patrec.2015.04.009},
	archiveprefix = {arXiv},
	arxivid      = {1502.02555},
	eprint       = {1502.02555}
}

@article{Croux1999,
	title        = {{Influence Function and Efficiency of the Minimum Covariance Determinant Scatter Matrix Estimator}},
	author       = {Croux, Christophe and Haesbroeck, Gentiane},
	year         = 1999,
	journal      = {Journal of Multivariate Analysis},
	volume       = 71,
	number       = 2,
	pages        = {161--190},
	doi          = {10.1006/jmva.1999.1839},
	issn         = {0047259X}
}

@article{Ishwaran2001a,
	title        = {{Gibbs Sampling Methods for Stick-Breaking Priors}},
	author       = {Ishwaran, Hemant and James, Lancelot F.},
	year         = 2001,
	month        = mar,
	journal      = {Journal of the American Statistical Association},
	volume       = 96,
	number       = 453,
	pages        = {161--173},
	doi          = {10.1198/016214501750332758},
	issn         = {0162-1459},
	url          = {http://www.tandfonline.com/doi/abs/10.1198/016214501750332758}
}

@article{Binder1978,
	title        = {{Bayesian Cluster Analysis}},
	author       = {Binder, D. A.},
	year         = 1978,
	month        = apr,
	journal      = {Biometrika},
	volume       = 65,
	number       = 1,
	pages        = 31,
	doi          = {10.2307/2335273},
	issn         = {00063444},
	url          = {https://www.jstor.org/stable/2335273?origin=crossref}
}

@article{Scarpa2009,
	title        = {{Bayesian hierarchical functional data analysis via contaminated informative priors}},
	author       = {Scarpa, Bruno and Dunson, David B.},
	year         = 2009,
	journal      = {Biometrics},
	volume       = 65,
	number       = 3,
	pages        = {772--780},
	doi          = {10.1111/j.1541-0420.2008.01163.x},
	issn         = {0006341X}
}

@article{Charytanowicz2010,
	title        = {{Complete gradient clustering algorithm for features analysis of X-ray images}},
	author       = {Charytanowicz, Ma{\l}gorzata and Niewczas, Jerzy and Kulczycki, Piotr and Kowalski, Piotr A. and {\L}ukasik, Szymon and Zak, S{\l}awomir},
	year         = 2010,
	journal      = {Advances in Intelligent and Soft Computing},
	volume       = 69,
	pages        = {15--24},
	doi          = {10.1007/978-3-642-13105-9_2},
	isbn         = 9783642131042,
	issn         = 18675662
}

@article{Bigelow2009,
	title        = {{Bayesian semiparametric joint models for functional predictors}},
	author       = {Bigelow, Jamie L. and Dunson, David B.},
	year         = 2009,
	journal      = {Journal of the American Statistical Association},
	volume       = 104,
	number       = 485,
	pages        = {26--36},
	doi          = {10.1198/jasa.2009.0001},
	issn         = {01621459}
}

@article{Rodriguez2014,
	title        = {{Functional clustering in nested designs: Modeling variability in reproductive epidemiology studies}},
	author       = {Rodriguez, Abel and Dunson, David B.},
	year         = 2014,
	journal      = {Annals of Applied Statistics},
	volume       = 8,
	number       = 3,
	pages        = {1416--1442},
	doi          = {10.1214/14-AOAS751},
	issn         = 19417330
}

@article{Petrone2009,
	title        = {{Hybrid dirichlet mixture models for functional data}},
	author       = {Petrone, Sonia and Guindani, Michele and Gelfand, Alan E.},
	year         = 2009,
	journal      = {Journal of the Royal Statistical Society. Series B: Statistical Methodology},
	volume       = 71,
	number       = 4,
	pages        = {755--782},
	doi          = {10.1111/j.1467-9868.2009.00708.x},
	issn         = 13697412
}

@article{canale2017,
	title        = {{On the Pitman-Yor process with spike and slab base measure}},
	author       = {Canale, A and Lijoi, A and Nipoti, B and Pr{\"{u}}nster, I},
	year         = 2017,
	journal      = {Biometrika},
	volume       = 104,
	number       = 3,
	pages        = {681--697},
	doi          = {10.1093/biomet/asx041},
	issn         = 14643510,
	url          = {https://academic.oup.com/biomet/article-abstract/104/3/681/4061289}
}

@article{Rigon,
archivePrefix = {arXiv},
arxivId = {1907.02493},
author = {Rigon, Tommaso},
eprint = {1907.02493},
issn = {23318422},
journal = {arXiv},
title = {{An enriched mixture model for functional clustering}},
url = {http://arxiv.org/abs/1907.02493},
year = {2019}
}

@article{Boudt2020,
	title        = {{The minimum regularized covariance determinant estimator}},
	author       = {Boudt, Kris and Rousseeuw, Peter J. and Vanduffel, Steven and Verdonck, Tim},
	year         = 2020,
	month        = feb,
	journal      = {Statistics and Computing},
	volume       = 30,
	number       = 1,
	pages        = {113--128},
	doi          = {10.1007/s11222-019-09869-x},
	issn         = {0960-3174},
	url          = {http://link.springer.com/10.1007/s11222-019-09869-x},
	archiveprefix = {arXiv},
	arxivid      = {1701.07086},
	eprint       = {1701.07086}
}

@article{Hubert2018,
	title        = {{Minimum covariance determinant and extensions}},
	author       = {Hubert, Mia and Debruyne, Michiel and Rousseeuw, Peter J.},
	year         = 2018,
	journal      = {Wiley Interdisciplinary Reviews: Computational Statistics},
	volume       = 10,
	number       = 3,
	pages        = {1--11},
	doi          = {10.1002/wics.1421},
	isbn         = {1939-0068},
	issn         = 19390068,
	archiveprefix = {arXiv},
	arxivid      = {1709.07045},
	eprint       = {1709.07045}
}

@inproceedings{Singh2019,
	title        = {{Comparison of Machine Learning Models in Food Authentication Studies}},
	author       = {Singh, Manokamna and Domijan, Katarina},
	year         = 2019,
	month        = jun,
	booktitle    = {2019 30th Irish Signals and Systems Conference (ISSC)},
	publisher    = {IEEE},
	pages        = {1--6},
	doi          = {10.1109/ISSC.2019.8904924},
	isbn         = {978-1-7281-2800-9},
	url          = {https://ieeexplore.ieee.org/document/8904924/},
	archiveprefix = {arXiv},
	arxivid      = {1905.07302},
	eprint       = {1905.07302}
}

@article{Crane2003,
	title        = {{Functional data analysis view of functional near infrared spectroscopy data}},
	author       = {Barati, Zeinab and Zakeri, Issa and Pourrezaei, Kambiz},
	year         = 2013,
	month        = nov,
	journal      = {Journal of Biomedical Optics},
	volume       = 18,
	number       = 11,
	pages        = 117007,
	doi          = {10.1117/1.JBO.18.11.117007},
	issn         = {1083-3668},
	url          = {http://biomedicaloptics.spiedigitallibrary.org/article.aspx?doi=10.1117/1.JBO.18.11.117007}
}

@article{Cappozzo2019e,
	title        = {{Anomaly and Novelty detection for robust semi-supervised learning}},
	author       = {Cappozzo, Andrea and Greselin, Francesca and Murphy, Thomas Brendan},
	year         = 2020,
	month        = sep,
	journal      = {Statistics and Computing},
	volume       = 30,
	number       = 5,
	pages        = {1545--1571},
	doi          = {10.1007/s11222-020-09959-1},
	issn         = {0960-3174},
	url          = {http://arxiv.org/abs/1911.08381 http://link.springer.com/10.1007/s11222-020-09959-1},
	archiveprefix = {arXiv},
	arxivid      = {1911.08381},
	eprint       = {1911.08381}
}

@inproceedings{Miller2003,
	title        = {{A mixture model and EM algorithm for robust classification, outlier rejection, and class discovery}},
	author       = {Miller, D.J. and Browning, John},
	year         = 2003,
	booktitle    = {2003 IEEE International Conference on Acoustics, Speech, and Signal Processing, 2003. Proceedings. (ICASSP '03).},
	publisher    = {IEEE},
	volume       = 2,
	number       = 11,
	pages        = {II--809--12},
	doi          = {10.1109/ICASSP.2003.1202490},
	isbn         = {0-7803-7663-3},
	url          = {http://ieeexplore.ieee.org/document/1318048/ http://ieeexplore.ieee.org/document/1202490/}
}

@article{Rousseau2011,
	title        = {{Asymptotic behaviour of the posterior distribution in overfitted mixture models}},
	author       = {Rousseau, Judith and Mengersen, Kerrie},
	year         = 2011,
	journal      = {Journal of the Royal Statistical Society. Series B: Statistical Methodology},
	volume       = 73,
	number       = 5,
	pages        = {689--710},
	doi          = {10.1111/j.1467-9868.2011.00781.x},
	issn         = 13697412
}

@article{Ferguson1973,
	title        = {{A Bayesian Analysis of Some Nonparametric Problems}},
	author       = {Ferguson, Thomas S.},
	year         = 1973,
	journal      = {The Annals of Statistics},
	publisher    = {The Annals of Statistics},
	volume       = 1,
	number       = 2,
	pages        = {209--230},
	doi          = {10.1214/aos/1176342360},
	isbn         = 9780874216561,
	issn         = {0090-5364},
	archiveprefix = {arXiv},
	arxivid      = {arXiv:1011.1669v3},
	eprint       = {arXiv:1011.1669v3},
	pmid         = 15991970
}

@article{Khan2014,
	title        = {{One-class classification: taxonomy of study and review of techniques}},
	author       = {Khan, Shehroz S. and Madden, Michael G.},
	year         = 2014,
	month        = jun,
	journal      = {The Knowledge Engineering Review},
	volume       = 29,
	number       = 3,
	pages        = {345--374},
	doi          = {10.1017/S026988891300043X},
	issn         = {0269-8889},
	url          = {https://www.cambridge.org/core/product/identifier/S026988891300043X/type/journal{\_}article},
	archiveprefix = {arXiv},
	arxivid      = {1312.0049},
	eprint       = {1312.0049}
}

@misc{Strumio1999,
	title        = {{Computer system for analysis of x-ray images of wheat grains (a preliminary announcement)}},
	author       = {Strumi{\l}{\l}o, A. and Niewczas, J. and Szczypi{\'{n}}ski, P. and Makowski, P. and Wo{\'{z}}niak, W.},
	year         = 1999,
	booktitle    = {International Agrophysics},
	volume       = 13,
	number       = 1,
	pages        = {133--140},
	issn         = {02368722}
}

@article{Abraham2003,
	title        = {{Unsupervised curve clustering using B-splines}},
	author       = {Abraham, C. and Cornillon, P. A. and Matzner-L{\o}ber, Eric and Molinari, N.},
	year         = 2003,
	journal      = {Scandinavian Journal of Statistics},
	volume       = 30,
	number       = 3,
	pages        = {581--595},
	doi          = {10.1111/1467-9469.00350},
	issn         = {03036898}
}

@article{Butler1993,
	title        = {{Asymptotics for the Minimum Covariance Determinant Estimator}},
	author       = {Butler, R. W. and Davies, P. L. and Jhun, M.},
	year         = 1993,
	month        = sep,
	journal      = {The Annals of Statistics},
	volume       = 21,
	number       = 3,
	pages        = {1385--1400},
	doi          = {10.1214/aos/1176349264},
	issn         = {0090-5364},
	url          = {http://projecteuclid.org/euclid.aop/1176996548 http://projecteuclid.org/euclid.aos/1176349264}
}

@article{Cator2012,
	title        = {{Central limit theorem and influence function for the MCD estimators at general multivariate distributions}},
	author       = {Cator, Eric A. and Lopuha{\"{a}}, Hendrik P.},
	year         = 2012,
	month        = may,
	journal      = {Bernoulli},
	volume       = 18,
	number       = 2,
	pages        = {520--551},
	doi          = {10.3150/11-BEJ353},
	issn         = {1350-7265},
	url          = {https://projecteuclid.org/euclid.bj/1334580723}
}

@article{Lau2007,
	title        = {{Bayesian model-based clustering procedures}},
	author       = {Lau, John W. and Green, Peter J.},
	year         = 2007,
	journal      = {Journal of Computational and Graphical Statistics},
	volume       = 16,
	number       = 3,
	pages        = {526--558},
	doi          = {10.1198/106186007X238855},
	issn         = 10618600
}

@article{MeilaM2007,
	title        = {{Comparing clusterings-an information based distance}},
	author       = {Meilǎ, Marina},
	year         = 2007,
	journal      = {Journal of Multivariate Analysis},
	volume       = 98,
	number       = 5,
	pages        = {873--895},
	doi          = {10.1016/j.jmva.2006.11.013},
	issn         = {0047259X}
}

@book{Gpras,
	title        = {Gaussian Processes for Machine Learning (Adaptive Computation and Machine Learning)},
	author       = {Rasmussen, Carl Edward and Williams, Christopher K. I.},
	year         = 2005,
	publisher    = {The MIT Press},
	isbn         = {026218253X}
}

@article{Yang2010,
	title        = {{Semiparametric Bayes hierarchical models with mean and variance constraints}},
	author       = {Yang, Mingan and Dunson, David B. and Baird, Donna},
	year         = 2010,
	journal      = {Computational Statistics and Data Analysis},
	volume       = 54,
	number       = 9,
	pages        = {2172--2186},
	doi          = {10.1016/j.csda.2010.03.025},
	issn         = {01679473}
}

@article{Maronna2017,
	title        = {{Robust and efficient estimation of multivariate scatter and location}},
	author       = {Maronna, Ricardo A and Yohai, Victor J},
	year         = 2017,
	journal      = {Computational Statistics and Data Analysis},
	volume       = 109,
	pages        = {64--75},
	doi          = {10.1016/j.csda.2016.11.006},
	issn         = {01679473},
	url          = {www.elsevier.com/locate/csda}
}

@article{Gutierrez2014,
	title        = {{Bayesian nonparametric classification for spectroscopy data}},
	author       = {Guti{\'{e}}rrez, Luis and Guti{\'{e}}rrez-Pe{\~{n}}a, Eduardo and Mena, Rams{\'{e}}s H.},
	year         = 2014,
	journal      = {Computational Statistics and Data Analysis},
	volume       = 78,
	pages        = {56--68},
	doi          = {10.1016/j.csda.2014.04.010},
	issn         = {01679473}
}

@article{DeBlasi2020,
	title        = {{On the inferential implications of decreasing weight structures in mixture models}},
	author       = {{De Blasi}, Pierpaolo and Mart{\'{i}}nez, Asael Fabian and Mena, Rams{\'{e}}s H. and Pr{\"{u}}nster, Igor},
	year         = 2020,
	journal      = {Computational Statistics and Data Analysis},
	volume       = 147,
	doi          = {10.1016/j.csda.2020.106940},
	issn         = {01679473}
}

@book{Boor2001,
	title        = {{A Practical Guide to Splines - Revised Edition}},
	author       = {de Boor, Carl},
	year         = 2001,
	booktitle    = {Springer-Verlag, New York},
	isbn         = 9780387953663
}

@article{Pitman1997,
	title        = {{The two-parameter Poisson-Dirichlet distribution derived from a stable subordinator}},
	author       = {Pitman, Jim and Yor, Marc},
	year         = 1997,
	journal      = {Annals of Probability},
	volume       = 25,
	number       = 2,
	pages        = {855--900},
	doi          = {10.1214/aop/1024404422},
	isbn         = {0091-1798},
	issn         = {00911798},
	pmid         = 17925410
}

@article{Pitman1995,
	title        = {{Exchangeable and partially exchangeable random partitions}},
	author       = {Pitman, Jim},
	year         = 1995,
	journal      = {Probability Theory and Related Fields},
	volume       = 102,
	number       = 2,
	pages        = {145--158},
	doi          = {10.1007/BF01213386},
	issn         = {01788051}
}

@article{Lo1984,
	title        = {{On a Class of Bayesian Nonparametric Estimates: I. Density Estimates}},
	author       = {Lo, Albert Y.},
	year         = 1984,
	journal      = {The Annals of Statistics},
	volume       = 12,
	number       = 1,
	pages        = {351--357},
	url          = {https://www.jstor.org/stable/pdf/2241054.pdf}
}

\title{Supplementary Material}
\author{}
\maketitle
In this Supplementary Material, we report proofs for the theoretical results reported in the main paper and some additional numerical experiments, for both the multivariate Brand and its functional extension.

\section{Proofs}

First, we derive the moments, the variance, and the covariance for the 
simpler case (A): $\tilde{p}_0 = \sum_{j=0}^J\pi_j\:\delta_{\bm{\Theta}_j}$. Then, we derive the same quantities starting from the discrete random measure that underlies Brand (B): $\tilde{p}=\sum_{j=1}^J\pi_j\:\delta_{\bm{\Theta}_j}+\pi_0\left(\sum_{l\geq 1} \omega_l\: \delta_{\bm{\Theta}^{nov}_l}\right)$.

\subsection{Case A:  \texorpdfstring{$\tilde{p}_0 = \sum_{j=0}^J\pi_j\:\delta_{\bm{\Theta}_j}$}{TEXT}}


\begin{equation*}
\begin{aligned}
\mathbb{E} \left[ \boldsymbol{\Theta}_m \right] &= \mathbb{E} \left[ \mathbb{E} \left[ \boldsymbol{\Theta}_m|\tilde{p}_0 \right]\right] =
\mathbb{E} \left[ \sum_{j=0}^J\pi_j\bm{\Theta}_j\right] =
\sum_{j=0}^J \mathbb{E} \left[ \pi_j\bm{\Theta}_j\right] =
\sum_{j=0}^J \mathbb{E} \left[ \pi_j\right]\mathbb{E} \left[\bm{\Theta}_j\right]
=\sum_{j=0}^J \frac{a_j}{a} \mu_j. \\
\mathbb{E} \left[ \boldsymbol{ \Theta}_m \right]^2 & = \sum_{j=0}^J \frac{a^2_j}{a^2} \mu_j^2+ 2\sum_{j>l\geq 0}\frac{a_j a_l}{a^2}\mu_j\mu_l.\\
\mathbb{E} \left[ \boldsymbol{ \Theta}^2_m \right] &=  \sum_{j=0}^J\frac{a_j}{a}\mathbb{E} \left[\bm{\Theta}^2_j\right] =\mathbb{E} \left[ \sum_{j=0}^J\pi_j\bm{\Theta}^2_j\right]= \sum_{j=0}^J \frac{a_j}{a} \mu_{j,2}.\\
\end{aligned}
\end{equation*}

\begin{equation*}
\begin{aligned}
\mathbb{E}\left[\boldsymbol{\Theta}_m\boldsymbol{\Theta}_{m'}\right]&=\mathbb{E}\left[\mathbb{E}\left[\boldsymbol{\Theta}_m\boldsymbol{\Theta}_{m'}|\tilde{p}\right]\right]=\mathbb{E}\left[\mathbb{E}\left[\boldsymbol{\Theta}_m|\tilde{p}\right]\mathbb{E}\left[\boldsymbol{\Theta}_{m'}|\tilde{p}\right]\right]=\mathbb{E}\left[\left(\sum_{j=0}^J\pi_j\bm{\Theta}_j\right)^2\right]\\
&=\sum_{j=0}^J\mathbb{E}\left[\pi_j^2\right]\mathbb{E}\left[\boldsymbol{\Theta}_j^2\right]+
2\sum_{j>l\geq 0}\mathbb{E}\left[\pi_j\pi_l\right]\mathbb{E}\left[\boldsymbol{\Theta}_j\right]\mathbb{E}\left[\boldsymbol{\Theta}_l\right]
\\&=\sum_{j=0}^J \frac{a_j(a_j+1)}{a(a+1)} \mu_{j,2}+ 2\sum_{j>l\geq 0}\frac{a_j a_l}{a(a+1)}\mu_j\mu_l.\\
\end{aligned}
\end{equation*}

\begin{equation*}
\begin{aligned}
\mathbb{V}\left[\boldsymbol{\Theta}_m\right]&=\sum_{j=0}^J \frac{a_j}{a} \mu_{j,2} -\sum_{j=0}^J \frac{a^2_j}{a^2} \mu_j^2- 2\sum_{j>l\geq 0}\frac{a_j a_l}{a^2}\mu_j\mu_l \\
&=\sum_{j=0}^J
    \frac{a_j}{a}\left(\mu_{j,2}-\frac{a_j}{a}\mu_{j}^2\right)-2\sum^J_{l>j\geq 0}\frac{a_ja_l}{a^2}\mu_l\mu_j.\\
Cov_0(\boldsymbol{\Theta}_m,\boldsymbol{\Theta}_{m'})&=\mathbb{E}\left[\boldsymbol{\Theta}_m\boldsymbol{\Theta}_{m'}\right]-\mathbb{E} \left[ \boldsymbol{\Theta}_m \right]\mathbb{E} \left[ \boldsymbol{\Theta}_{m'} \right]\\&=\sum_{j=0}^J \frac{a_j(a_j+1)}{a(a+1)} \mu_{j,2}+ 2\sum_{j>l\geq 0}\frac{a_j a_l}{a(a+1)}\mu_j\mu_l-
\sum_{j=0}^J \frac{a^2_j}{a^2} \mu_j^2- 2\sum_{j>l\geq 0}\frac{a_j a_l}{a^2}\mu_j\mu_l\\&=
\sum_{j=0}^J \left(\frac{a_j(a_j+1)}{a(a+1)} \mu_{j,2} - \frac{a^2_j}{a^2} \mu_j^2\right) - 2\sum_{j>l\geq 0}\frac{a_j a_l}{a^2(a+1)}\mu_j\mu_l.
\end{aligned}
\end{equation*}

\subsection{Case B: \texorpdfstring{$\tilde{p}=\sum_{j=0}^J\pi_j\:\delta_{\bm{\Theta}_j}+\pi_0\left(\sum_{l\geq 1} \omega_l\: \delta_{\bm{\Theta}^{nov}_l}\right)$}{TEXT}}

\begin{equation*}
\begin{aligned}
\mathbb{E} \left[ \boldsymbol{\Theta}_m \right] &= \sum_{j=0}^J \frac{a_j}{a} \mu_j. \\
\mathbb{E} \left[ \boldsymbol{ \Theta}_m \right]^2 &= \sum_{j=0}^J \frac{a^2_j}{a^2} \mu_j^2+ 2\sum_{j>l\geq 0}\frac{a_j a_l}{a^2}\mu_j\mu_l.\\
\mathbb{E} \left[ \boldsymbol{ \Theta}^2_m \right] & =\mathbb{E} \left[ \sum_{j=0}^J\pi_j\bm{\Theta}^2_j\right]=   \sum_{j=0}^J\frac{a_j}{a}\mathbb{E} \left[\bm{\Theta}^2_j\right]=\sum_{j=0}^J \frac{a_j}{a} \mu_{j,2}.\\
\end{aligned}
\end{equation*}

\begin{equation*}
\begin{aligned}
\mathbb{E}\left[\boldsymbol{\Theta}_m\boldsymbol{\Theta}_{m'}\right]&=\mathbb{E}\left[\mathbb{E}\left[\boldsymbol{\Theta}_m\boldsymbol{\Theta}_{m'}|\tilde{p}\right]\right]=\mathbb{E}\left[\mathbb{E}\left[\boldsymbol{\Theta}_m|\tilde{p}\right]\mathbb{E}\left[\boldsymbol{\Theta}_{m'}|\tilde{p}\right]\right]=\mathbb{E}\left[\left( \sum_{j=0}^J\pi_j\:{\bm{\Theta}_j}+\pi_0\sum_{l\geq 1} \omega_l\:{\bm{\Theta}^{nov}_l}\right)^2\right]\\&=%
\mathbb{E}\left[\left( \sum_{j=0}^J\pi_j\:{\bm{\Theta}_j}\right)^2+\left(\pi_0\sum_{l\geq 1} \omega_l\:{\bm{\Theta}^{nov}_l}\right)^2 + 
2\left( \sum_{j=0}^J\pi_j\:{\bm{\Theta}_j}\right)\left(\pi_0\sum_{l\geq 1} \omega_l\:{\bm{\Theta}^{nov}_l}\right)
\right]
\\&=\sum_{j=1}^J \frac{a_j(a_j+1)}{a(a+1)} \mu_{j,2}+ 2\sum_{j>l\geq 1}\frac{a_j a_l}{a(a+1)}\mu_j\mu_l \\&+ \frac{a_0(a_0+1)}{a(a+1)}\left[\frac{\mu_{0,2}}{1+\gamma} + \frac{\gamma\mu_{0}^2 }{1+\gamma}\right]+ 2\mu_0 \sum_{j=1}^J \frac{a_0 a_j}{a(a+1)}\mu_j \\ &=
\sum_{j=0}^J \frac{a_j(a_j+1)}{a(a+1)} \mu_{j,2} + 2\sum_{j>l\geq 0}\frac{a_j a_l}{a(a+1)}\mu_j\mu_l - \frac{a_0(a_0+1)}{a(a+1)} \frac{\gamma}{1+\gamma}\left(\mu_{0,2}-\mu_0\right)^2.\\
\end{aligned}
\end{equation*}

\begin{equation*}
\begin{aligned}
\mathbb{V}\left[\boldsymbol{\Theta}_m\right]&=\sum_{j=0}^J \frac{a_j}{a} \mu_{j,2} -\sum_{j=0}^J \frac{a^2_j}{a^2} \mu_j^2- 2\sum_{j>l}\frac{a_j a_l}{a^2}\mu_j\mu_l \\
&=\sum_{j=0}^J
    \frac{a_j}{a}\left(\mu_{j,2}-\frac{a_j}{a}\mu_{j}^2\right)-2\sum^J_{l>j\geq 0}\frac{a_ja_l}{a^2}\mu_l\mu_j.\\
Cov_\gamma(\boldsymbol{\Theta}_m,\boldsymbol{\Theta}_{m'})&=\mathbb{E}\left[\boldsymbol{\Theta}_m\boldsymbol{\Theta}_{m'}\right]-\mathbb{E} \left[ \boldsymbol{\Theta}_m \right]\mathbb{E} \left[ \boldsymbol{\Theta}_{m'} \right]\\&=
\sum_{j=0}^J \frac{a_j(a_j+1)}{a(a+1)} \mu_{j,2} + 2\sum_{j>l\geq 0}\frac{a_j a_l}{a(a+1)}\mu_j\mu_l - \frac{a_0(a_0+1)}{a(a+1)} \frac{\gamma}{1+\gamma}\left(\mu_{0,2}-\mu_0^2\right)
\\ &-
\sum_{j=0}^J \frac{a^2_j}{a^2} \mu_j^2- 2\sum_{j>l\geq 0}\frac{a_j a_l}{a^2}\mu_j\mu_l\\
&=Cov_0(\boldsymbol{\Theta}_m,\boldsymbol{\Theta}_{m'})  - \frac{a_0(a_0+1)}{a(a+1)} \frac{\gamma}{1+\gamma}\sigma_0^2.
\end{aligned}
\end{equation*}

Let us represent all the possible values of $\boldsymbol{\Theta}$ in one vector $\tilde{\boldsymbol{\Theta}}=\left(\boldsymbol{\Theta}_1,\ldots,\boldsymbol{\Theta}_J,\boldsymbol{\Theta}^{nov}_1,\ldots\right)$

\begin{equation}
\begin{aligned}
\mathbb{P}(\boldsymbol{\Theta}_m =\boldsymbol{\Theta}_{m'})&=
\mathbb{E}\left[\mathbb{P}\left(\boldsymbol{\Theta}_m=\boldsymbol{\Theta}_{m'}|\tilde{p}\right)\right]\\ &=  \sum_{j\geq 0} \mathbb{E}\left[\mathbb{P}\left(\boldsymbol{\Theta}_m=\tilde{\boldsymbol{\Theta}}_j|\tilde{p}\right)\cdot \mathbb{P}\left(\boldsymbol{\Theta}_{m'}=\tilde{\boldsymbol{\Theta}}_j|\tilde{p}\right)\right]\\
&=\sum_{j= 1}^{J} \mathbb{E}\left[\mathbb{P}\left(\boldsymbol{\Theta}_m=\tilde{\boldsymbol{\Theta}}_j|\tilde{p}\right)\cdot \mathbb{P}\left(\boldsymbol{\Theta}_{m'}=\tilde{\boldsymbol{\Theta}}_j|\tilde{p}\right)\right]\\&+\sum_{j\geq J+1} \mathbb{E}\left[\mathbb{P}\left(\boldsymbol{\Theta}_m=\tilde{\boldsymbol{\Theta}}_j|\tilde{p}\right)\cdot \mathbb{P}\left(\boldsymbol{\Theta}_{m'}=\tilde{\boldsymbol{\Theta}}_j|\tilde{p}\right)\right]\\
&=\sum_{j= 1}^{J} \mathbb{E}\left[\pi_j^2\right]+\sum_{j\geq J+1} \mathbb{E}\left[\pi_0^2\omega_j^2\right]\\
&=\sum_{j= 1}^{J} \mathbb{E}\left[\pi_j^2\right]+\sum_{j\geq J+1} \mathbb{E}\left[\pi_0^2\right]\mathbb{E}\left[\omega_j^2\right]\\&=
\sum_{j= 1}^{J}
\frac{a_j(a_j+1)}{a(a+1)}+\frac{a_0(a_0+1)}{a(a+1)}\cdot\frac{1}{1+\gamma}.
\end{aligned}
\end{equation}
\clearpage

\section{Multivariate Brand - Additional experiments}
The present section integrates and extends the simulation study reported in Section 6.1 of the manuscript. Particularly, the multivariate Brand is applied to a real $13$-dimensional dataset (Section \ref{sec:wine_data}), and to two additional simulation studies: the former dealing with non-Gaussian shaped components (Section \ref{sec:non_gaussian}) and the latter encompassing higher-dimensional, sparse covariance structures (Section 	\ref{sec:high_dim}).

\subsection{Wine dataset} \label{sec:wine_data}
The dataset, publicly available in the University of California Irvine Machine Learning repository,  comprises $13$ chemical measurements from $178$ wine samples from the Piedmont region, Italy \citep{forina1986multivariate}. The samples arise from three different cultivars: Barolo, Grignolino, and Barbera. We randomly select $66$ samples from the first two varieties to build the training set, whereas the remaining $112$ samples, including $48$ wines from the third cultivar, defines the test set. The multivariate Brand methodology is fitted to the datasets to detect the third unobserved wine type. The model hyper-parameters are set as follows. First, $h_{MCD}=0.95$ induces robust priors elicitation for the two observed  classes. Fairly uninformative priors are used for the base measure $H$, namely $\bm{m}_0=\boldsymbol{0}, \: \lambda_0=0.01, \: \nu_0=15$ and $\bm{S}_0=\boldsymbol{I}_{13}$, where with $\boldsymbol{0}$ we denote the $13$-dimensional zero vector. Such priors agree with the ones employed in Section 6.2.1 in the main paper for the seed dataset, underlying their general applicability in scenarios where no initial information is available. After initiating the MCMC with $20,000$ iterations for the burn-in phase, $20,000$ dependent samples are retained from the target posterior distribution. The resulting classification is reported in Table \ref{tab:wine}: our semi-parametric model almost perfectly recovers the underlying partition, efficiently identifying the novel wine type. The derived accuracy is particularly high, with performance comparable to those obtained in fully-supervised experiments \citep{Aeberhard1993}.

\begin{table}[th!]
 \caption{Confusion matrix for multivariate Brand on the test set, wine dataset. The label ``New'' indicates observations that are estimated to have arisen from the novelty component.}
 \centering
\begin{tabular}{ |c|ccc| } 
\hline
&  & Truth & \\
Classification & Barolo & Grignolino & Barbera \\ 
  \hline
  Barolo   &  29 &   0 &   0 \\  
  Grignolino   &   1 &  32 &   2 \\  
  New &   0 &   0 & 48 \\
\hline
\end{tabular}
\label{tab:wine}
\end{table}

\subsection{Non-Gaussian shaped classes} \label{sec:non_gaussian}
\begin{figure}[t]
  \centering
  \includegraphics[scale=.6]{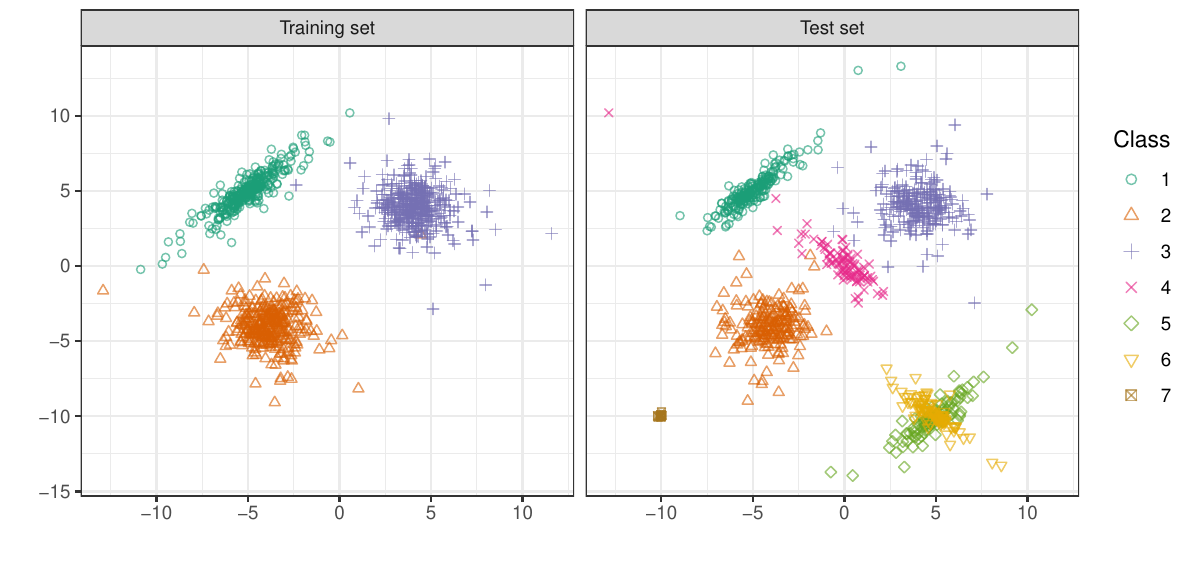}
  \caption{Synthetic data for the simulated experiments of Section \ref{sec:non_gaussian}. Classes $1$ to $5$ are generated according to a multivariate $t$ distribution with $5$ degrees of freedom, while classes $6$ and $7$ are realizations from a multivariate Laplace distribution.}
  \label{fig:non_gaussian_learning}
\end{figure}

\begin{figure*}[t]
  \centering
  \includegraphics[width=\textwidth, keepaspectratio]{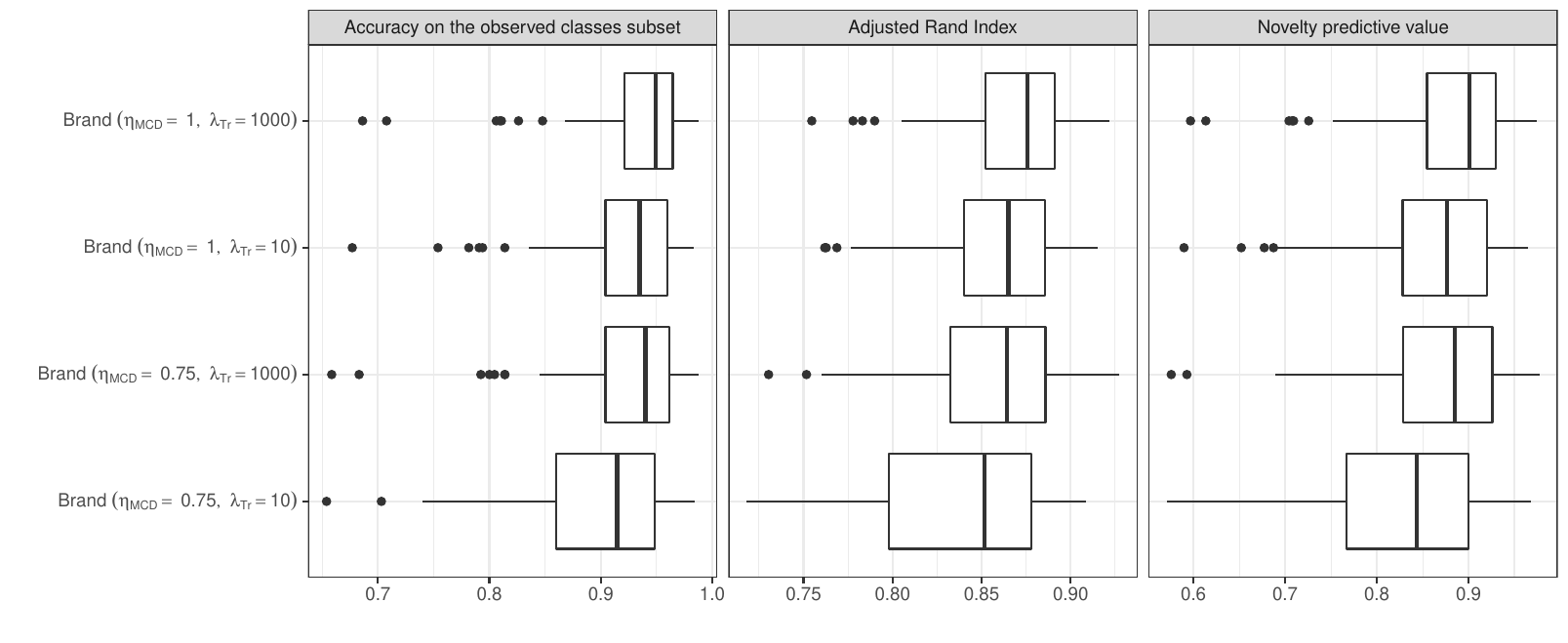}
  \caption{Box plots for (from left to right) accuracy on the known classes,  ARI and  novelty predictive value 
 metrics for $B = 100$ repetitions of the simulated experiment of Section \ref{sec:non_gaussian}.}
  \label{fig:non_gaussain_APPENDIX}
\end{figure*}

In this section, we employ the same hyperparameters configuration considered for the experimental setup in Section 6.1.1 of the main paper to generate samples from elliptical distributions that differ from the Gaussian, with the final aim of validating Brand performance under model misspecification. 
As discussed, the general structure reported in Equation (1) allows for any distributional kernel specification. However, departures from Gaussianity would induce the loss of the conjugacy properties, worsening the computational cost. Therefore, we now want to test how robust the Gaussian kernel is in capturing symmetric components with heavier tails. To this extent, we repeat the experiment for the \texttt{Label noise = False} and \texttt{Novelty size = Not small} scenario, with resulting sample sizes being equal to
$$
    \begin{aligned}
n_1=300, \: n_2=300, \: n_3=400,
    \end{aligned}
$$
and
$$
    \begin{aligned}
m_1=200, \: m_2=200, \: m_3=250, \: m_4=90,\quad m_5=100, \: m_6=100, \: m_7=10  ,  
    \end{aligned}
$$
for the training and test sets, respectively. In contrast to the simulations reported in the main paper, we generate groups $1$ to $5$ via a multivariate $t$ distribution with $5$ degrees of freedom, while classes $6$ and $7$ are realizations from a multivariate Laplace distribution. The resulting learning framework is displayed in Figure \ref{fig:non_gaussian_learning}.
The simulation results, for the same hyper-parameters specification employed in the main paper, are reported in Figure \ref{fig:non_gaussain_APPENDIX}. We immediately notice that different prior settings do not substantially influence the overall model performance, with satisfactorily good results showcased for all the considered metrics. Notwithstanding, when we compare the simulation outcome in Figure \ref{fig:non_gaussain_APPENDIX} with the one displayed in Figure 4 of the main paper, it is apparent that the results are in some measure influenced by model misspecification.\\ In detail, the novelty term tends to absorb all those units sampled from the (heavy) tails of the known components,  as they present very low Gaussian density, particularly when $\eta_{MCD}=0.75$. Consequently, the clustering induced by the DPMM shows many more groups a posteriori in this scenario, with singletons trying to accommodate patterns that the main classes cannot explain.
While clearly accuracy measures for assessing the goodness of a clustering procedure shall be application-dependent \citep{Hennig2015}, we argue here that, in principle, the outcome showcased by our method could still be relevant in contexts that violate the Gaussianity assumption. More specifically, if the sought classes are assumed to be unimodal and elliptical, both stages in the brand methodology concur to achieve this result. Stage I trims the group-wise most outlying values, and Stage II flexibly captures all the unexplained variability with Gaussian-like shapes. 
Once this has been accomplished, one can employ the output evaluation in terms of trimmed units and a posteriori assignment to determine which assumptions were not met by the dataset at hand. As an example, the identification of small novel clusters in the vicinity of bigger ones may be an indication that components with heavier tails are needed to properly account for the true underlying partition. For a general overview on the difficult problem of finding groups in data and associated clustering validity measures, the interested reader is refered to \citet{Akhanli2020}. 

\FloatBarrier

\subsection{High-dimensional heteroscedastic classes} \label{sec:high_dim}
\begin{figure}[h]
  \centering
    \vspace*{-2cm}
  \includegraphics[width=\linewidth, keepaspectratio]{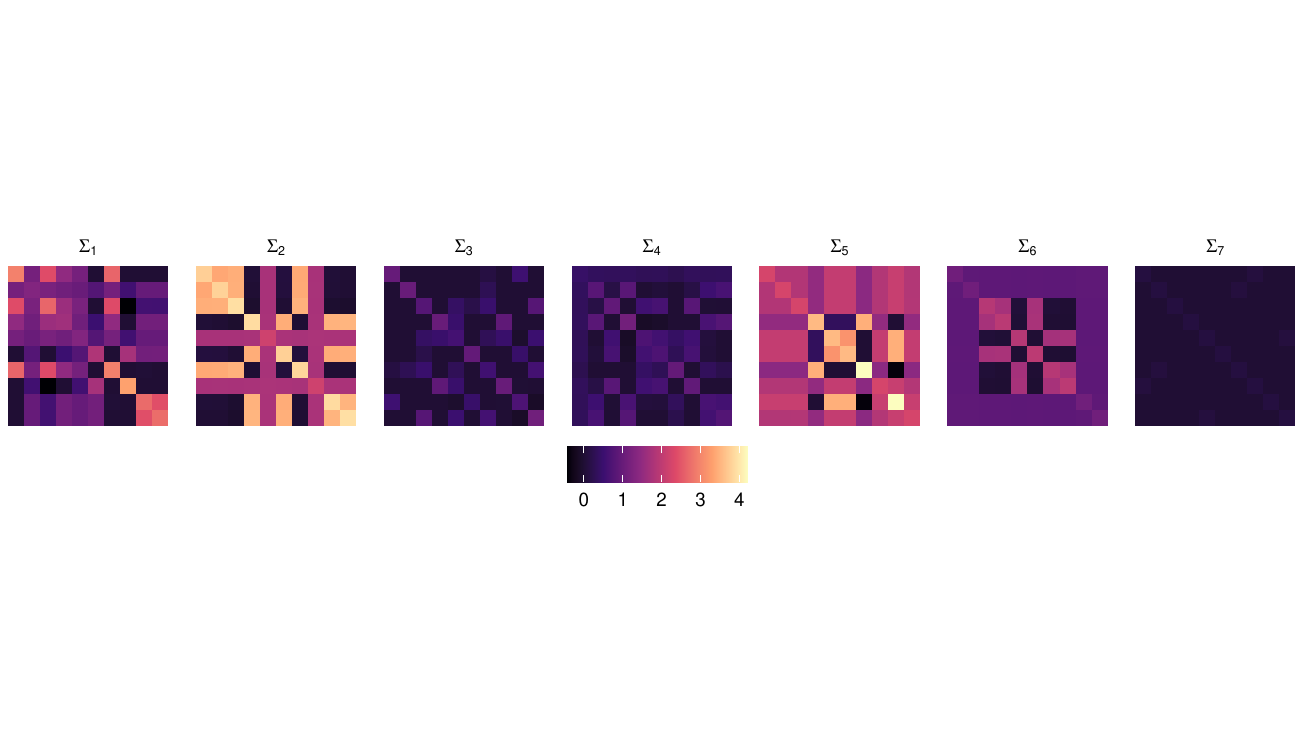}
  \vspace*{-3cm}
  \caption{Covariance matrices for the simulated experiments of Section \ref{sec:high_dim}.}
  \label{fig:sparse_cov}
\end{figure}

\begin{figure}[h!]
  \centering
  \includegraphics[width=\linewidth, keepaspectratio]{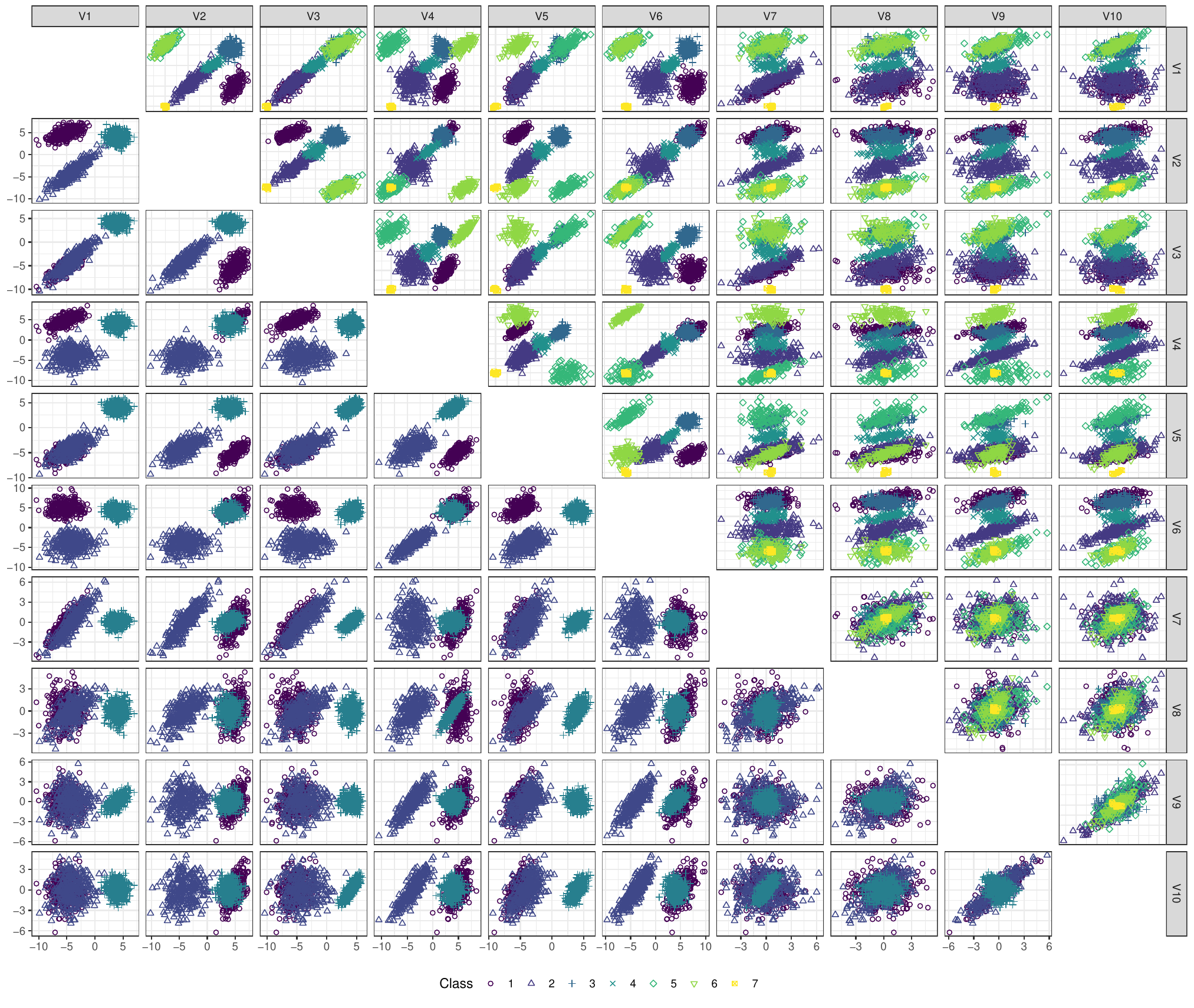}

  \caption{Learning scenario for the simulated experiment of Section \ref{sec:high_dim}. Plots below the main diagonal
represent the training set. Plots above the main diagonal represent the test set.}
  \label{fig:high_dim_scenario}
\end{figure}

For this experiment, we consider the same number of classes (known and hidden ones) and sample sizes for training and test sets introduced in the previous section. Each component is distributed according to a multivariate Normal density with mean vectors equal to:
\[  \boldsymbol{\mu}_1=(-5, 5, -5, 5, -5, 5, 0, 0, 0, 0)', \quad \quad \quad \:\:\boldsymbol{\mu}_2=(-4, -4, -4, -4, -4, -4, 0, 0, 0, 0)',\]
\[  \boldsymbol{\mu}_3=(4, 4, 4, 4, 4, 4, 0, 0, 0, 0)', \quad \boldsymbol{\mu}_4=(0, 0, 0, 0, 0, 0, 0, 0, 0, 0)',\]
\[ \boldsymbol{\mu}_5=(5, -10, 5, -10, 5, -10, 0, 0, 0, 0)',  \quad \boldsymbol{\mu}_6=(5, -10, 5, 10, -5, -10, 0, 0, 0, 0)', \]
\[\boldsymbol{\mu}_7=(-10, -10, -10, -10, -10, -10, 0, 0, 0, 0)',\]
and covariance matrices exhibiting different degrees of sparsity, as illustrated in Figure \ref{fig:sparse_cov}.
We display the resulting learning framework in Figure \ref{fig:high_dim_scenario}, where the training and test samples are reported in the lower and in the upper diagonal plots, respectively. Notice that the generative mechanism induces a quite challenging novelty detection problem,  as the last four dimensions are irrelevant for group separation.

\begin{figure*}[t]
  \centering
  \includegraphics[width=\textwidth, keepaspectratio]{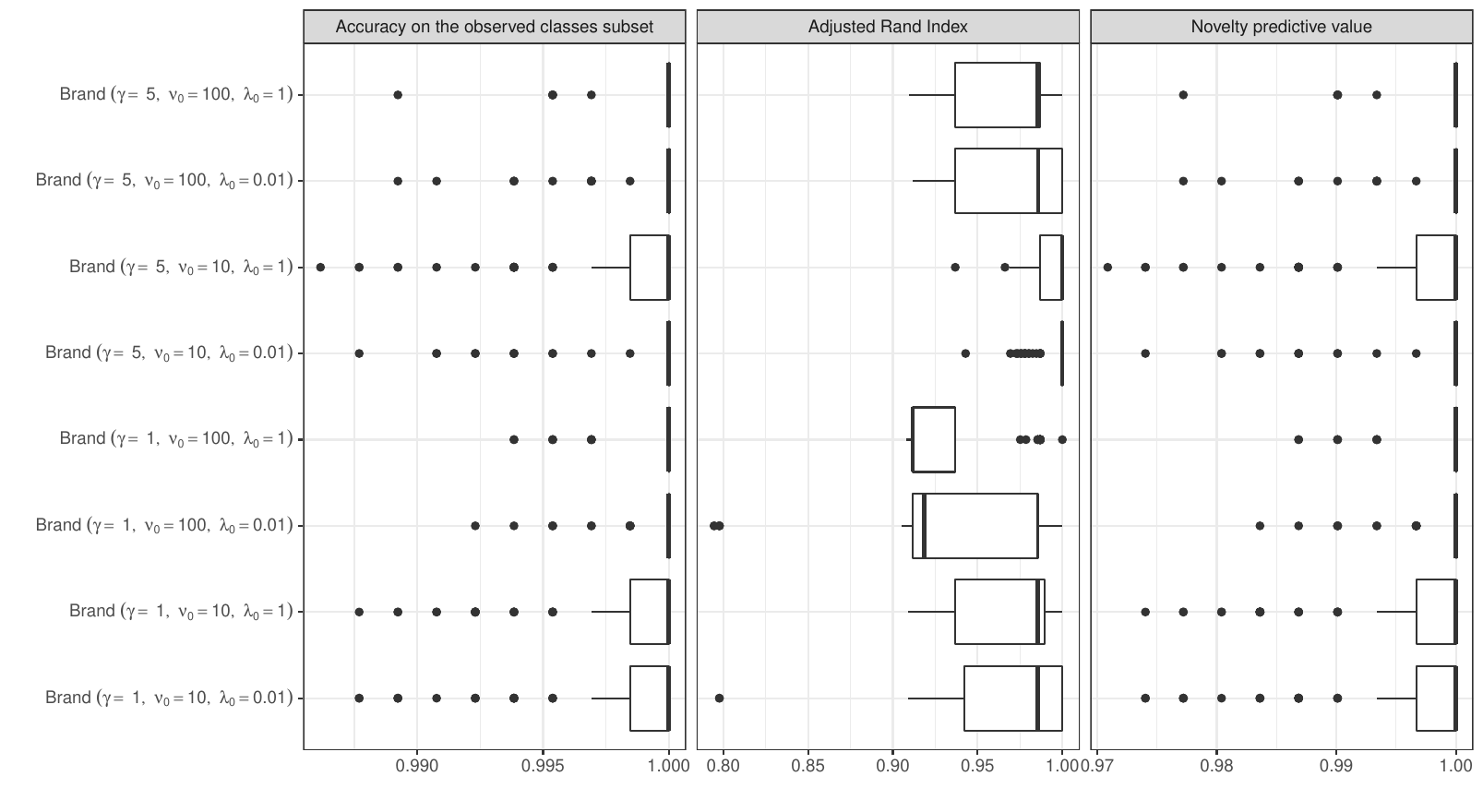}
  \caption{Box plots for (from left to right) accuracy on the known classes, ARI, and novelty predictive value 
 metrics for $B = 100$ repetitions of the simulated experiment of Section \ref{sec:high_dim}.}
  \label{fig:sim_high_dim}
\end{figure*}

By again monitoring the metrics defined in the main manuscript, we aim at investigating the Brand performance when dealing with a $10$-dimensional dataset with heterogeneous covariance patterns.  Given the well-established efficacy of both the MCD and RMCD estimators in high-dimensional settings \citep{Driessen1999, Hubert2018, Boudt2020}, we focus here on the second stage of the Brand method, studying its sensitivity under different hyper-parameters specifications. In detail, a total of $8$ models are fitted to $B=100$ simulated datasets varying:
\begin{itemize}
\item the concentration parameter $\gamma$ of the Dirichlet Process prior, letting it be equal to $1$ or $5$,
\item the degrees of freedom $\nu_0$ associated with the novelty components, letting it be equal to $10$ or $100$,
\item the precision parameter $\lambda_0$ associated with the novelty components, letting it be equal to $0.01$ or $1$.
\end{itemize}
Simulation results are reported in Figure \ref{fig:sim_high_dim}. We immediately notice that the overall performance showcased by our methodology is satisfactorily good, regardless of the hyperparameters specification. Both known and novel patterns are correctly identified as such, with results mirroring the ones obtained in the bi-dimensional experiments reported in Section 6.1.2 of the main paper. In particular, eliciting very flat priors for the base measure $H$ ($\lambda_0=0.01$, $\nu_0=10$) still produces excellent outcomes. The only appreciable difference is in terms of ARI, induced by the DP concentration parameter $\gamma$. Specifically, increasing $\gamma$ favors a priori the creation of more clusters. As a result, the model is more prone to accommodate components with a limited sample size, like the seventh group in the present experiment. On the other hand, when $\gamma=1$  the $20$ test units arisen by the last density are usually merged within some other components, producing the slight difference in ARI visible in the middle plot of Figure \ref{fig:sim_high_dim}.

\FloatBarrier

\clearpage

\section{Functional Brand - Controlled experiment}
\begin{figure}[ht!]
    \centering
    \includegraphics[scale=.5]{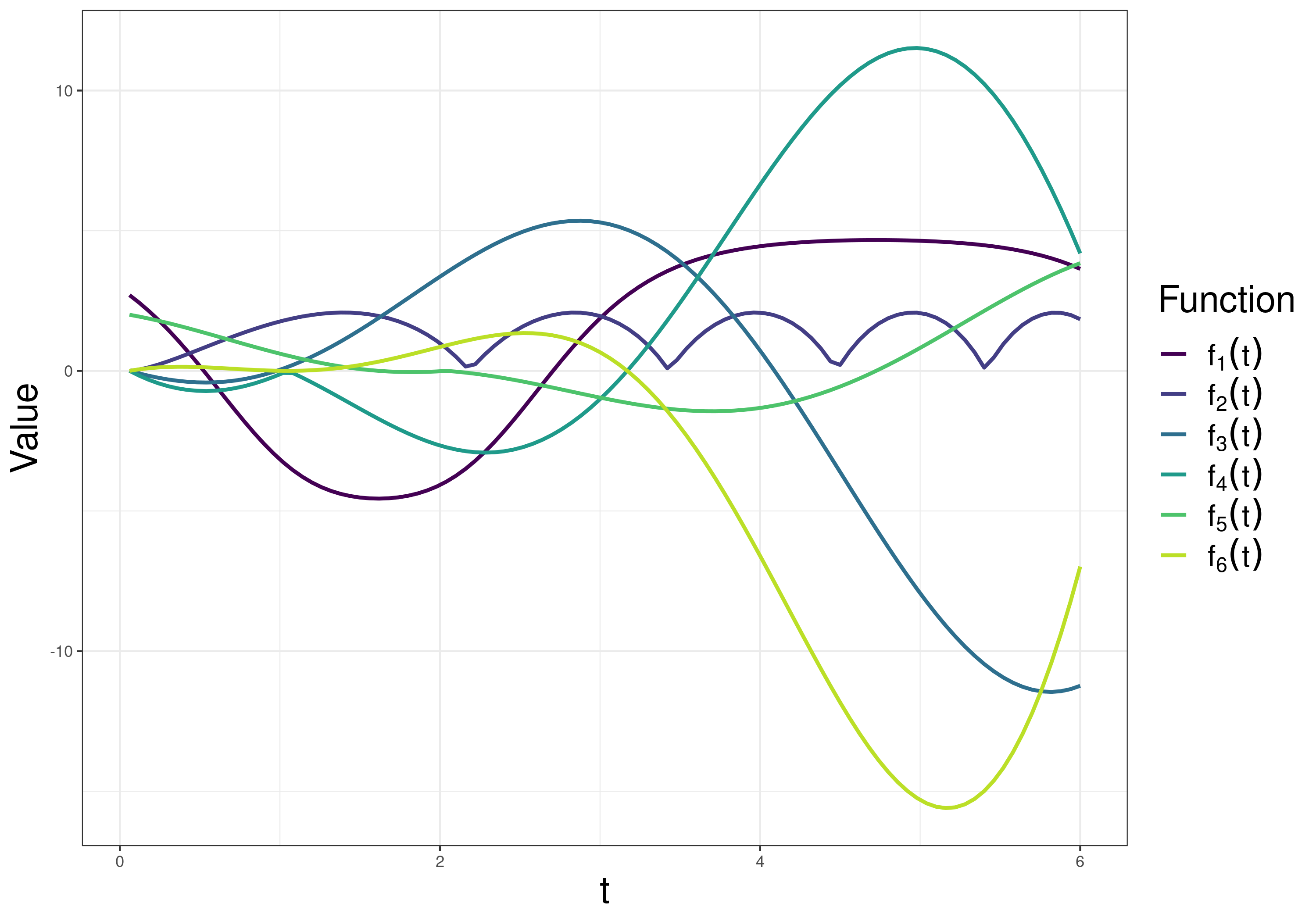}
    \caption{The six different functions considered in this experiment.}
    \label{fig:fs}
\end{figure}

This section investigates how functional Brand behaves according to various prior specifications and different levels of noise in the data. 
We aim to provide the researchers and practitioners with guidelines for the hyperparameter specification to exploit the flexibility of our model at its fullest. In the next experiments, we will consider the following six functions, evaluated on the interval $t\in\left[0,6\right]$:
\begin{equation*}
    \begin{aligned}
    f_1(t) &= 5\:\cos(\exp(\sin(t))), \\
    f_3(t) &= 2t\cos(t-2.5), \\
    f_5(t) &= |t-2|\cos(t), \\
    \end{aligned}
    \quad \quad
    \begin{aligned}
    f_2(t) &= 3\log(\sin(t^{1.5})+1), \\
    f_4(t) &= -3|t-1|\sin(t),\\
    f_6(t) &= |t-1|^2\sin(t). \\
\end{aligned}
\end{equation*}
Figure \ref{fig:fs} provides a visual representation of the six functions. Each of these functions presents its distinctive peculiarities. Simultaneously, they overlap, especially in the left half of the support, which could make the classification more difficult conditioning on the considered noise level.
These six functions constitute the functional means of different groups we are going to study.

\begin{figure}[ht!]
    \centering
    \includegraphics[width=\linewidth]{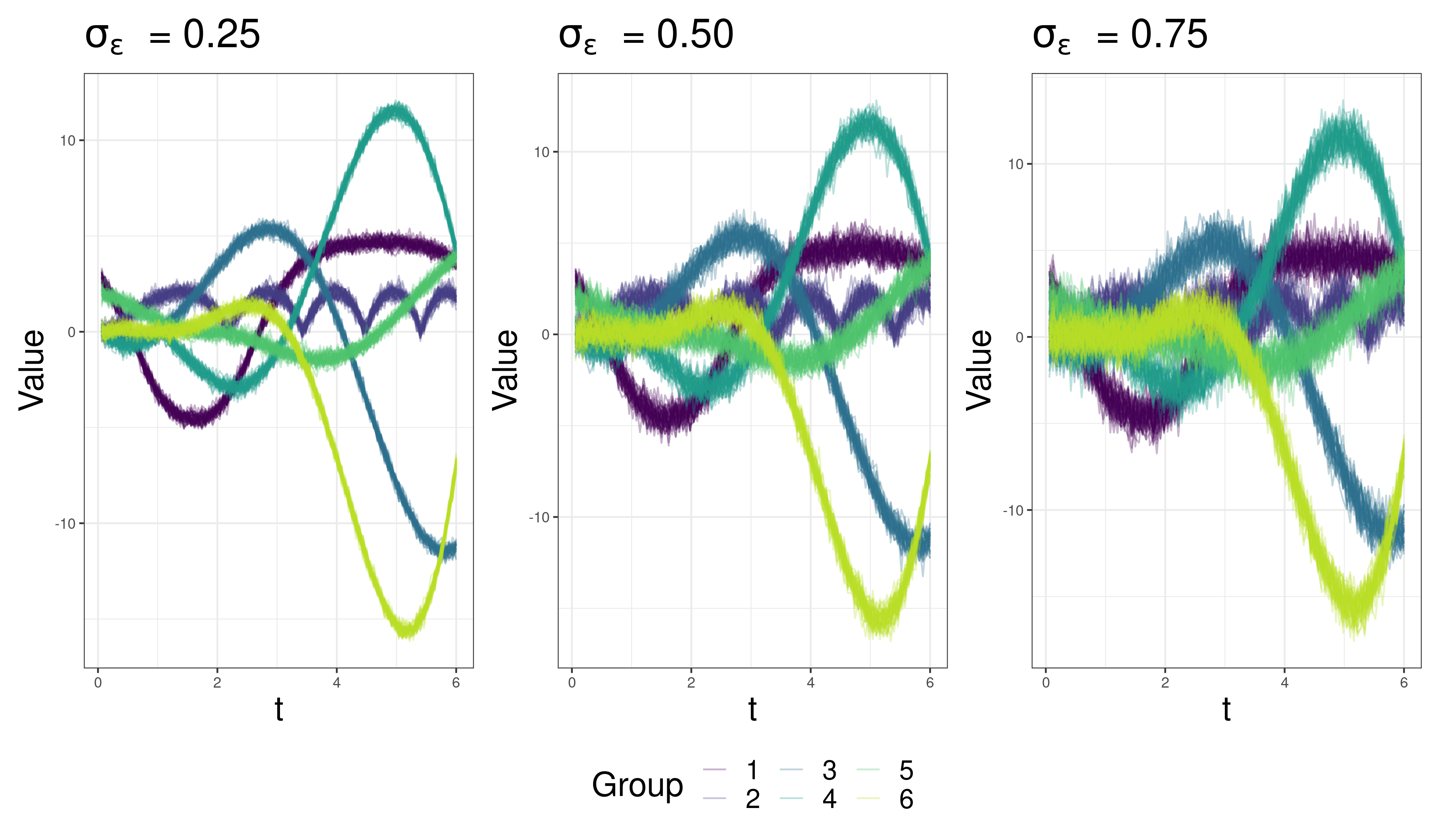}
    \caption{The test sets considered in the analysis stratified by the noise level used to generate the functional data. Left panel: $\bm{Y}_1$; central panel: $\bm{Y}_2$; right panel: $\bm{Y}_3$.}
    \label{fig:fs2}
\end{figure}

We suppose that every function is observed on a discretized collection of 100 time points between $0$ and $6$. 
In all our simulation settings, we fix the number of functions sampled from each group to $50$. We assume that samples from the first three groups are contained in both the training set $\bm{X}$ and in the test set $\bm{Y}$, while samples from the last three groups represent the novelties we want to detect, and therefore they appear only in $\bm{Y}$. We are left with $150$ data objects in the training sets and $300$ in the test sets. More formally, for each group in the training set ntained in both the training set $\bm{X}$, we have 
$$x_j(t) = f_j(t)+\epsilon_j(t) \:\:\text{ with } j=1,\ldots,3, \:\:\text{ and }\:\: \epsilon_j(t)\sim {N}(0,\sigma^2_\epsilon) \:\: \forall t=1,\ldots,100.$$
For each group in the test set $\bm{Y}$, we have 
$$y_j(t) = f_j(t)+\epsilon_j(t)\:\:\text{ with } j=1,\ldots,6, \:\:\text{ and }\:\: \epsilon_j(t)\sim {N}(0,\sigma^2_\epsilon) \:\: \forall t=1,\ldots,100.$$
Finally, we assume three different levels of noise in the data generating process, considering $\sigma_\epsilon\in\{0.25,0.50,0.75\}$, yielding three simulated cases of training and test sets ($\bm{X}_1,\bm{Y}_1$), ($\bm{X}_2,\bm{Y}_2$) and ($\bm{X}_3,\bm{Y}_3,$).\\
Figure \ref{fig:fs2} shows the three generated test sets, stratified by the increasing noise level. 
As already mentioned, a higher level of noise induces a more evident overlap among the functional objects. We expect this overlap to make the classification more challenging.

\subsection{The robust estimation of the mean function}
\begin{figure}[ht!]
    \centering
    \includegraphics[width=\linewidth]{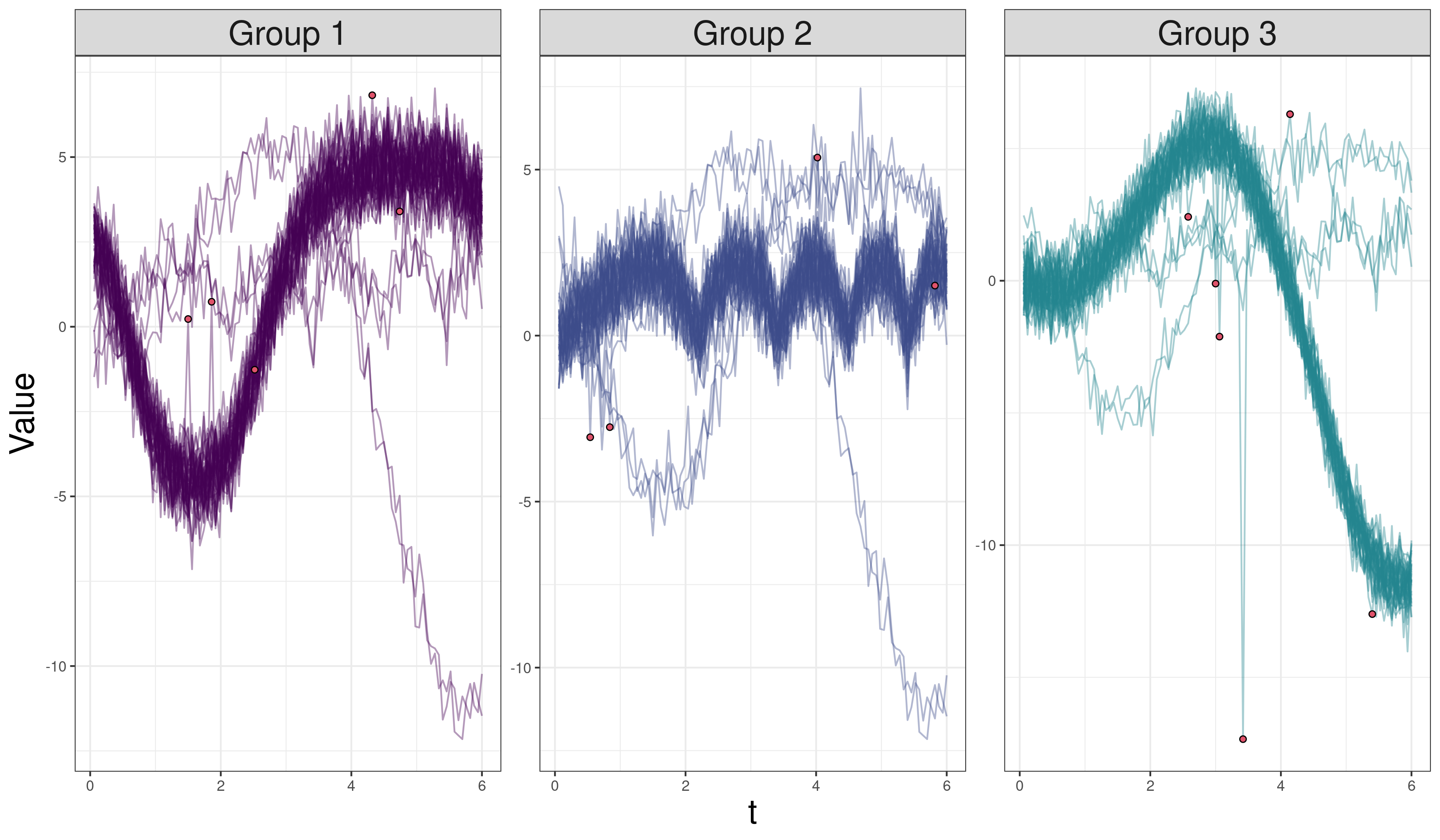}
    \caption{Functional objects contained in $\bm{X}_3$ - generated specifying $\sigma_\epsilon=0.75$. Each known class is characterized by label noise and time-specific outliers (highlighted by red dots).}
    \label{fig:Xnoise}
\end{figure}

We focus on Stage I, and we investigate the robust extraction of prior information in the functional case. In the main paper, we discuss how we smooth the training functions using B-splines bases. Similarly, for this study, we use 100 bases of order 5, which provide accurate results for a reasonable computational cost.\\
Our functional robust estimation works as follows. We first collect the spline coefficients, and we interpret them as multivariate objects. We then apply the MCD estimator on the spline coefficients. We therefore recover, for each group, the robust mean coefficients that, once convoluted with the bases, will produce our mean functions. 
A crucial quantity is the percentage of extreme coefficients to remove in our robust estimation. This percentage  significantly affects also the recovery of the training noise $\bar{\sigma}^2_{j}(t)$. In the following, we demonstrate how the ratio of trimmed values $\eta_{MCD}$ affects Stage I in the functional case. \\
To showcase the effect of the MCD estimator, we contaminate the known classes in the training sets with time-point specific outliers and label noise. In detail, we add noise sampled from $\varepsilon(t)\sim N(0,25)$ to fifteen randomly selected functions at fifteen random timestamps.
We contaminate five functions from the first group, four in the second, and six in the third. Then, we shuffle $10\%$ of the functions across groups to generate label noise. To illustrate this step, we report in Figure \ref{fig:Xnoise} a visual depiction of the contaminated version of the training dataset $\bm{X}_3$, stratified by class.

We consider three different specifications of $\eta_{MCD}$: 
\begin{itemize}
    \item \texttt{no trimming} by setting $\eta_{MCD}=1$: all the estimated spline coefficients are considered;
    \item \texttt{mild trimming}, $\eta_{MCD}=0.95$: only 5\% of the estimated spline coefficients are removed for the estimation of the group centroids and functional noise;
    \item  \texttt{strong trimming}. $\eta_{MCD}=0.75$: a fourth of the estimated spline coefficients is removed before computing the group centroids and functional noise.
\end{itemize}

\begin{figure}[ht!]
    \centering
    \includegraphics[width=.9\linewidth]{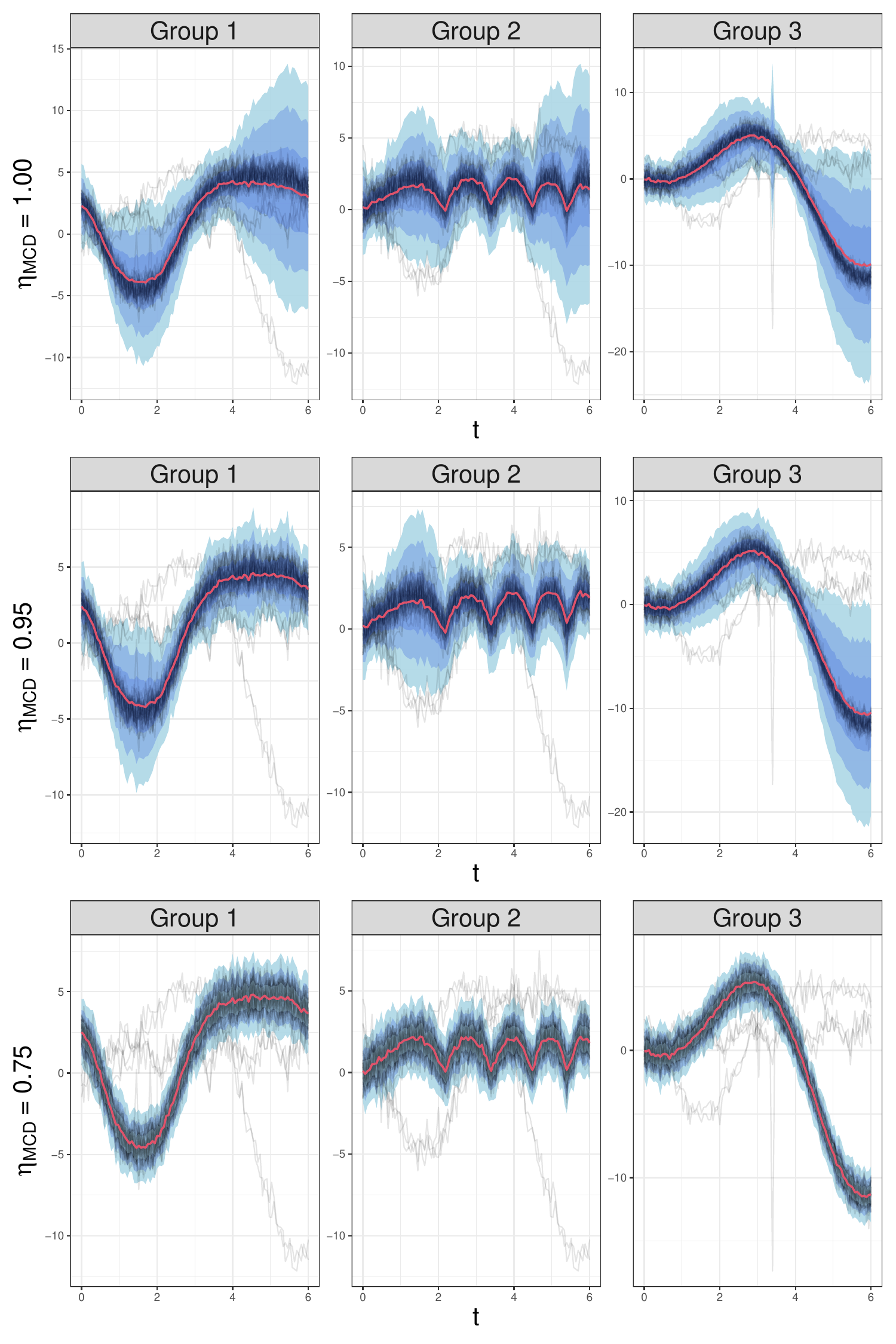}
    \caption{Estimated representative function (in red) for each of the known groups, characterized by the presence of label noise and time-specific outliers. Each row displays a different level of trimming, while groups are reported column-wise.}
    \label{fig:XnoiseMCD}
\end{figure}

We report the extracted mean functions and the estimated noise in the panels of Figure \ref{fig:XnoiseMCD}. Specifically, the group centroids are depicted in red, superimposed onto the observations of every group. We also plot in blue the intervals of variation computed as $\bar{f}_{j}(t) \pm q\cdot \bar{\sigma}^2_{j}(t)$ to give an idea of the estimated noise, for $j,q=1,2,3$. Each column of the plot shows the evolution of the mean and noise functions when considering different trimming levels, stratified by group. We can observe how employing trimming benefits the estimation of the representative functions in each group.\\
For example, in the first row of Figure \ref{fig:XnoiseMCD}, we see how the red line fails to represent the group characteristics. This behavior is exacerbated close to the boundaries in all the groups. It is evident how even a small percentage of trimming yields considerable benefits. In our experience, we obtain very satisfactory results in the functional case when considering a \texttt{strong trimming}. Results in a case of \texttt{extreme trimming}, obtained by setting $\eta_{MCD}=0.5$ were explored but not reported, since similar to the case $\eta_{MCD}=0.75$. In conclusion, if enough data are available, we suggest considering strong trimming as the safest option when dealing with the functional case. In this way, we can eradicate the influence of potential outliers on both the mean and the variance.  We will use the mean and noise functions recovered with $\eta_{MCD}=0.75$ for the subsequent analyses.

\subsection{Model performance according to different prior specifications}
We can summarise the hyperparameters specific of functional Brand in three main subsets:
\begin{itemize}
    \item \texttt{Subset 1}: hyperparameters controlling the dispersion around the extracted mean functions of the known groups: $\varphi_j$ and $\nu_j$;
    \item \texttt{Subset 2}: hyperparameters controlling the dispersion around the novel spline coefficients, representing the unknown mean functions: $a_\tau$, $b_\tau$, and $s^2$;
    \item \texttt{Subset 3}: hyperparameters controlling the dispersion of the noise around the novel mean functions: $a_H$ and $b_H$.
\end{itemize}
We investigate the sensitivity of functional Brand to various hyperparameter specifications while considering the different noise levels in the observed data, as previously discussed. As a first consideration, we point out that setting the hyperparameters controlling the variance around the known classes and/or the novel ones requires the most care. A poorly elicited prior for the error terms could lead to losing Brand hierarchical structure, letting the novelty components take over the known groups in lieu of their flexibility. The opposite issue may also arise, since larger mixture weights are assigned to the known components. 
In the following, we will provide some considerations regarding the tuning of the hyperparameters listed in \texttt{Subset 1} and \texttt{Subset 2}. For the sake of brevity, we omit the results obtained tweaking the parameters that belong to \texttt{Subset 3}, which control the noise level in the novelty term $\sigma^{2\: nov}_h(t), \:\: \forall h$. Given the insights collected in 
several applications, we suggest eliciting an Inverse Gamma prior that favors low noise values, promoting more sensitive estimation of the random functions $f^{nov}_h (t),\:\: \forall h$. Thus, we fix $a_H=5$ and $b_H=1$ in the following, as well as in the functional application in the main paper.

\FloatBarrier

Let us first discuss the hyperparameter specification that we adopted in the main text. We are going to assume these values as our default setting. The values are

$$ \varphi_j\approx 0, \:\:\:\nu_j\approx 0, \:\:\: a_\tau=3
, \:\:\: b_\tau=1, \:\:\: s^2=1\, \:\:\: a_H=5, \:\:\: b_H=1. $$

First, let us describe and justify our choice of hyperparamters. 
The parameters in \texttt{Subset 1} are set to negligible values. In this way, we are implicitly enforcing an \texttt{inductive} learning approach: the information extracted from the training set for the known groups is trustworthy and need not be updated. 
We calibrate the parameters in \texttt{Subset 2} to induce a reasonably tight prior over the novel splines coefficients to help the model identifiability. One can claim that an uninformative specification would also be reasonable to reflect our ignorance about the novel groups. However, as we will show, letting the parameters to vary too freely may negatively affect the results.
Finally, as already discussed, the informative specification of the parameters in \texttt{Subset 3} bounds the variance of the errors around the novel functions: this will avoid limiting cases where all the information contained in the data is explained by the noise term, helping the identification of the latent mean functions.\\

In the following, we tweak one subset of parameters at a time to investigate its effect, starting from this prior configuration. In all the subsequent experiments, we run 7,500 MCMC iterations after a burn-in period of the same length.

\subsubsection{Tuning hyperparameters: \texttt{Subset 1}}
For this first experiment, we consider the three hyperprior configurations devised to increase the uncertainty around the functional means and variance extracted from the training set. 
In other words, we want to understand how, ceteris paribus, the concentration of the priors for $f_j(t)$ and $\sigma^2_j(t)$, $j=1,\ldots,J$ affects the estimation of the mean functions \emph{of the known groups}.\\
As already mentioned, a proper elicitation of the prior noises $\sigma^2_j(t)$ for the known classes  is crucial for carrying out sensible inference.
Throughout many simulation studies, we observed that $\sigma^2_j(t)$ plays a fundamental role in preserving the specific hierarchy that characterizes Brand. 
In fact, in scenarios where the prior variance of the measurement errors around the training mean is too wide, the known components lose their ``priority'' over the novel ones. In other words, the robust estimates extracted from the training set are deemed as less representative of the truth. Simultaneously, the flexibility of the novelty terms leaves the new estimated functions free to adapt to the dataset, ending up absorbing the known classes. In short, if the variances of the known components are not carefully tuned, Brand loses the ability to distinguish between known groups and novelties, hindering the resulting classification. 
To provide an idea on how different choices of $\varphi_j$, and $\nu_j$ affect the results, we consider:
\begin{itemize}
    \item[HPC 1 -] Inductive setting: $\varphi_j\approx 0, \:\:\:\nu_j\approx 0$,
    \item[HPC 2 -] Informative specification: $\varphi_j= 0.01, \:\:\:\nu_j= 1/100000$,
    \item[HPC 3 -] ``Uninformative'' specification: $\varphi_j= 1, \:\:\:   \nu_j= 1/100$. The variance around the mean function is set to 1.
\end{itemize}

\begin{figure}[ht!]
    \centering
    \includegraphics[width=.85\linewidth]{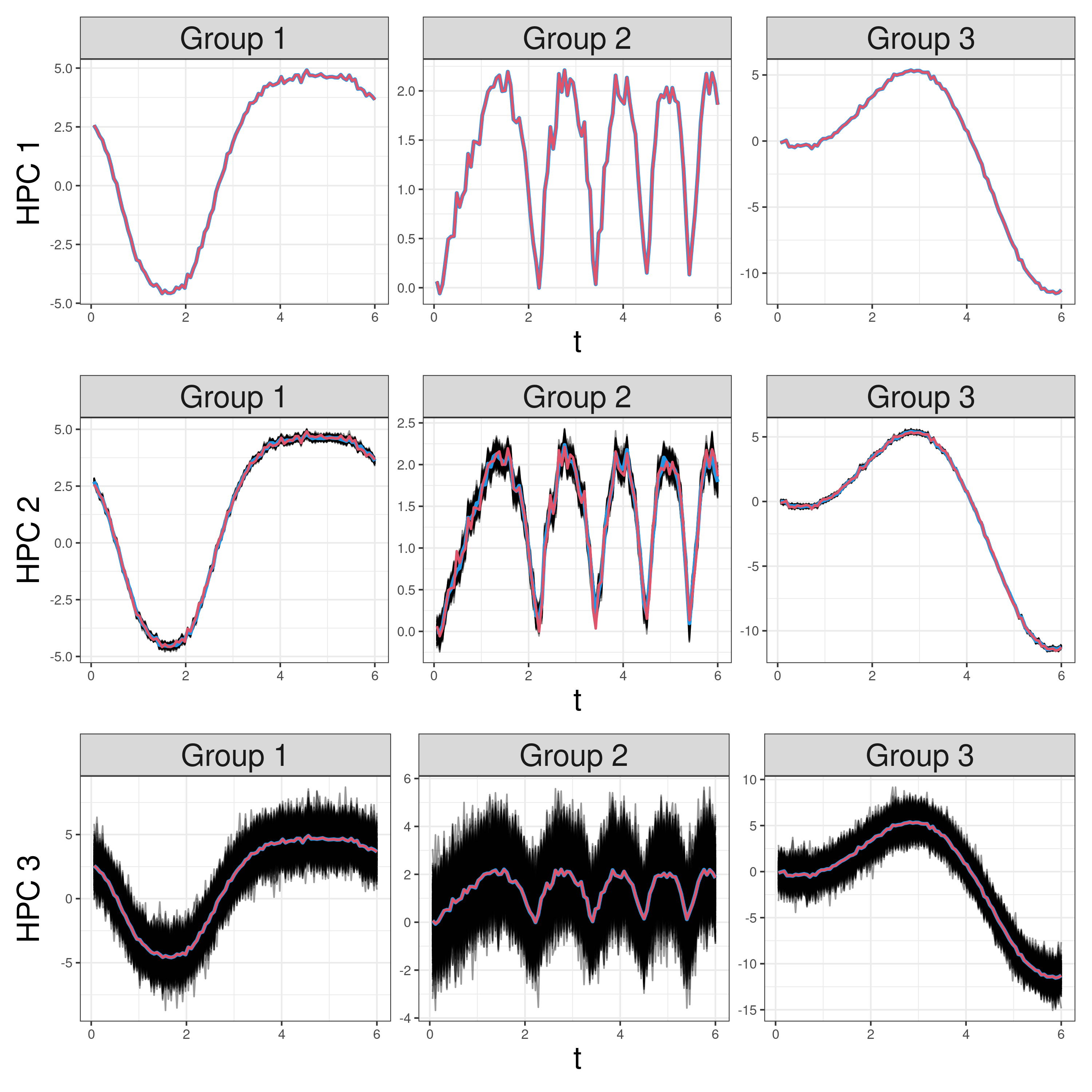}
    \caption{\texttt{Subset 1}, Dataset $\bm{Y}_3$. The known groups are presented column-wise, while the various hyperprior specification are reported row-wise. The panels contain the extracted robust mean function estimates (in red), the simulated posterior mean
functions (computed pointwise, in blue) on top of the MCMC iterations (in black).}
    \label{fig:G1_3}
\end{figure}

For each configuration, we estimate functional Brand on the three different test sets $\bm{Y}_1$, $\bm{Y}_2$, and $\bm{Y}_3$. We show the confusion matrices displaying the assignation of the functional objects contained in the test set into known classes or novelties in Table \ref{tab:G1}. 
Moreover, to better exemplify the model's behavior, we display the functional estimates of the known mean functions for $\bm{Y}_3$ in Figure \ref{fig:G1_3}. For each known group, we collect the robustly extracted mean functions (in red), the simulated posterior mean
functions (computed pointwise, in blue) on top of the MCMC iterations (in black).\\ 

We can detect a recurring pattern across different datasets. 
On the one hand, the first and second prior configurations (HPC 1 and HPC 2) lead to perfect classification of the functional data into known components (Classes 1, 2, and 3 into Cluster 1, 2, and 3, respectively) and novelties (Classes 4, 5, and 6 into Cluster 0) regardless the level of noise in the dataset. 
Looking at the different panels of Figure \ref{fig:G1_3}, we see how HPC 2 allows more uncertainty in the mean estimates (represented by a wider range of variation of the MCMC simulations) and a difference between the starting robust means (in red) and the MCMC posterior means (in blue). As expected, in the \texttt{inductive} setting, the red and blue lines coincide.\\
On the other hand, we can see what happens when tight constraints are not placed on the variances $\sigma^2_j(t)$ by focusing on the last block of columns in Table \ref{tab:G1} and the last row of panels in Figure \ref{fig:G1_3}.
In all the three cases $\bm{Y}_1$, $\bm{Y}_2$, and $\bm{Y}_3$ we see how the novelty component takes over and ruins the classification process. For example, in the bottom row of Figure \ref{fig:G1_3} we see that in the three panels the estimated posterior mean function coincides with the prior since no data have been assigned to those components. In the first two datasets, we see how the higher variances lead Brand to mislead a novel group as a known one and viceversa.\\ 
We can conclude that, for the functional setting, care is needed when tuning the discussed hyperprior parameters. Thus, when enough data are available, we suggest either to rely on the \texttt{inductive} learning process or, otherwise, to induce small a priori variations.

\begin{table}[ht!]
\small
    \centering
    \hspace{-0.5cm}
    \begin{tabular}{cccccccccccccccccccc}
    \toprule
         &  &\multicolumn{6}{c}{HPC 1} & \multicolumn{6}{c}{HPC 2} & \multicolumn{6}{c}{HPC 3}\\
        \cmidrule(lr){3-8} \cmidrule(lr){9-14} \cmidrule(lr){15-20}
\multicolumn{2}{c}{Ground Truth} & \textbf{1} &\textbf{2} &\textbf{3} &\textbf{4} &\textbf{5} &\textbf{6} & \textbf{1} &\textbf{2} &\textbf{3} &\textbf{4} &\textbf{5} &\textbf{6}  & \textbf{1} &\textbf{2} &\textbf{3} &\textbf{4} &\textbf{5} &\textbf{6}\\
\midrule
\multirow{4}{*}{$\bm{Y}_1$}
&\textbf{0} & 0  & 0  & 0  & 50 & 50 & 50   & 0  & 0  & 0  & 50 & 50 & 50     & 50  & 50  & 50  & 0 & 50 & 50 \\
&\textbf{1} & 50 & 0  & 0  & 0  & 0  & 0    & 50 & 0  & 0  & 0  & 0  & 0      & 0 & 0  & 0  & 50  & 0  & 0  \\
&\textbf{2} & 0  & 50 & 0  & 0  & 0  & 0    & 0  & 50 & 0  & 0  & 0  & 0      & 0  & 0 & 0  & 0  & 0  & 0  \\
&\textbf{3} & 0  & 0  & 50 & 0  & 0  & 0    & 0  & 0  & 50 & 0  & 0  & 0      & 0  & 0  & 0 & 0  & 0  & 0  \\
    \cmidrule(lr){1-2}    \cmidrule(lr){3-8} \cmidrule(lr){9-14} \cmidrule(lr){15-20}\multirow{4}{*}{$\bm{Y}_2$}
&\textbf{0} & 0  & 0  & 0  & 50 & 50 & 50   & 0  & 0  & 0  & 50 & 50 & 50  & 50  & 50  & 50  & 0 & 50 & 50 \\
&\textbf{1} & 50 & 0  & 0  & 0  & 0  & 0    & 50 & 0  & 0  & 0  & 0  & 0   & 0 & 0  & 0  & 50  & 0  & 0   \\
&\textbf{2} & 0  & 50 & 0  & 0  & 0  & 0    & 0  & 50 & 0  & 0  & 0  & 0   & 0  & 0 & 0  & 0  & 0  & 0   \\
&\textbf{3} & 0  & 0  & 50 & 0  & 0  & 0    & 0  & 0  & 50 & 0  & 0  & 0   & 0  & 0  & 0 & 0  & 0  & 0   \\
    \cmidrule(lr){1-2}    \cmidrule(lr){3-8} \cmidrule(lr){9-14} \cmidrule(lr){15-20}
\multirow{4}{*}{$\bm{Y}_3$}
&\textbf{0} & 0  &0  &0  &50 &50 &50    & 0  &0  &0  &50 &50 &50      &50 &50 &50 &50 &50 &50 \\
&\textbf{1} & 50 &0  &0  &0 &0 &0       & 50 &0  &0  &0  &0  &0       & 0  &0  &0  &0  &0  &0 \\
&\textbf{2} & 0  &50 &0  &0 &0 &0       & 0  &50 &0  &0  &0  &0       & 0  &0  &0  &0  &0  &0 \\
&\textbf{3} & 0  &0  &50 &0 &0 &0       & 0  &0  &50 &0  &0  &0       & 0  &0  &0  &0  &0  &0\\
\bottomrule
    \end{tabular}
    \caption{Confusion matrices obtained in the first part of the simulation study. The different datasets are reported along the rows. The different hyperprior specifications are reported along the columns. }
    \label{tab:G1}
\end{table}

\subsubsection{Tuning hyperparameters: \texttt{Subset 2}}

The discussion for this second set of hyperparameters is structured similarly to the one reported in the previous Subsection. We now investigate Brand's ability to identify and correctly cluster the novelty terms, while tweaking the variability of the prior distributions of $\psi_h$ and $ \tau_h^2$ (and, therefore, of $f_{h}^{nov}(t)$). We consider the following three hyperparameters configurations:

\begin{itemize}
    \item[HPC 1 -] Adopted specification: $a_\tau=3$, $b_\tau=1$, $s^2=1$
    \item[HPC 2 -] Low variability: $a_\tau=10$, $b_\tau=1$, $s^2=0.001$
    \item[HPC 3 -] High variability: $a_\tau=0.1$, $b_\tau=0.
    1$, $s^2=10$. \\ We have also investigated a more extreme case,  $a_\tau=0.01$, $b_\tau=0.
    01$, $s^2=1000$, for which we omit the results since this extremely uninformative specification resulted in numerical instabilities and poor classification performance.
\end{itemize}

The first and second hyperprior specifications lead to perfect clustering for all three datasets. We instead report in Table \ref{G2Y3} the results obtained on $\bm{Y}_3$. A tighter prior for the novelty components leads to the misclassification of only few observations (2/300) and the correct detection of the overall cluster structure. Conversely, a wide prior can overrule the hierarchical structure of the model, absorbing known classes into the novelty term. 

\begin{table}[ht!]
\small
    \centering
    \hspace{-0.5cm}
    \begin{tabular}{cccccccccccccccccccc}
    \toprule
        \multicolumn{2}{c}{$\bm{Y}_3$} &\multicolumn{6}{c}{HPC 1} & \multicolumn{6}{c}{HPC 2} & \multicolumn{6}{c}{HPC 3}\\
        \cmidrule(lr){3-8} \cmidrule(lr){9-14} \cmidrule(lr){15-20}
\multicolumn{2}{c}{Ground Truth} & \textbf{1} &\textbf{2} &\textbf{3} &\textbf{4} &\textbf{5} &\textbf{6} & \textbf{1} &\textbf{2} &\textbf{3} &\textbf{4} &\textbf{5} &\textbf{6}  & \textbf{1} &\textbf{2} &\textbf{3} &\textbf{4} &\textbf{5} &\textbf{6}\\
\midrule
\multirow{3}{*}{CKC}
&\textbf{1} & 50 & 0  & 0  & 0  & 0  & 0    & 50 & 0  & 0    & 0  & 0  & 0      & 50  & 0   & 0      & 0  & 0  & 0 \\
&\textbf{2} & 0  & 50 & 0  & 0  & 0  & 0    & 0  & 50 & 0    & 0  & 0  & 0      & 0   & 1   & 0      & 0  & 0  & 0 \\
&\textbf{3} & 0  & 0  & 50 & 0  & 0  & 0    & 0  & 0  & 49   & 0  & 0  & 0      & 0   & 47  & 0      & 0  & 0  & 0 \\
    \cmidrule(lr){1-2}    \cmidrule(lr){3-8} \cmidrule(lr){9-14} \cmidrule(lr){15-20}
\multirow{3}{*}{CNC}
&\textbf{4} & 0  & 0  & 0 & 50 & 0  & 0     & 0  & 0  & 1 & 50 & 0  & 0       & 0  & 1  & 0 & 50  & 0  & 0  \\
&\textbf{5} & 0  & 0  & 0 & 0  & 50 & 0     & 0  & 0  & 0 & 0  & 50 & 0       & 0  & 0  & 50 & 0  & 47  & 0  \\
&\textbf{6} & 0  & 0  & 0 & 0  & 0  & 50    & 0  & 0  & 0 & 0  & 0  & 49      & 0  & 0  & 0 & 0  & 0  & 50  \\
\cmidrule(lr){1-2}    \cmidrule(lr){3-8} \cmidrule(lr){9-14} \cmidrule(lr){15-20}
\multirow{3}{*}{ENC}
&\textbf{7} & 0  & 0  & 0 & 0  & 0  & 0    & 0  & 0  & 0 & 0  & 0  & 1        &  0  & 0  & 0 & 0  & 1  & 0  \\
&\textbf{8} & 0  & 0  & 0 & 0  & 0  & 0    & 0  & 0  & 0 & 0  & 0  & 0        &  0  & 0  & 0 & 0  & 1  & 0  \\
&\textbf{9} & 0  & 0  & 0 & 0  & 0  & 0    & 0  & 0  & 0 & 0  & 0  & 0        &  0  & 1  & 0 & 0  & 1  & 0  \\
\bottomrule
\end{tabular}
\caption{Classification results for dataset $\bm{Y}_3$ under the three different hyperparameter configurations. The table displays the ground truth across the columns (1-3: known groups, 4-6 novel groups). We divided the resulting labels from the clustering process (on the rows) into three subgroups: CKC, correct known clusters (1-3); CNC, correct novel clusters (4-6); and ENC, extra novel clusters.}
\label{G2Y3}
\end{table}

Overall, we can conclude that Brand is more robust to different hyperprior settings regarding the specification of the novelty components. Our specification (HPC 1) provides a nice balance between the flexibility required by the novelty term and the stability of the results.

\clearpage  
\FloatBarrier
\section{Additional Figures}
\subsection{Meat variety dataset analysis}

\begin{figure}[ht!]
    \centering
    \includegraphics[scale=.8]{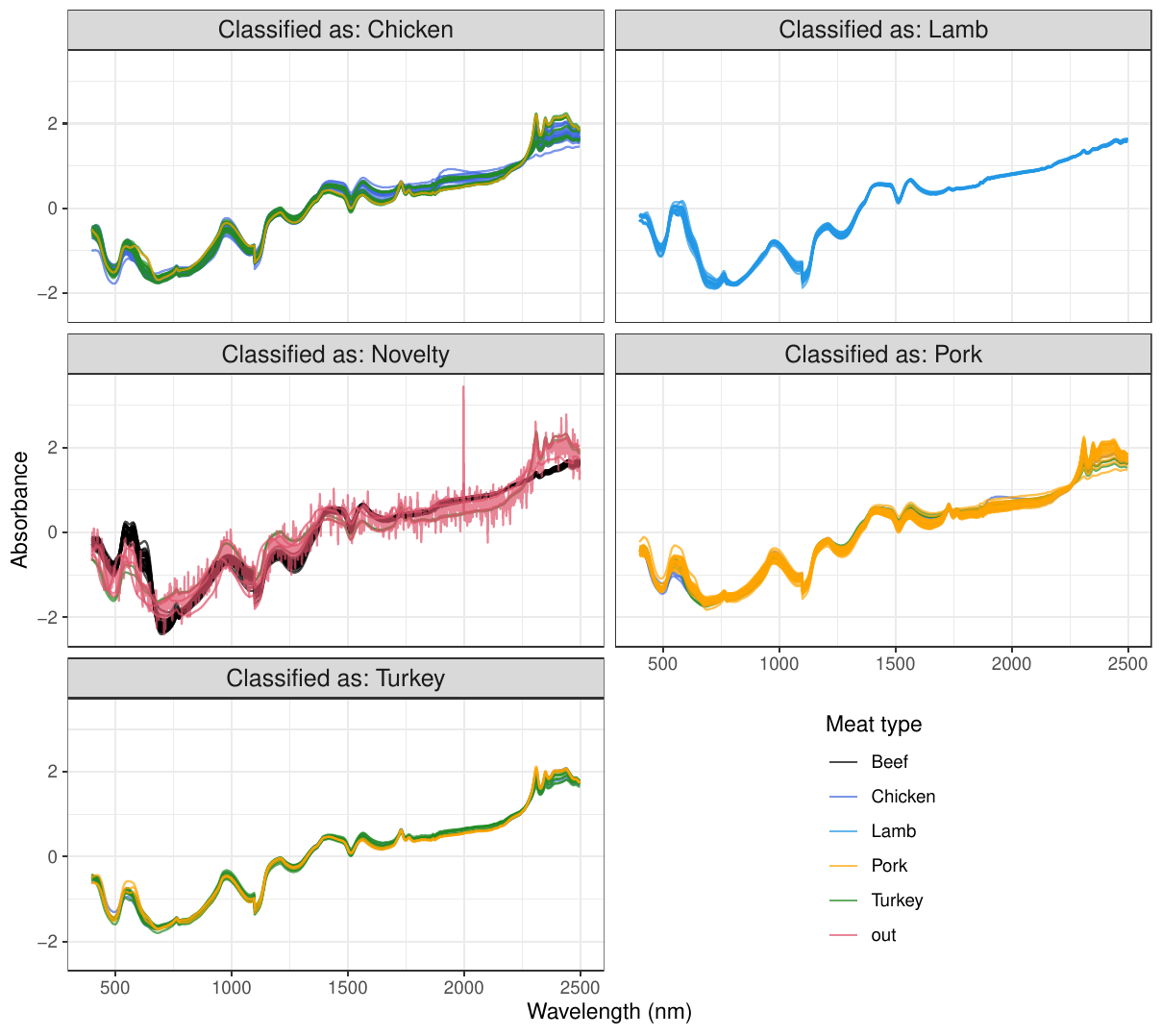}
    \caption{Summary of the resulting classification given by the model considered in Section 6.3. Each functional object is colored according to its correct data type, while each panel contains the meat spectra that Brand classifies together.}
    \label{fig:summary}
\end{figure}

\begin{figure}[ht!]
    \centering
    \includegraphics[scale=.6]{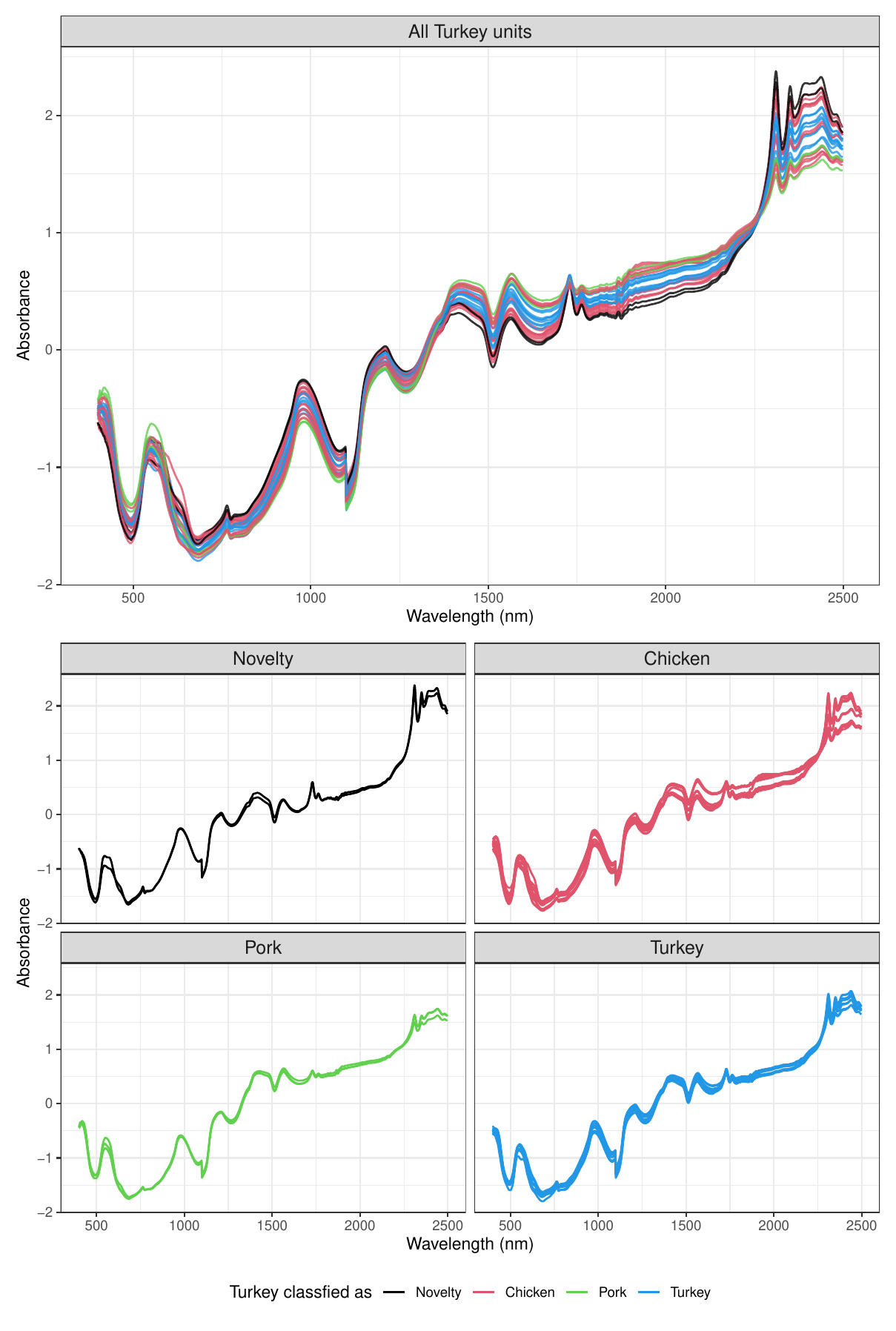}
    \caption{Classification of the turkey units according to Brand. The different colors indicate the estimated labels. The top panel shows all the functional data, while the bottom four panels break down the data objects in the four classes, highlighting their different characteristics.}
    \label{fig:summary}
\end{figure}

\begin{figure}
    \centering
    \includegraphics[width=\linewidth]{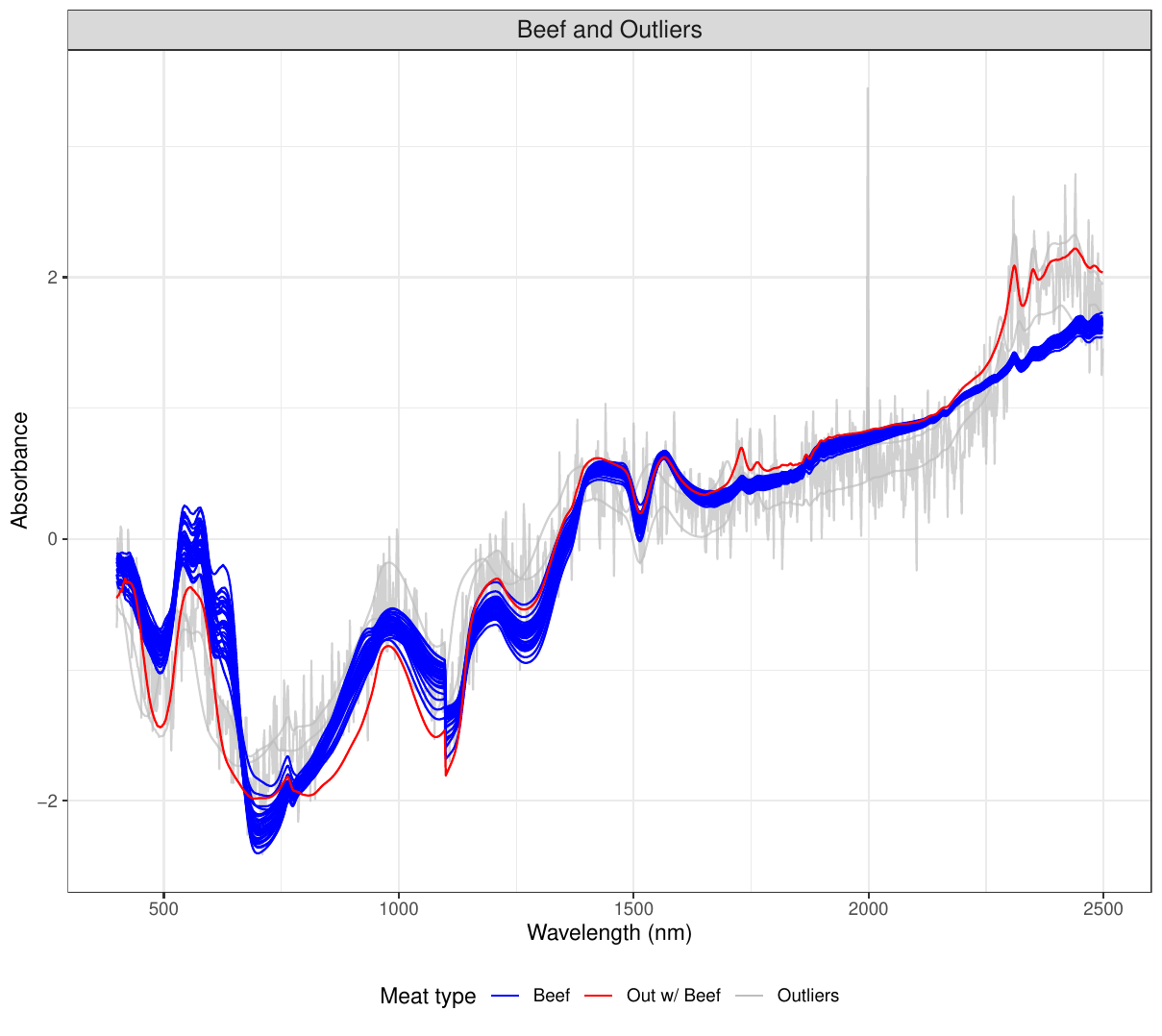}
    \caption{Functional objects classified as Novelties, colored according to their assigned cluster (Beef, in blue and Outliers, in gray). We highlighted in red the outlying functional data that is assigned to the Beef cluster.}
    \label{fig:Beef and outliers}
\end{figure}

\FloatBarrier

\newpage

\section{Additional Table for the Simulation Study of Section 6.1}

\begin{table*}[ht!]
\caption{ Accuracy  on  the  known classes, Adjusted Rand Index and Novelty predictive value metrics  for $B= 100$ repetitions  of  the  simulated experiment, varying data contamination level and test set sample size. Standard errors are reported in parentheses.}
\begin{footnotesize}
\label{tab:sim_results}
\centering
\begin{tabular}{lllllll}
    \toprule
  & \multicolumn{3}{c}{Label noise = False} & \multicolumn{3}{c}{Label noise = True} \\  
\cmidrule(lr){2-4}\cmidrule(lr){5-7}
 & Accuracy & ARI & Precision & Accuracy & ARI & Precision\\ 
 Novelty Size = Not small &  &  &  &  &  & \\
 \cmidrule(lr){1-1}
$AMDA$ & 0.999 & 0.932 & 0.998 & 0.562 & 0.814 & 0.519 \\ 
   & (0.002) & (0.014) & (0.004) & (0.223) & (0.03) & (0.215) \vspace*{.25cm}\\ 
  $RAEDDA$ & 0.966 & 0.885 & 0.934 & 0.996 & 0.924 & 0.992 \\ 
 & (0.029) & (0.041) & (0.051) & (0.003) & (0.003) & (0.005) \vspace*{.25cm}\\ 
  $Brand$ & 0.998 & 0.93 & 0.997 & 0.309 & 0.931 & 0.4 \\ 
  $(\eta_{MCD}=1,\lambda_{Tr}=1,000)$& (0.001) & (0.005) & (0.002) & (0.001) & (0.002) & (0.001) \vspace*{.25cm}\\ 
  $Brand$ & 0.995 & 0.927 & 0.992 & 0.465 & 0.931 & 0.481 \\ 
  $(\eta_{MCD}=1,\lambda_{Tr}=10)$ & (0.001) & (0.005) & (0.002) & (0.206) & (0.01) & (0.13) \vspace*{.25cm}\\ 
  $Brand$ & 0.99 & 0.928 & 0.985 & 0.997 & 0.93 & 0.995 \\ 
  $(\eta_{MCD}=0.75,\lambda_{Tr}=1,000)$ & (0.007) & (0.007) & (0.013) & (0.004) & (0.008) & (0.009) \vspace*{.25cm}\\ 
  $Brand$ & 0.997 & 0.928 & 0.994 & 0.994 & 0.927 & 0.99 \\  
  $(\eta_{MCD}=0.75,\lambda_{Tr}=10)$ & (0.001) & (0.005) & (0.002) & (0.001) & (0.007) & (0.001) \vspace*{.5cm}\\ 
   \midrule
    Novelty Size = Small &  &  &  &  &  & \\ 
     \cmidrule(lr){1-1}
$AMDA$ & 0.999 & 0.986 & 0.996 & 0.69 & 0.905 & 0.459 \\ 
  & (0.002) & (0.004) & (0.011) & (0.29) & (0.026) & (0.353) \vspace*{.25cm}\\ 
  $RAEDDA$ & 0.969 & 0.947 & 0.855 & 0.996 & 0.981 & 0.977 \\ 
   & (0.015) & (0.019) & (0.06) & (0.003) & (0.003) & (0.014) \vspace*{.25cm}\\ 
  $Brand$ & 0.998 & 0.986 & 0.994 & 0.413 & 0.986 & 0.23 \\ 
  $(\eta_{MCD}=1,\lambda_{Tr}=1,000)$ & ($<$ 0.01) & (0.001) & (0.002) & (0.003) & (0.003) & (0.001) \vspace*{.25cm}\\ 
  $Brand$ &  0.999 & 0.986 & 0.996 & 0.504 & 0.986 & 0.289  \\ 
  $(\eta_{MCD}=1,\lambda_{Tr}=10)$ & (0.01) & (0.001) & (0.002) & (0.222) & (0.003) & (0.285) \vspace*{.25cm}\\ 
  $Brand$ & 0.992 & 0.985 & 0.979 & 0.996 & 0.985 & 0.986 \\ 
  $(\eta_{MCD}=0.75,\lambda_{Tr}=1,000)$ & (0.022) & (0.006) & (0.056) & (0.001) & (0.002) & (0.007) \vspace*{.25cm}\\ 
  $Brand$ & 0.999 & 0.986 & 0.995 & 0.999 & 0.986 & 0.997 \\ 
  $(\eta_{MCD}=0.75,\lambda_{Tr}=10)$ & ($<$ 0.01) & ($<$ 0.01) & (0.002) & (0.001) & (0.001) & (0.004) \\ 
   \bottomrule
\end{tabular}
   \end{footnotesize}
\end{table*}


\end{document}